\documentclass[aps,prx,reprint,superscriptaddress,nofootinbib,longbibliography,floatfix]{revtex4-2}
\usepackage[T1]{fontenc}
\usepackage{lmodern}
\usepackage{microtype}
\usepackage{amsmath,amssymb,amsfonts,mathtools,bm,mathrsfs}
\usepackage{amsthm}
\usepackage{graphicx}
\usepackage{xcolor}
\usepackage{booktabs,multirow,longtable}
\usepackage{enumitem}
\usepackage{hyperref}
\usepackage{bookmark}

\hypersetup{pdftitle={Non-Hermitian Statistical Geometry},pdfsubject={Reviewed manuscript and supplemental material},colorlinks=true,linkcolor=blue!45!black,citecolor=blue!45!black,urlcolor=blue!45!black}

\theoremstyle{definition}

\providecommand{\func}[1]{\operatorname{#1}}

\begin{document}

\title{Non-Hermitian Statistical Geometry}

\author{Xi Hu}
\email{202421101046@mail.bnu.edu.cn}
\affiliation{School of Physics and Astronomy,
Beijing Normal University, Beijing 100875, China}

\author{Yan-Qiong Ma}
\email{202631101036@mail.bnu.edu.cn}
\affiliation{School of Physics and Astronomy,
Beijing Normal University, Beijing 100875, China}

\author{Su-Peng Kou}
\email[Corresponding author: ]{spkou@bnu.edu.cn}
\affiliation{School of Physics and Astronomy,
Beijing Normal University, Beijing 100875, China}

\begin{abstract}
Particle exchange in two dimensions gives rise to quantum statistics beyond
the boson--fermion dichotomy, with central roles in fractional quantum Hall
physics, topological quantum computation, and strongly correlated quantum
matter. In conventional topological phases, microscopic exchange geometry is
compressed into rigid braid and fusion data. We develop a theory of
non-Hermitian statistical geometry in which relaxing Hermiticity softens
these exchange data into a larger complex space of spectral and
relative-geometric degrees of freedom. For two adjacent braid generators, a
single projective exchange spectrum is insufficient: their relative
eigenline geometry supplies an additional gauge-invariant modulus. On the generic irreducible two-channel branch, braid consistency locks these coordinates through an exact spectrum--geometry relation. Using Riesz projectors and Kato transport, we realize this geometry in a five-arm Kitaev--Majorana network and test spectral and relative-geometric responses in an independent two-dimensional four-vortex Kitaev model.
Finite-time dynamics tests the exchange operation, microscopic fusion sewing tests its composition across fusion descriptions, and matched-spectrum controls isolate the role of relative geometry. Nonzero projective Kato curvature shows that braid-consistent, fusion-coherent exchange can depend on the control path. Non-Hermiticity
thus exposes exchange geometry, braid consistency re-locks it, fusion
coherence further constrains it, and projective flatness determines when
geometric exchange becomes intrinsically topological.
\end{abstract}

\maketitle

\section{Introduction}

\label{sec:intro}

Particle exchange in two dimensions connects topology to quantum mechanics. For ordinary bosons and
fermions, exchange is represented by one-dimensional unitary phases, whereas
for anyons it can generate nontrivial Abelian or non-Abelian representations
of the braid group~\cite%
{LeinaasMyrheim1977,Wilczek1982,Arovas1984,MooreRead1991,Kitaev2003,Nayak2008,FreedmanKitaevLarsenWang2003}%
. In a topologically ordered phase, the exchange operation depends on the braid class when the anyons remain well separated and the gap stays open. This path independence is the foundation of topological quantum computation and the categorical description of anyon statistics. For a unitary modular tensor category, the
surviving information is encoded algebraically by fusion multiplicities, $F$%
-moves, $R$-symbols, and their braid, Pentagon, and Hexagon consistency
equations~\cite{BakalovKirillov2001,TuraevBook,ENO2005}.

This algebraic description also has a geometric realization. Conventional
anyon theory already contains configuration spaces, vector bundles of
degenerate states, Berry--Wilczek--Zee connections, conformal-block
monodromies, and Chern--Simons geometry~\cite%
{Berry1984,WilczekZee1984,Witten1989,Nayak2008}. What topological protection
does is compress much of the continuous exchange geometry into robust
algebraic data. The exchange depends on the homotopy class of the particle path~\cite{FadellNeuwirth1962}. 

Non-Hermitian band topology has established the concepts of point and line gaps, exceptional points, non-Bloch bulk--boundary correspondence, and the skin effect~\cite{Ashida2020,Gong2018,YaoWang2018,Kunst2018,Kawabata2019,Bergholtz2021}. The
familiar algebraic structure of quantum statistics then raises a sharper
question once the exchange dynamics is allowed to become non-Hermitian:
What is quantum statistics for a non-Hermitian microscopic Hamiltonian?
Equivalently, one may ask what becomes of the rigidity of topological
statistics when Hermiticity is relaxed.

Garrison and Wright showed that cyclic non-Hermitian evolution replaces a unit-modulus geometric phase with a complex multiplier~\cite{GarrisonWright1988,Dattoli1990,ChuWuLayton1989,CuiZheng2012,Brody2014,Singhal2023}, $e^{i\theta}\to e^{\chi+i\theta}$. The
real exponent $\chi $ may encode amplification, while $\theta $ retains the
compact phase information. This was the first clear indication that
non-Hermiticity can enlarge geometric transport beyond unit-modulus phases.
A more direct step toward NH anyon statistics was taken by Zhao \emph{et al.}%
, who studied NH generalizations of Majorana zero modes and found
continuously tunable braiding Berry phases~\cite{Zhao2021}. Related proposals use reciprocal circuits~\cite{Ezawa2019} or environment-assisted Majorana migration and braiding through symmetry breaking or the non-Hermitian skin effect~\cite{Tang2025}.  The result is highly suggestive, but it also exposes the next
question: does a tunable single-exchange phase automatically define a
consistent family of adjacent exchanges on the same logical space? In
general, it does not.

A complementary direction concerns nonunitary categorical statistics. Sanno \emph{et al.} proposed a dissipative Majorana platform for
Yang--Lee anyons, which share the Fibonacci fusion rule but possess
nonunitary non-Abelian braiding data~\cite{Sanno2022}.  Yang--Lee specifies discrete fusion and braiding data; continuous non-Hermitian deformation instead varies microscopic exchange geometry. A fixed braided fusion category imposes additional constraints on that continuous geometry.

We connect complex exchange spectra, relative eigenline geometry, and fusion constraints in a common framework.  Non-Hermitian evolution enriches the exchange
response by allowing relative channel amplification alongside relative
phase. When neighboring exchanges act on the same protected space, their relative eigenline geometry is a statistical observable. This raises a physical question: how do exchange data
extend continuously away from a unitary topological reference, and which
relations govern their assembly into consistent quantum statistics?

We call this continuous extension of rigid unitary exchange data into complex spectral and relative-geometric coordinates non-Hermitian softening.
 On the generic irreducible two-channel
branch, the Artin relation fixes the relative eigenline geometry once the
projective exchange spectrum is specified, yielding exact
spectrum--geometry locking (SGL).

Fusion moves connect descriptions based on different fusion trees. The Pentagon equation makes reassociation consistent, and the Hexagon equations make braiding compatible with fusion. These conditions select categorical statistics within the softened exchange geometry. Fusion coherence and microscopic path independence are tested separately.

To extract statistical observables from a Hamiltonian, we use Kato transport of an isolated spectral subspace. We realize this structure both in a finite-arm Kitaev--Majorana
junction~\cite{Kitaev2001,Alicea2011} and in the two-dimensional chiral
Kitaev model. The five-arm network gives control over relative geometry, finite-time dynamics, fusion sewing, and control-path curvature. The two-dimensional model independently tests spectral and relative-geometric responses. 

 Section~\ref%
{sec:statistical_geometry} develops single-exchange spectra,
neighboring-generator geometry, spectrum--geometry locking, and fusion
consistency. Section~\ref{sec:Kato} constructs the Hamiltonian-to-statistics
map and the projective-flatness criterion. Section~\ref{sec:fivearm}
presents the five-arm realization, matched-spectrum controls, finite-time
dynamics, the independent two-dimensional test, fusion sewing, and the exact
curvature obstruction. Section~\ref{sec:interpretation} assembles the
physical implications of non-Hermitian softening, braid re-locking,
categorical coherence, and path dependence, while Sec.~\ref{sec:outlook}
concludes. Derivations, higher-rank extensions, numerical checks, and reproducibility details are given in the Supplemental Material~\cite{SupplementalMaterial}.

% ============================================================

\section{Statistical Geometry of Non-Hermitian Exchange}

\label{sec:statistical_geometry}
% ============================================================

\begin{figure*}[t]
\centering
\includegraphics[width=0.98\textwidth]{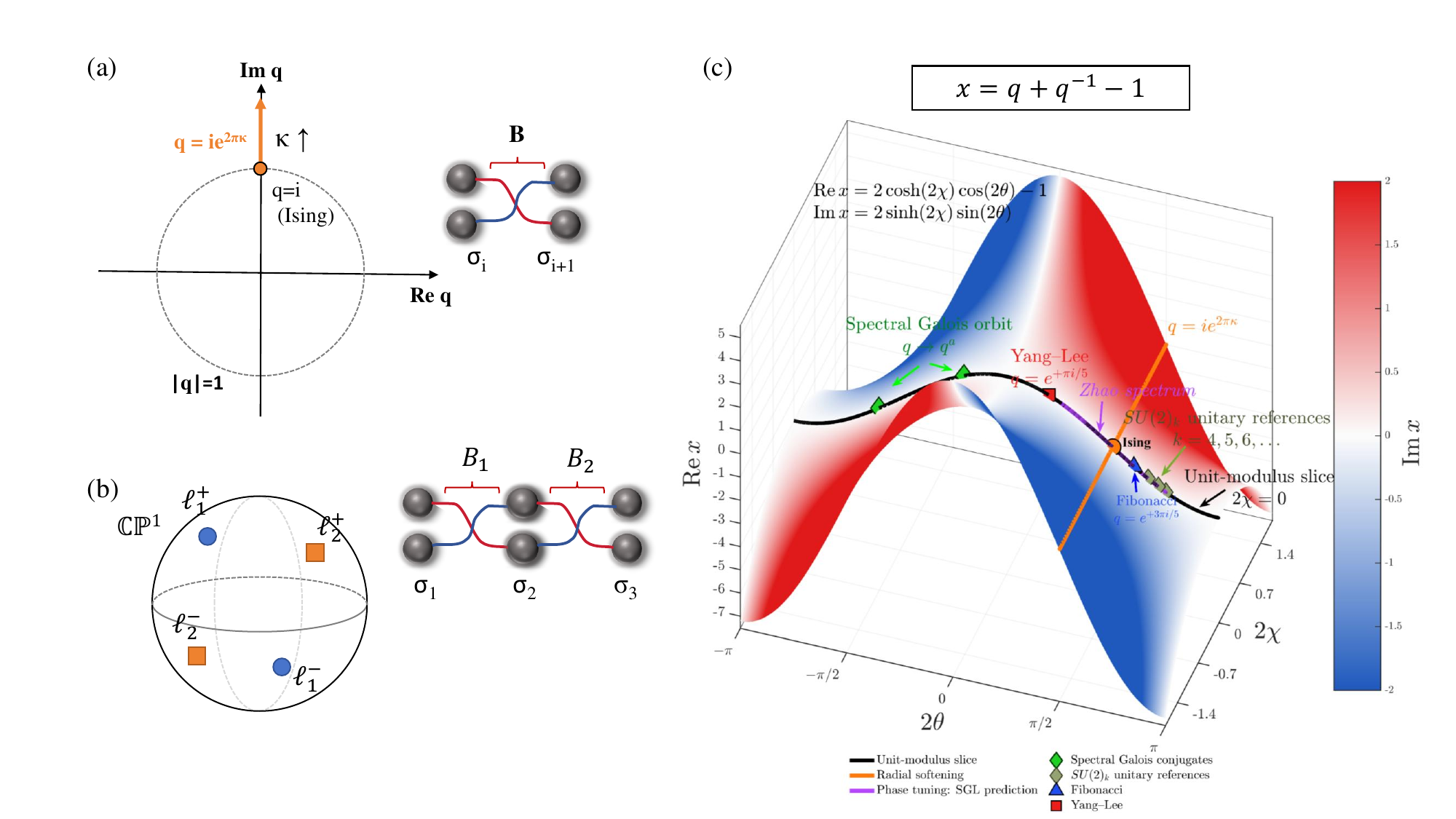}
\caption{\textbf{Non-Hermitian softening and spectrum--geometry locking.}
(a) The projective eigenvalue ratio $q=\lambda_+/\lambda_-$ characterizes
a single two-channel exchange. The radial continuation $q=ie^{2\pi\kappa}$
extends the Ising reference $q=i$ by changing relative channel amplification
at fixed relative phase.
(b) Two neighboring exchanges acting in a common logical space supply four
ordered right-eigenlines. Their cross ratio
$\mathfrak{x}=\operatorname{cr}(\ell_1^+,\ell_1^-;\ell_2^+,\ell_2^-)$
specifies the relative geometry; the sphere is schematic.
(c) On the generic irreducible two-channel branch, Artin consistency imposes
$\mathfrak{x}=q+q^{-1}-1$. For $q=e^{2\chi+2i\theta}$, the horizontal
coordinates are $2\theta$ and $2\chi$, the height is
$\operatorname{Re}\mathfrak{x}$, and color encodes
$\operatorname{Im}\mathfrak{x}$.
The black curve is the unit-modulus slice $\chi=0$; the orange curve is
the radial Ising branch $\theta=\pi/4$.
The purple segment shows the phase-tuned Zhao spectrum and its SGL-predicted relative geometry.
Green diamonds, together with the named Fibonacci and Yang--Lee markers,
represent the spectral Galois orbit $\{e^{ia\pi/5}:a=1,3,7,9\}$.
The positive-phase representatives are $q_{\mathrm{Fib}}=e^{3\pi i/5}$
and $q_{\mathrm{YL}}=e^{\pi i/5}$.
Olive diamonds show selected unitary $SU(2)_k$ references with $k\geq4$
and $q_k=e^{i\pi k/(k+2)}$; the $k=2,3$ coordinates are represented by
the Ising and Fibonacci markers.
The two sets distinguish spectral Galois conjugation from variation of the unitary fusion level.}
\label{fig:moduli}
\end{figure*}

The statistical structure unfolds naturally with particle number. Two anyons
already define a single-exchange projective spectrum. A third anyon introduces two neighboring braid generators on the same protected space, with a relative eigenline geometry. Different fusion trees give descriptions of the same many-anyon state space that must be connected consistently.Figure~\ref{fig:moduli} connects single-exchange softening to relative eigenline geometry and spectrum--geometry locking.

% ============================================================

\subsection{Single-exchange projective spectrum}

\label{subsec:statistical_moduli}
% ============================================================

A one-dimensional statistical space carries a nonzero exchange multiplier,
\begin{equation}
r=e^{\chi+i\theta}\in \mathbb{C}^{\times}.  \label{eq:abelian_multiplier}
\end{equation}
Its phase $\theta$ generalizes the conventional exchange phase, while $\chi$
describes relative amplification or attenuation.  In a multichannel space, neighboring exchanges also carry a gauge-invariant relative eigenline geometry.

We now consider a pair exchange acting on a two-channel protected or fusion
space $V\simeq \mathbb{C}^{2}$. For Ising fusion, the two local outputs are $%
1 $ and $\psi$; a coherent logical doublet is supplied by $V^{\sigma \sigma
\sigma}_{\sigma}$, or by four $\sigma$ anyons with fixed total charge.  Let $B\in \mathrm{GL}(2,\mathbb{C})$ be diagonalizable
with nonzero ordered eigenvalues $\lambda_\pm$. In the eigenbasis ordered as
$(-,+)$, and with a continuous square-root branch, we separate the common
scalar component from the relative channel action:
\begin{equation}
\begin{gathered} B=gD(q),\qquad D(q)= \begin{pmatrix} q^{-1/2}&0\\ 0&q^{1/2}
\end{pmatrix}, \\ q=\lambda_+/\lambda_-\in \mathbb{C}^{\times}.
\end{gathered}  \label{eq:projective_spectrum}
\end{equation}
Here $g\in \mathbb{C}^{\times}$ multiplies both channels equally, whereas $q$
survives projectivization and characterizes the relative exchange action.
We choose $g=\sqrt{\lambda_+\lambda_-}$, so that $\lambda_\pm=gq^{\pm1/2}$. In a general basis, $B=S[gD(q)]S^{-1}$. 

We write
\begin{equation}
q=e^{2\chi +2i\theta },\qquad D(q)=%
\begin{pmatrix}
e^{-\chi -i\theta } & 0 \\
0 & e^{\chi +i\theta }%
\end{pmatrix}
\label{eq:D-chi-theta}
\end{equation}%
to separate relative phase and amplification. The phase $2\theta =\arg q$ measures the relative
exchange phase, whereas $2\chi =\log |q|$ measures relative amplification. A
unitary exchange has $\chi =0$; the conventional Ising reference corresponds
to $\theta =\pi /4$, $q=i$. We define \emph{exchange-spectrum softening} as
the continuous extension of these relative channel weights away from their
unitary reference values into $\mathbb{C}^{\times }$. The coordinate $\chi$ measures its radial component. Braid and fusion relations determine how these single-exchange data combine into multigenerator statistics. Figure~\ref{fig:moduli}(a) illustrates this
single-exchange layer.

We can express the projective spectrum through its fusion channels.

Suppose $a\times a=c_{+}\oplus c_{-}$, and let $P_{c}$ project onto channel $%
c$. For the $SU(2)$ labels used below, we introduce
\begin{equation}
Q_{C}=\sum_{c}C_{2}(c)P_{c},\qquad C_{2}(j)=j(j+1),
\label{eq:casimir-channel-operator}
\end{equation}%
with the standard Casimir normalization~\cite%
{BakalovKirillov2001,Witten1989,ReshetikhinTuraev1991,KauffmanLins1994}.
Starting from a channel-diagonal reference exchange $T_{0}$, we parameterize
the relative channel dressing as
\begin{equation}
T(z)=T_{0}e^{zQ_{C}}.  \label{eq:casimir-dressing-main}
\end{equation}%
The Hamiltonian fixes the channel weights; the Casimir operator expresses their relative contribution to the spectrum. Channel $c$
acquires the factor $e^{zC_{2}(c)}$, leading to
\begin{equation}
q(z)=q_{0}e^{z\Delta C_{2}},\qquad \Delta C_{2}=C_{2}(c_{+})-C_{2}(c_{-}).
\label{eq:casimir-q-main}
\end{equation}%
With $z=\xi +i\delta \theta $ we have
\begin{equation}
\begin{aligned} \ln|q/q_0|&=\xi \Delta C_2,\\ \arg(q/q_0)&=\delta \theta \,
\Delta C_2 \quad(\mathrm{mod}\ 2\pi). \end{aligned}
\label{eq:casimir-phase-modulus-main}
\end{equation}%
These equations relate the channel coordinate $\xi$ to the spectral coordinate $\chi=\tfrac12\ln|q|$ through $\Delta C_2$ and $q_0$.

For the fundamental two-channel sector of $SU(2)_{k}$, $k\geq 2$, we have
\begin{equation}
\tfrac{1}{2}\times \tfrac{1}{2}=0\oplus 1,\qquad Q_{C}=0P_{0}+2P_{1},
\label{eq:add-su2-casimir}
\end{equation}%
and choosing the orientation and channel ordering that give $q=i$ at $k=2$,
\begin{equation}
q_{0}^{(k)}=e^{i\pi k/(k+2)},\qquad q^{(k)}(z)=q_{0}^{(k)}e^{2z}.
\label{eq:add-su2-channel-law}
\end{equation}%
The level $k$ labels the discrete reference fusion theory, whereas $z$
describes the relative complex channel dressing. Neighboring-exchange geometry and braid constraints enter next.

For the Ising sector, $\sigma \times \sigma =1+\psi $, with labels $%
1\leftrightarrow j=0$, $\sigma \leftrightarrow j=1/2$, and $\psi
\leftrightarrow j=1$. Choosing $q=\lambda _{\psi }/\lambda _{1},$ we have
\begin{equation}
\Delta C_{2}=C_{2}(\psi )-C_{2}(1)=2,\qquad q(z)=ie^{2z}.
\label{eq:ising-casimir-q}
\end{equation}%
For a radial continuation of this reference, $z=\xi =\chi $. More generally,
channel weights $T=T_{0}\exp (f_{1}P_{1}+f_{\psi }P_{\psi })$ give $%
z=(f_{\psi }-f_{1})/2$.

%% Radial continuation from the Ising reference.
We use the Ising sector ($k=2$) as the reference for the Majorana exchange protocols. Starting from its
unitary projective spectrum $q=i$, we consider the radial
non-Hermitian continuation
\begin{equation}
q(\kappa)=ie^{2\pi\kappa},
\qquad \kappa\in\mathbb{R}.
\label{eq:radial_q}
\end{equation}
At $\kappa=0$, this trajectory reduces to the unitary Ising
point. For finite $\kappa$, the corresponding projective
eigenvalues are
\begin{equation}
\widetilde{\lambda}_{\pm}
=q^{\pm1/2}
=e^{\pm\pi\kappa}e^{\pm i\pi/4},
\label{eq:soft-channel-eigenvalues}
\end{equation}
Their modulus ratio is
\begin{equation}
|q|
=\frac{|\widetilde{\lambda}_{+}|}
{|\widetilde{\lambda}_{-}|}
=e^{2\pi\kappa}.
\label{eq:soft-channel-modulus}
\end{equation}
The parameter $\kappa$ changes relative channel amplification at fixed exchange phase.
We derive this spectrum from five-arm Riesz--Kato transport in Sec.~\ref{subsec:fivearm-architecture}.

% ============================================================

\subsection{Neighboring exchanges and spectrum--geometry locking}

\label{subsec:SGL}
% ============================================================

 A third particle introduces two neighboring braid generators on the same encoded space, with a relative eigenline geometry. For adjacent exchanges of identical anyons in the
same two-channel sector, the braid representation assigns the generators a
common ordered projective spectrum.

By choosing $B_1=D(q)$ as the reference frame, we write
\begin{equation}
B_2=G_{\mathrm{br}}D(q)G_{\mathrm{br}}^{-1}, \qquad G_{\mathrm{br}}\in%
\mathrm{PGL}(2,\mathbb{C}).  \label{eq:relative_braid_frame}
\end{equation}
Here $G_{\mathrm{br}}$ relates the eigenframes of neighboring exchanges within one fusion chart. The categorical $F$ move introduced in Sec.~\ref%
{subsec:global_coherence} instead connects different fusion-tree charts of
the same state space.

We denote the ordered right-eigenlines by $\ell_1^\pm,\ell_2^\pm \in \mathbb{%
CP}^{1}$, with labels inherited from the exchange eigenvalues. A
simultaneous change of logical basis acts on all four lines through the same
projective transformation and preserves their relative configuration.
Let $\mathrm{Conf}_4(\mathbb{CP}^1)$ denote ordered configurations of four distinct lines. Identifying configurations related by the same projective basis transformation gives
\begin{equation}
\frac{\mathrm{Conf}_{4}(\mathbb{CP}^{1})} {\mathrm{PGL}(2,\mathbb{C})}
\simeq \mathcal{M}_{0,4},  \label{eq:M04}
\end{equation}
with coordinate
\begin{equation}
\mathfrak{x} = \func{cr} (\ell_1^+,\ell_1^-;\ell_2^+,\ell_2^-).
\label{eq:cross_ratio}
\end{equation}
$\mathcal{M}_{0,4}$ describes eigenlines in the logical space; particle trajectories lie in real-space configuration space. This distinction is illustrated in Fig.~\ref{fig:moduli}(b).

Before imposing braid consistency, $q$ and $\mathfrak{x}$ are independent
kinematic coordinates. The generic candidate space for a common-spectrum
pair is therefore
\begin{equation}
\mathcal{M}_{\mathrm{stat}}^{\mathrm{cand}} = (\mathbb{C}_{q}^{\times}%
\setminus \{1\}) \times \mathcal{M}_{0,4}.
\label{eq:candidate_statistical_moduli}
\end{equation}
The coordinates $q$ and $\mathfrak{x}$ describe the exchange spectrum and relative eigenline geometry, respectively. Non-Hermitian softening thus enlarges the exchange-data
space; braid consistency determines the allowed submanifold inside this
space.

The elementary generators of the three-particle braid group obey
\begin{equation}
B_1B_2B_1=B_2B_1B_2.  \label{eq:Artin}
\end{equation}
 Microscopic path independence additionally requires projective Kato flatness, defined in Sec.~\ref{sec:Kato}. On the
generic irreducible two-channel branch, Artin consistency fixes the relative
geometry once the projective spectrum is specified.

Using the residual projective freedom that preserves the reference
eigenlines of $B_1$, we choose
\begin{equation}
G_{\mathrm{br}} =
\begin{pmatrix}
c & s \\
s & -c%
\end{pmatrix}%
, \qquad c^2+s^2=1.  \label{eq:Gbr_symmetric}
\end{equation}
The normalization is algebraic and admits complex $c,s$. Substitution into
Eq.~\eqref{eq:Artin} gives
\begin{equation}
c^2 = -\frac{q}{(q-1)^2} = \frac{1}{2-q-q^{-1}}.  \label{eq:c2_lock}
\end{equation}
The resulting generators can be written explicitly as
\begin{equation}
B_1 =
\begin{pmatrix}
q^{-1/2} & 0 \\
0 & q^{1/2}%
\end{pmatrix}%
, \qquad B_2 =
\begin{pmatrix}
d_{11} & d_{12} \\
d_{12} & d_{22}%
\end{pmatrix}%
,  \label{eq:add-explicit-braids}
\end{equation}
where
\begin{equation}
\begin{aligned} d_{11}&=c^2q^{-1/2}+s^2q^{1/2},\\
d_{22}&=s^2q^{-1/2}+c^2q^{1/2},\\ d_{12}&=cs(q^{-1/2}-q^{1/2}),\\
s^2&=\frac{q^2-q+1}{(q-1)^2}. \end{aligned}
\label{eq:add-explicit-braid-entries}
\end{equation}
The square-root branches continue from $q=i$, where $c=s=1/\sqrt2$ and $q^{1/2}=e^{i\pi/4}$. In this convention,
\begin{equation}
B_2(i) = \frac{1}{\sqrt{2}}
\begin{pmatrix}
1 & -i \\
-i & 1%
\end{pmatrix}%
.  \label{eq:add-b2-ising}
\end{equation}
We compare a microscopic pair with this Artin representative using one simultaneous change of logical basis and separate scalar normalizations of the generators.

For the ordering in Eq.~\eqref{eq:cross_ratio}, the cross ratio satisfies
\begin{equation*}
\mathfrak{x}=-\frac{s^2}{c^2}.
\end{equation*}
Together with $s^2=1-c^2$, this gives the exact spectrum--geometry locking
relation
\begin{equation}
\boxed{\mathfrak{x}=q+q^{-1}-1}.  \label{eq:SGL}
\end{equation}
The Artin-consistent locus is therefore
\begin{equation}
\Gamma_{\mathrm{SGL}} = \{(q,\mathfrak{x})\in \mathcal{M}_{\mathrm{stat}}^{%
\mathrm{cand}} : \mathfrak{x}=q+q^{-1}-1\}.  \label{eq:SGL_locus}
\end{equation}

These expressions apply on the generic distinct-line branch, where
$q\neq1$ and $q^{2}-q+1\neq0$.
The first condition keeps the projective spectrum nondegenerate, while the
second keeps the four eigenlines distinct. The exact matrix Artin relation
uses compatible $\mathrm{GL}(2,\mathbb{C})$ lifts. The coordinates $q,%
\mathfrak{x}$ and the locked locus remain invariant under individual nonzero
scalar rescalings and simultaneous changes of logical basis. Building on the known rigidity of low-dimensional braid representations~\cite{TubaWenzl2001}, we recast this structure as an exact locking relation between separately accessible spectral and relative-geometric observables. 

Before imposing Artin consistency, $q$ and $\mathfrak{x}$ are independent
coordinates; the resulting locked locus is shown in
Fig.~\ref{fig:moduli}(c).

To test SGL microscopically, we extract the spectrum from exchange eigenvalues and the relative geometry from neighboring eigenlines. We evaluate both in the same logical space.

For the canonical radial Ising branch of Eq.~\eqref{eq:radial_q}, the
locking law gives
\begin{equation}
\mathfrak{x}(\kappa )=-1+2i\sinh (2\pi \kappa ).  \label{eq:radial_x}
\end{equation}%
Its real and imaginary parts are
\begin{equation}
\mathrm{Re}\, \mathfrak{x}=-1,\qquad \mathrm{Im}\, \mathfrak{x}=2\sinh (2\pi
\kappa ),
\end{equation}%
and near $\kappa=0$,
\begin{equation}
\mathfrak{x}(\kappa )=-1+4\pi i\kappa +O(\kappa ^{3}).
\end{equation}%
The single-exchange spectrum $q=ie^{2\pi\kappa}$ fixes the relative eigenline geometry through Artin consistency.

% ============================================================

\subsection{Geometry of the locked manifold and unitary embedding}

\label{subsec:unitary_embedding}
% ============================================================

\subsubsection{Shifted Joukowski geometry}

The locking law may be written as
\begin{equation}
\mathfrak{x}+1=q+q^{-1}.  \label{eq:Joukowski}
\end{equation}%
It is therefore a shifted Joukowski map from the projective exchange
spectrum to the relative eigenline geometry. The spectra $q$ and $q^{-1}$ have the same relative geometry because $\mathfrak{x}(q)=\mathfrak{x}(q^{-1})$. The inverse map is
\begin{equation}
q_{\pm }(\mathfrak{x})=\frac{\mathfrak{x}+1\pm \sqrt{(\mathfrak{x}-1)(%
\mathfrak{x}+3)}}{2},  \label{eq:Joukowski_inverse}
\end{equation}%
with branch points
\begin{equation}
\mathfrak{x}=1,-3,\qquad q=1,-1.  \label{eq:Joukowski_branch}
\end{equation}%
The point $q=1$ also corresponds to projective spectral degeneracy and lies
outside the generic distinct-spectrum sector considered above.

%% NEW: directly connect to Fig. 1(c)
For $q=e^{2\chi+2i\theta}$, SGL gives
\begin{equation}
\begin{aligned} \mathrm{Re}\, \mathfrak{x} &=
2\cosh(2\chi)\cos(2\theta)-1,\\ \mathrm{Im}\, \mathfrak{x} &=
2\sinh(2\chi)\sin(2\theta). \end{aligned}  \label{eq:SGL-parametric-surface}
\end{equation}
Equation~\eqref{eq:SGL-parametric-surface} gives the surface in Fig.~\ref{fig:moduli}(c). The unit-modulus spectral slice is $\chi=0$.
Within this slice, a common positive-definite metric identifies the
simultaneously unitarizable sector, as discussed below.
The radial Ising branch corresponds to $\theta=\pi/4$ and $\chi=\pi\kappa$.

 The SGL relation determines the relative geometry associated with
a projective spectrum, while identifying $q\leftrightarrow q^{-1}$ with a
physical orientation or chirality reversal requires additional microscopic
information.

\subsubsection{Unit-modulus spectra and a common positive metric}

On the unit-modulus spectral slice, $q=e^{i\vartheta}$. Equation~\eqref{eq:SGL} then becomes
\begin{equation}
\mathfrak{x}=2\cos \vartheta-1,  \label{eq:unit_circle_SGL}
\end{equation}
whose image has closure $-3\leq \mathfrak{x}\leq1$.

 The generators are simultaneously unitarizable if their projective representatives $\widetilde B_i$ preserve a common positive-definite Hermitian metric $h$,
\begin{equation}
\widetilde B_i^\dagger h\, \widetilde B_i=h, \qquad h>0.
\label{eq:common_metric}
\end{equation}
This reduces the projective structure group from $PGL(2,\mathbb{C})$ to a
compact form. 

On the generic locked unit-circle branch, the positive-metric condition is
equivalent to $\mathfrak{x}<0$, so its image is $[-3,0)$. Indeed, $%
c^2=1/[4\sin^2(\vartheta/2)]$ and $s^2=1-c^2$ can both be positive exactly
when $\sin^2(\vartheta/2)>1/4$. The remaining part $(0,1)$ of the
unit-modulus image lies outside this common-positive-metric sector.

We distinguish unit-modulus spectra, simultaneous unitarizability, braid--fusion coherence, and projective path independence.
Projective curvature diagnoses path dependence for both Hermitian and
non-Hermitian connections. The topological limit combines braid--fusion
coherence and flatness on the admissible control domain with locality and
protection. Spectral softening describes channel weights; control-path curvature describes path dependence.

\subsubsection{Conventional anyons as a constrained locus}

The unitary \(SU(2)_k\) sequence provides a reference. For the
fundamental two-channel sector ($k\geq2$)~\cite%
{Kitaev2006,Nayak2008,KauffmanLins1994,BakalovKirillov2001,
RowellStongWang2009}, the projective braid ratio may be written, up to
inversion, as
\begin{equation}
q_k=e^{\pm i\pi k/(k+2)},  \label{eq:q_SU2k}
\end{equation}
while the corresponding quantum dimension is
\begin{equation}
d_k=2\cos \frac{\pi}{k+2}.  \label{eq:d_SU2k}
\end{equation}
Substituting Eq.~\eqref{eq:q_SU2k} into the SGL relation gives
\begin{equation}
\mathfrak{x}_k=1-d_k^2.  \label{eq:x_SU2k}
\end{equation}
The relative eigenline geometry on this coherent unitary sequence is therefore fixed by the quantum dimension.

Combining this reference sequence with the channel law in Eq.~%
\eqref{eq:add-su2-channel-law} gives, on the Artin-consistent branch,
\begin{equation}
\mathfrak{x}^{(k)}(z) = q_0^{(k)}e^{2z} + [q_0^{(k)}]^{-1}e^{-2z} -1.
\label{eq:add-su2-sgl-continuation}
\end{equation}
At $z=0$ this expression gives $1-d_k^2$; at $k=2$ and $z=\pi \kappa$ it
gives the radial Ising trajectory. 

For example,
\begin{equation}
k=2:\qquad d_{2}=\sqrt{2},\qquad \mathfrak{x}_{2}=-1,
\end{equation}%
which gives the Ising value, while
\begin{equation}
k=3:\qquad d_{3}=\varphi ,\qquad \mathfrak{x}_{3}=-\varphi
\end{equation}%
gives the Fibonacci two-channel value, where $\varphi=(1+\sqrt5)/2$. As $k\rightarrow
\infty $,
\begin{equation}
d_{k}\rightarrow 2,\qquad \mathfrak{x}_{k}\rightarrow -3.
\end{equation}

Figure~\ref{fig:moduli}(c) distinguishes this level-dependent reference
sequence from spectral Galois conjugation.
For $\zeta_{10}=e^{i\pi/5}$, the automorphisms
$\zeta_{10}\mapsto\zeta_{10}^{\,a}$, $a\in\{1,3,7,9\}$, generate the
spectral orbit
\begin{equation}
\mathcal O_{10}=\{e^{\pm i\pi/5},e^{\pm3i\pi/5}\}.
\label{eq:spectral-galois-orbit}
\end{equation}
Its SGL image consists of two relative-geometric values,
\begin{equation}
\begin{aligned}
q=e^{\pm i\pi/5}&:\quad \mathfrak{x}=\varphi^{-1},\\
q=e^{\pm3i\pi/5}&:\quad \mathfrak{x}=-\varphi,
\end{aligned}
\label{eq:spectral-galois-image}
\end{equation}

The inverse spectral pairs share the same $\mathfrak{x}$ because the
locking map is invariant under $q\mapsto q^{-1}$.
The positive-phase Fibonacci representative coincides with the $k=3$
reference in these two-channel projective coordinates.
Galois conjugation changes the algebraic embedding of fixed cyclotomic spectral data; varying $k$ changes the unitary fusion theory.
The coordinate coincidence refers to this two-channel sector; the full fusion and braiding data specify each category.

 Fusion equations select the categorically compatible points on the continuation in Eq.~\eqref{eq:add-su2-sgl-continuation}.

% ============================================================

\subsection{From local braid consistency to global coherence}

\label{subsec:global_coherence}
% ============================================================

Spectrum--geometry locking solves a local multi-generator problem: it tells
us how neighboring braid generators may sit relative to one another within a
fixed fusion description. A many-anyon theory also connects different fusion trees describing the same physical space.

 The $F$ move connects fusion descriptions. The Pentagon equation makes reassociation consistent, and the Hexagon equations make exchange compatible with fusion.

\subsubsection{Fusion-tree sewing}

Let $\mathcal{H}_L$ and $\mathcal{H}_R$ denote two fusion-tree descriptions
of the same multi-anyon state space. The reassociation map
\begin{equation}
F:\mathcal{H}_L\longrightarrow \mathcal{H}_R  \label{eq:F_sewing}
\end{equation}
changes the fusion frame without exchanging particles.

$G_{\mathrm{br}}$ relates neighboring exchange eigenframes within one fusion chart; $F$ connects different fusion charts.

The same physical braid must transform covariantly between the two fusion
frames,
\begin{equation}
B_i^{(R)} = F B_i^{(L)}F^{-1},  \label{eq:fusion_covariance}
\end{equation}
up to any explicitly separated central normalization. Equation~\eqref{eq:fusion_covariance} expresses the same braid in both fusion charts.

\subsubsection{Pentagon coherence: path independence in fusion-tree space}

Different sequences of reassociations can connect the same initial and final
fusion trees. 
For four anyons, the two sequences of reassociations around the five fusion trees satisfy
\begin{equation}
P_1=P_2,  \label{eq:pentagon_paths}
\end{equation}
or equivalently the closed reassociation holonomy obeys
\begin{equation}
W_{\mathrm{assoc}}=I.  \label{eq:pentagon_flatness}
\end{equation}

Pentagon coherence makes the identification between fusion trees
independent of the reassociation path.

Pentagon coherence concerns the sequence of fusion reassociations; projective Kato flatness concerns smooth deformations of the Hamiltonian control path. The two conditions are independent.

\subsubsection{Hexagon coherence: compatibility of braiding and fusion}

The Hexagon equations make exchange consistent with a change of fusion description: exchanging before reassociation gives the same operation as reassociating first and then applying the corresponding exchange.

For the fixed Ising fusion data and our projective conventions, the Hexagon condition reduces to
\begin{equation}
q^2+1=0,  \label{eq:Ising_hexagon}
\end{equation}
and therefore
\begin{equation}
q=\pm i, \qquad \mathfrak{x}=-1.  \label{eq:Ising_point}
\end{equation}
Artin consistency permits a continuous complex family, while the fixed Ising fusion data select the discrete roots $q=\pm i$.

  Other fusion data
carry their corresponding Hexagon equations.

% ============================================================

\subsection{Statistical hierarchy and the microscopic question}

\label{subsec:statistical_hierarchy}
% ============================================================

Artin consistency constrains neighboring exchanges. The $F$ moves and Pentagon equation make fusion reassociation consistent, and the Hexagon equations make exchange compatible with fusion.
These algebraic layers define coherent statistics. Intrinsic topological
statistics adds projective path independence together with locality and
spectral protection. 
Starting from complex spectral and relative-geometric coordinates, Artin consistency selects braid-consistent geometry. Fusion sewing and the Pentagon and Hexagon equations make these data coherent. Intrinsic topological statistics additionally requires projective flatness, locality, and protection.

A microscopic Hamiltonian generates these spectral and geometric observables and determines their response to admissible control-path deformations. Section~\ref{sec:Kato} constructs this
Hamiltonian-to-statistics map.

% ============================================================

\section{Kato Transport: From Hamiltonian Control to Statistical Geometry}

\label{sec:Kato}
% ============================================================

To realize statistical geometry, we transport an isolated low-energy or protected spectral subspace with a Hamiltonian exchange protocol.

For a non-Hermitian Hamiltonian, this transport must be defined without
choosing a particular instantaneous eigenvector normalization. We therefore
use Riesz projectors and Kato parallel transport~\cite%
{Kato1950,KatoBook,NenciuRasche1992}. The resulting map is
\begin{equation}
H[\lambda(t)] \longrightarrow P(t) \longrightarrow W_\gamma \longrightarrow %
\bigl( q_{\mathrm{micro}}, \mathfrak{x}_{\mathrm{micro}}, F_{\mathrm{micro}%
}, \ldots \bigr),  \label{eq:Kato-realization-chain}
\end{equation}
which lets us test SGL and fusion coherence in the Hamiltonian model.

 Artin consistency
establishes
\begin{equation}
\mathfrak{x}=q+q^{-1}-1,
\end{equation}
whereas Kato transport determines the spectral and geometric observables
produced by a specified Hamiltonian path. Their comparison connects the
algebraic locking relation to microscopic controls. 

% ============================================================

\subsection{Isolated spectral sectors and Kato transport}

\label{subsec:riesz-kato}
% ============================================================

Consider a smooth family of finite-dimensional non-Hermitian Hamiltonians $%
H(\lambda)$ over a control manifold $\mathcal{M}_{\mathrm{ctrl}}$. Suppose
that a cluster of $r$ eigenvalues remains spectrally isolated from the rest
of the spectrum throughout the region explored by the exchange protocol. The
corresponding Riesz projector is
\begin{equation}
P(\lambda) = \frac{1}{2\pi i} \oint_\Gamma \bigl(z-H(\lambda)\bigr)^{-1}\,dz,
\label{eq:riesz}
\end{equation}
where $\Gamma$ encloses the selected spectral cluster.

Its image
\begin{equation}
\mathcal{E}_{\lambda }=\func{Im}P(\lambda )
\end{equation}%
defines a rank-$r$ spectral bundle $\mathcal{E}\longrightarrow \mathcal{M}_{%
\mathrm{ctrl}}$. For a non-Hermitian Hamiltonian, the Riesz projector is
generally nonorthogonal ($P^{\dagger}\neq P$), so the corresponding transport
need not be unitary.

The Kato connection follows this moving spectral subspace directly,
\begin{equation}
\nabla_K=P\,d,  \label{eq:Kato-connection}
\end{equation}
or, along a control path $\lambda(t)$,
\begin{equation}
\frac{d}{dt}U_K(t) = [\dot P(t),P(t)]\,U_K(t), \qquad U_K(0)=I.
\label{eq:Kato-generator}
\end{equation}
It obeys
\begin{equation}
P(t)U_K(t)=U_K(t)P(0),  \label{eq:Kato-intertwining}
\end{equation}
so the initial spectral subspace is transported into the instantaneous one
without choosing an eigenvector gauge.

For a closed control loop $\gamma$, after any required endpoint
identification, the resulting holonomy is
\begin{equation}
W_\gamma = \func{Hol}_{\nabla_K}(\gamma) \in GL(r,\mathbb{C}).
\label{eq:Kato-holonomy}
\end{equation}

For strongly non-normal Hamiltonians, we also check Riesz-projector conditioning and transport convergence to assess numerical and dynamical stability.

% ============================================================

\subsection{From Kato holonomy to exchange statistics}

\label{subsec:Kato-statistics}
% ============================================================

We interpret Kato holonomies as exchange data by identifying the control protocols with particle exchanges. Consider two neighboring
exchange protocols represented on the same reference fiber. They generate
\begin{equation}
B_1^{\mathrm{micro}}, \qquad B_2^{\mathrm{micro}} \in GL(2,\mathbb{C}).
\label{eq:micro-braids}
\end{equation}

Their projective spectrum and relative eigenline geometry are then obtained
directly as
\begin{equation}
q_{\mathrm{micro}} = \frac{\lambda_+}{\lambda_-}, \qquad \mathfrak{x}_{%
\mathrm{micro}} = \func{cr} \bigl( \ell_1^+,\ell_1^-; \ell_2^+,\ell_2^- %
\bigr).  \label{eq:micro-qx}
\end{equation}
We test SGL without fit parameters by comparing
\begin{equation}
\mathfrak{x}_{\mathrm{micro}} \overset{?}{=} q_{\mathrm{micro}} + q_{\mathrm{%
micro}}^{-1} -1.  \label{eq:micro-SGL}
\end{equation}

We extract the spectrum and relative geometry independently from the same microscopic holonomies.

The same construction applies to an open protocol connecting two fusion-tree
descriptions. Its Kato transport defines a microscopic sewing map
\begin{equation}
F_{\mathrm{micro}} : \mathcal{E}_L \longrightarrow \mathcal{E}_R,
\label{eq:Fmicro}
\end{equation}
which can subsequently be tested against fusion covariance, Pentagon
coherence, and Hexagon compatibility.

Let $\mathfrak P_{\rm ctrl}$ denote the control protocols and $\mathcal D_{\rm stat}$ their extracted statistical data. Kato transport defines the map
\begin{equation}
\begin{gathered} \Phi_K: \mathfrak P_{\rm ctrl} \longrightarrow \mathcal
D_{\rm stat}, \\ \Gamma \longmapsto (q_{\rm micro}, \mathfrak x_{\rm micro},
F_{\rm micro}, \ldots). \end{gathered}  \label{eq:PhiK}
\end{equation}
The statistical consistency conditions derived in Sec.~\ref%
{sec:statistical_geometry} then determine whether these microscopic data
form a coherent exchange structure.

%% REVISED: generic Kato statement for the Ising radial branch.
%% No five-arm solid-angle relation is assumed here.
For a microscopic exchange whose projective spectrum lies on the radial
Ising ray, the Kato holonomy satisfies
\begin{equation}
\arg q_{\mathrm{micro}}=\frac{\pi }{2}\quad (\mathrm{mod}\ 2\pi ).
\label{eq:Kato-Ising-ray}
\end{equation}%
Its radial coordinate can then be defined directly from the observable
projective spectrum as
\begin{equation}
\kappa _{K}=\frac{1}{2\pi }\log |q_{\mathrm{micro}}|,  \label{eq:Kato-kappa}
\end{equation}%
so that
\begin{equation}
q_{\mathrm{micro}}=ie^{2\pi \kappa _{K}}.  \label{eq:Kato-radial-q}
\end{equation}%
The same coordinates are
\begin{equation}
\chi _{\mathrm{micro}}=\frac{1}{2}\log |q_{\mathrm{micro}}|=\pi \kappa
_{K},\qquad \theta _{\mathrm{micro}}=\frac{1}{4}\pi .
\label{eq:Kato-radial-coordinates}
\end{equation}

We relate this measured radial coordinate to junction controls in the five-arm model.

% ============================================================

\subsection{Projective flatness: when does exchange become topological?}

\label{subsec:physical-flatness}
% ============================================================

Intrinsic topological statistics requires the projective exchange action to remain unchanged under smooth control-path deformations that preserve the braid class.

This condition is controlled locally by the Kato curvature
\begin{equation}
\mathcal{F}_K = P\,(dP\wedge dP)\,P.  \label{eq:Kato-curvature}
\end{equation}
The projective action removes the common scalar contribution, so we use the traceless curvature
\begin{equation}
\mathcal{F}_{\mathrm{proj}} = \mathcal{F}_K - \frac{1}{r} \func{Tr}(\mathcal{%
F}_K)I_r.  \label{eq:projective-curvature}
\end{equation}

Local projective flatness requires
\begin{equation}
\mathcal{F}_{\mathrm{proj}}=0.  \label{eq:projective-flatness}
\end{equation}
On a regular simply connected control region, this condition ensures that a
smooth deformation of a path with fixed endpoints cannot change its $PGL(r,%
\mathbb{C})$ transport, although the full $GL(r,\mathbb{C})$ holonomy may
still acquire a common scalar factor.

Pentagon coherence concerns fusion reassociation paths, whereas $\mathcal F_{\rm proj}=0$ concerns smooth deformations of Hamiltonian control paths.
Neither condition implies the other.

Braid consistency and fusion coherence organize coherent exchange data. Topological transport further combines
projective flatness on the admissible control domain with locality, spectral
isolation, and stability, together with the identification of control loops
with physical braid classes.

The five-arm dark bundle has generically nonzero projective Kato
curvature, distinguishing braid and fusion constraints from control-path
flatness.

% ============================================================

\subsection{Similarity covariance and transport convention}

\label{subsec:similarity}
% ============================================================

A parameter-dependent isospectral transformation can change Kato transport.

Suppose
\begin{equation}
\begin{aligned} H'(\lambda) &= S(\lambda)H(\lambda)S^{-1}(\lambda), \\
P'(\lambda) &= S(\lambda)P(\lambda)S^{-1}(\lambda). \end{aligned}
\label{eq:similarity}
\end{equation}
A control-independent $S$ changes the basis. Closed Kato holonomies transform by conjugation,
\begin{equation}
W^{\prime }_\gamma = S W_\gamma S^{-1},  \label{eq:constant-similarity}
\end{equation}
and the projective spectrum and cross ratio are unchanged.

For a parameter-dependent $S$, define $\Omega=S^{-1}dS$. The transformed Kato connection is
\begin{equation}
S^{-1}\nabla _{K}^{\prime }S=\nabla _{K}+P\Omega P\qquad \text{on }\func{Im}%
P.  \label{eq:variable-similarity}
\end{equation}%
The projective transport can therefore change even though the instantaneous
spectra remain identical.

We define microscopic exchange and sewing maps by Riesz--Kato transport, independently of instantaneous eigenvector normalization.

We are now ready to ask the question directly: can a local non-Hermitian Hamiltonian generate the predicted spectrum--geometry locking, and what part of that statistical structure survives changes of the exchange protocol? We address both questions in the five-arm Kitaev--Majorana network.

% ============================================================
% ============================================================

% ============================================================

\section{Microscopic Realizations and Independent Tests}

\label{sec:fivearm}
% ============================================================

 We first use a five-arm Kitaev--Majorana network to control neighboring exchange operations. We extract their spectra and relative eigenline geometry, vary the geometry at fixed spectra, and test the operations at finite time.

We then test spectral and relative-geometric responses in a two-dimensional four-vortex Kitaev protocol whose control paths and bond texture are fixed before the non-Hermitian coupling is varied.  We return to the five-arm network to test fusion sewing and control-path curvature.
In what follows, we address four questions. First, can a local Hamiltonian realize the spectrum--geometry locking predicted by the statistical theory? Second, does the distinction between exchange spectrum and relative geometry persist in an independent two-dimensional setting? Third, can the resulting exchange data be sewn consistently between fusion descriptions? Fourth, does such coherent exchange become path independent, and hence topological?

% ============================================================

\subsection{Five-arm Kitaev--Majorana realization}

\label{subsec:fivearm-realization}
% ============================================================

The five-arm network lets us control and extract the exchange spectrum and relative eigenline geometry separately. 

% ------------------------------------------------------------

\subsubsection{Architecture and exchange protocols}

\label{subsec:fivearm-architecture}
% ------------------------------------------------------------

The device consists of five finite Kitaev arms coupled through a central
Majorana mode. The microscopic Hamiltonian is
\begin{equation}
H_{5Y}(s)=\sum_{a=1}^{5}H_{K}^{(a)}+i\gamma _{0}\sum_{a=1}^{5}\lambda
_{a}(s)\alpha _{a,1},  \label{eq:H5Y}
\end{equation}%
where the arms are Kitaev chains~\cite{Kitaev2001}, with junction control
motivated by wire-network braiding~\cite{Alicea2011}. Here $\gamma _{0}$ is
the central Majorana operator and $\alpha _{a,1}$ is the Majorana mode
adjacent to the junction on arm $a$. We vary the complex couplings $\lambda_a(s)$ while maintaining an isolated low-energy spectral subspace. Riesz--Kato transport gives $B_1^{\mathrm{micro}}$ and $B_2^{\mathrm{micro}}$ in the same logical space.

Each arm has an outer termination Majorana $\eta_a$ and Hamiltonian
\begin{equation}
\begin{aligned} H_K^{(a)}={}&-\frac{i\mu}{2}
\sum_{j=1}^{L}\alpha_{a,j}\beta_{a,j}
+it\sum_{j=1}^{L-1}\beta_{a,j}\alpha_{a,j+1}\\ &+iM\eta_a\beta_{a,L}.
\end{aligned}  \label{eq:add-full-arm}
\end{equation}
With $\{ \gamma_I,\gamma_J\}=2\delta_{IJ}$, we write
\begin{equation}
H_{5Y} = \frac{i}{4}\boldsymbol{\gamma}^{T} A\boldsymbol{\gamma}, \qquad
A^{T}=-A, \qquad h_M=iA.  \label{eq:add-majorana-matrix}
\end{equation}
We diagonalize $h_M=iA$.

For identical arms, the junction term can be projected analytically onto the
five endpoint Majorana zero modes $\zeta_a$. The common endpoint weight sets the energy scale $\Delta$ of the projected junction,
\begin{equation}
\begin{aligned} H_{\rm low}(s) &= i\Delta \,
\gamma_0\,n^{T}(s)\boldsymbol{\zeta}, \\ n(s) &=
\frac{\boldsymbol{\lambda}(s)}
{\sqrt{\boldsymbol{\lambda}^{T}(s)\boldsymbol{\lambda}(s)}}, \qquad n^{T}n=1
. \end{aligned}  \label{eq:fivearm-low-projected}
\end{equation}
The control vector $n$ satisfies the complex quadratic constraint $n^Tn=1$. The Majorana combination parallel to $n$ couples to $\gamma_0$ and is bright. The four combinations satisfying $n^Tv=0$ decouple from $\gamma_0$ and form the dark subspace, with projector
\begin{equation}
P_0=I-nn^{T}.
\end{equation}
Real $n$ gives the compact Hermitian control sphere; allowing complex $n$
opens noncompact projector geometry without requiring the protecting gap to
close. This complex extension realizes the spectral softening defined in Sec.~\ref{subsec:statistical_moduli}.

For the radial family, the exchange path is a complex-geodesic triangle in a
three-dimensional active subspace. With $\beta=2\pi \kappa$, one convenient
choice is
\begin{equation}
\begin{gathered} a=e_5,\qquad b=e_1,\\ c=\cosh \beta \,e_2+i\sinh \beta
\,e_5, \qquad c^Tc=1. \end{gathered}  \label{eq:fivearm-radial-triangle}
\end{equation}
For normalized vertices $v^{T}v=w^{T}w=1$, define
\begin{equation}
\begin{aligned} g(u;v,w) &= \frac{\sin[(1-u)\vartheta]}{\sin \vartheta}v +
\frac{\sin(u\vartheta)}{\sin \vartheta}w, \\ \cos \vartheta&=v^{T}w, \qquad
0\leq u\leq1 , \end{aligned}  \label{eq:add-geodesic}
\end{equation}
with branches continued from the Hermitian protocol. The first exchange loop
is
\begin{equation}
\mathcal{C}_1: \quad e_5 \longrightarrow e_1 \longrightarrow \cosh \beta
\,e_2+i\sinh \beta \,e_5 \longrightarrow e_5 .
\end{equation}
On segment $j$, $u\in[0,1]$ measures progress between $v_j$ and $v_{j+1}$, and the junction controls are
\begin{equation}
\boldsymbol{\lambda}^{(1)}(s;\kappa) = \Lambda \,g(u;v_j,v_{j+1}), \qquad
\beta=2\pi \kappa .  \label{eq:add-fivearm-control}
\end{equation}

 The
analytically continued spherical-excess formula gives
\begin{equation}
\tan \frac{\Omega _{C}}{2}=\frac{a\cdot (b\times c)}{1+a\cdot b+b\cdot
c+c\cdot a}=\tan \left( \frac{\pi }{4}-\frac{i\beta }{2}\right) ,
\label{eq:fivearm-complex-solid-angle}
\end{equation}%
so on the branch continuous from the Hermitian Ising loop,
\begin{equation}
 \Omega_C = \frac{\pi}{2}-i\beta = \frac{\pi}{2}-2i\pi \kappa .
\label{eq:fivearm-OmegaC}
\end{equation}%
The control path determines the complex braid angle,
\begin{equation}
\alpha =\frac{\Omega _{C}}{2}=\frac{\pi }{4}-i\pi \kappa .
\label{eq:fivearm-alpha-from-path}
\end{equation}%
The corresponding action on Majorana states, called the Spin lift, is~\cite{Ivanov2001,Alicea2011}
\begin{equation}
U_{12}^{\mathrm{NH}}=\exp \left[ \frac{\Omega _{C}}{2}\gamma _{1}\gamma _{2}%
\right] .  \label{eq:UNH}
\end{equation}%
With $\gamma_1\gamma_2|\pm\rangle=\pm i|\pm\rangle$, we obtain
\begin{equation}
\lambda _{+}(\kappa )=e^{\pi \kappa +i\pi /4},\qquad \lambda _{-}(\kappa
)=e^{-\pi \kappa -i\pi /4},  \label{eq:single-exchange-eigs}
\end{equation}%
and therefore
\begin{equation}
q = \frac{\lambda_+}{\lambda_-} = e^{i\Omega_C} = i e^{2\pi \kappa}.
  \label{eq:single-exchange-q-derived}
\end{equation}

In the five-arm model, the complex solid angle of the control path fixes
the radial coordinate:
\begin{equation}
\kappa[\boldsymbol{\lambda}(\cdot)] = -\frac{\func{Im}\Omega_C[n(\cdot)]}{%
2\pi}.  \label{eq:kappa-path-functional-main}
\end{equation}
The radial Ising branch thus follows from the junction geometry rather
than an inserted nonunitary factor in the logical gate.

% ------------------------------------------------------------

\subsubsection{Microscopic spectrum--geometry locking}

\label{subsec:micro-SGL}
% ------------------------------------------------------------

The first loop fixes the exchange spectrum; the second realizes the relative eigenline geometry required by Artin consistency. Let $%
c_{q},s_{q}$ denote the branch-continuous coefficients appearing in Eqs.~%
\eqref{eq:c2_lock} and \eqref{eq:add-explicit-braid-entries}, evaluated at $%
q=ie^{2\pi \kappa }.$ We define
\begin{equation}
\begin{aligned} a_q&=c_q^2-s_q^2, \qquad b_q=2c_qs_q, \\
h_q&=-a_qe_1+b_qe_3, \qquad m_q=b_qe_1+a_qe_3 . \end{aligned}
\label{eq:add-second-axis}
\end{equation}%
The second exchange path is then chosen as
\begin{equation}
\begin{aligned} \mathcal C_2:\quad e_5 &\longrightarrow m_q \longrightarrow
e_2 \\ &\longrightarrow \cosh \beta \,h_q+i\sinh \beta \,m_q \longrightarrow
m_q \longrightarrow e_5 . \end{aligned}  \label{eq:add-second-loop}
\end{equation}%
At $\kappa=0$, $\mathcal C_2$ gives the neighboring Ising exchange.
For nonzero $\kappa$, both its spectrum and eigenframe vary.

To express both exchanges in the same logical space, we use the initial Majorana frame $Q_0$ and endpoint identification $\mathcal G_r$ to close the full-chain Kato transport:
\begin{equation}
O_r = Q_0^{T} \mathcal{G}_r U_{K,r}(1) Q_0 \in SO(4,\mathbb{C}), \qquad
r=1,2 ,  \label{eq:add-closed-majorana-map}
\end{equation}
and a continuous Spin lift
\begin{equation}
\mathcal{S}_r^{-1} \widehat \gamma_a \mathcal{S}_r =
\sum_b(O_r)_{ab}\widehat \gamma_b, \qquad B_r^{\mathrm{micro}} = \left.%
\mathcal{S}_r\right|_{\mathrm{even}}.  \label{eq:add-logical-spin-map}
\end{equation}

For a $2\times2$ exchange matrix $B$, we evaluate the spectral SGL prediction directly from its trace and determinant:
\begin{equation}
\mathfrak{x}_{\mathrm{spec}} = \frac{(\func{Tr}B)^2}{\det B}-3.
\label{eq:xspec}
\end{equation}
Independently, if $v_i^\pm$ span the right eigenlines of $B_i^{\mathrm{micro}%
}$, the relative geometry is
\begin{equation}
\mathfrak{x}_{\mathrm{geom}} = \frac{ [v_1^+,v_2^+]\,[v_1^-,v_2^-] }{
[v_1^+,v_2^-]\,[v_1^-,v_2^+] }, \qquad [u,v]=\det(u,v).  \label{eq:xgeom}
\end{equation}
We extract these quantities separately from the spectrum and eigenlines of the microscopic holonomies.

Along the radial family, we have
\begin{equation}
\mathfrak{x}_{\mathrm{SGL}}(\kappa )=-1+2i\sinh (2\pi \kappa ).
\label{eq:radialSGL-fivearm}
\end{equation}%
Figure~\ref{fig:fivearm}(b) shows that the eigenvalue and eigenline extractions agree with the radial SGL relation.

% ------------------------------------------------------------

\subsubsection{Matched-spectrum and geometry-detuning controls}

\label{subsec:geometry-negative-control}
% ------------------------------------------------------------

To isolate the role of relative eigenline geometry, we compare protocols satisfying
\begin{equation}
q_A=q_B, \qquad \Delta_{\mathrm{iso}}^A = \Delta_{\mathrm{iso}}^B,
\label{eq:matched-spectrum-gap}
\end{equation}
while changing
\begin{equation}
\mathfrak{x}_A\neq \mathfrak{x}_B .  \label{eq:geometry-control}
\end{equation}
We vary the relative eigenline geometry at fixed projective spectrum and isolation gap.

For the braid-consistent radial protocol, we have
\begin{equation}
\mathfrak{x}_{\mathrm{SGL}}=-1+2i\sinh (2\pi \kappa ),
\end{equation}%
whereas a fixed-axis control retains
\begin{equation}
\mathfrak{x}_{\mathrm{ctrl}}=-1.
\end{equation}%
At nonzero $\kappa $, the latter develops a finite Artin defect even though
its local exchange spectrum can be matched to that of the SGL protocol. More
generally, continuously detuning the relative geometry at fixed spectral
data gives
\begin{equation}
\epsilon _{\mathrm{braid}}^{\mathrm{proj}}\sim C\, \delta _{\mathrm{geom}%
},\qquad \delta _{\mathrm{geom}}=|\mathfrak{x}_{\mathrm{geom}}-\mathfrak{x}_{%
\mathrm{spec}}|,  \label{eq:linear-artin-defect}
\end{equation}%
near the locked locus.

\begin{figure*}[t]
\centering
\includegraphics[width=0.98\textwidth]{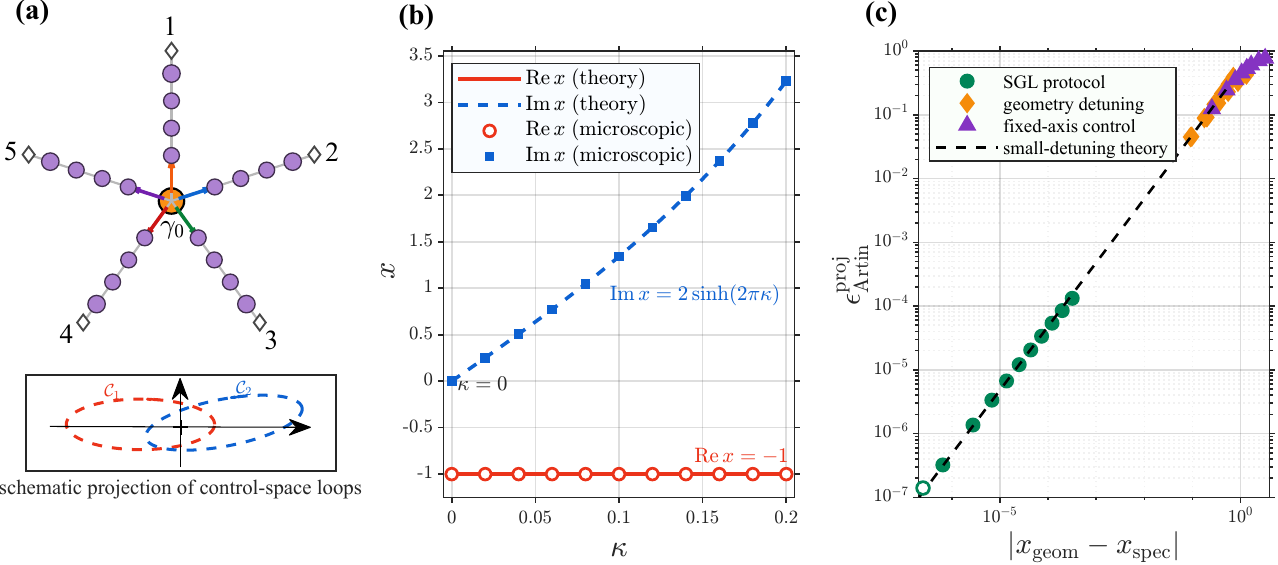}
\caption{\textbf{Microscopic realization and geometric control in a
five-arm Majorana network.}
(a) Five finite Kitaev arms coupled to a central Majorana $\gamma_0$.
The inset shows schematic projections of the junction-control loops $\mathcal C_1$ and $\mathcal C_2$ for neighboring exchanges.
(b) Relative eigenline geometry extracted from full-Hamiltonian Kato
holonomies along the designed radial branch. Open red circles and filled
blue squares show the microscopic real and imaginary parts of
$\mathfrak{x}$; the corresponding lines give
$\operatorname{Re}\mathfrak{x}=-1$ and
$\operatorname{Im}\mathfrak{x}=2\sinh(2\pi\kappa)$.
(c) Projective Artin residual versus the independently evaluated
spectral--geometric mismatch
$|\mathfrak{x}_{\mathrm{geom}}-\mathfrak{x}_{\mathrm{spec}}|$.
Green circles denote the SGL protocol, orange diamonds the
geometry-detuned controls, and purple triangles the fixed-axis controls.
The dashed line gives the small-detuning prediction in its domain of
validity. The controls isolate the role of relative eigenline geometry in Artin consistency.}
\label{fig:fivearm}
\end{figure*}

Figure~\ref{fig:fivearm}(c) shows the growth of Artin residuals with relative eigenline mismatch in the geometry-detuned and fixed-axis controls.

Protocols with the same exchange spectrum and isolation gap can have different braid actions.
SGL constrains a separately measurable relative eigenline geometry.

% ------------------------------------------------------------

\subsubsection{Finite-time realization of geometric exchange}

\label{subsec:fivearm-finite-time}
% ------------------------------------------------------------

To connect Kato transport to finite-time exchange, we compare it with full microscopic Schr\"odinger evolution over a range of durations.

The complete quadratic Hamiltonian is propagated as
\begin{equation}
\widehat H(t) = \frac{i}{4}\Gamma^{T}A[t/T]\Gamma, \qquad A^{T}=-A ,
\label{eq:finite-time-majorana-H}
\end{equation}
along the same geometric control paths used for the Kato construction.
Within each segment we use the smooth schedule
\begin{equation}
\begin{aligned} \boldsymbol{\lambda}_r(t) &= \Lambda
\,g(\rho(u);v_j,v_{j+1}), \\ \rho(u)&=3u^2-2u^3 . \end{aligned}
\label{eq:add-time-schedule}
\end{equation}
We propagate the full microscopic state throughout the protocol.

\begin{figure*}[t]
\centering
\includegraphics[width=0.98\textwidth]{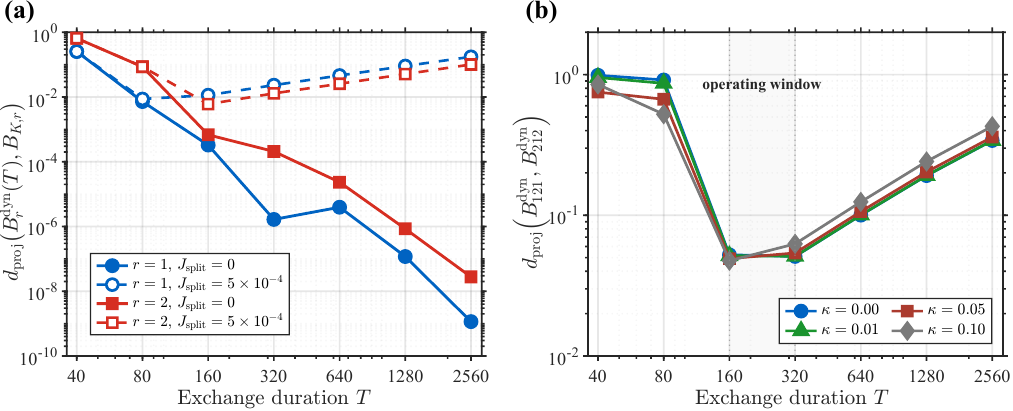}
\caption{\textbf{Finite-time realization of geometric exchange.}
(a) At $\kappa=0.10$, projective distance
$d_{\mathrm{proj}}(B_r^{\mathrm{dyn}}(T),B_{K,r})$
between full finite-time logical operations and their Kato references
for neighboring exchanges $r=1$ (blue circles) and $r=2$ (red squares).
Filled symbols and solid lines correspond to $J_{\mathrm{split}}=0$;
open symbols and dashed lines correspond to the persistent splitting
$J_{\mathrm{split}}=5\times10^{-4}$.
With zero splitting, longer protocols approach the Kato references. Persistent splitting selects an intermediate window that balances nonadiabatic corrections and internal dynamical evolution.
(b) Projective distance between the uninterrupted microscopic exchange
words $121$ and $212$ for $\kappa=0,0.01,0.05,0.10$.
Panel (b) uses persistent $J_{\mathrm{split}}=5\times10^{-4}$;
the displayed calculations use $\Delta t=0.025$.
Here $T$ is the duration of each elementary exchange, so each word lasts
$3T$. The sampled minima occur at $T=160$ or $320$, identifying an
intermediate operating interval across the displayed deformations.}
\label{fig:finite-time-dynamics}
\end{figure*}

We compare the full dynamical operation with the Kato reference using
\begin{equation}
d_{\mathrm{proj}}(X,Y) = \min_{z\in \mathbb{C}} \frac{\|X-zY\|_F}{\|X\|_F}.
\label{eq:finite-time-projective-distance}
\end{equation}
In the exact-dark limit, the projected Hamiltonian inside the protected
manifold is projectively trivial, so increasing the exchange duration
suppresses nonadiabatic leakage and drives $B_r^{\mathrm{dyn}}(T)$ toward
the Kato operation. Figure~\ref{fig:finite-time-dynamics}(a) shows this convergence.

A weak persistent splitting adds an internal dynamical time scale. Short protocols are nonadiabatic, while long protocols accumulate a dynamical phase within the protected subspace. The two mechanisms select an intermediate operating window,
\begin{equation}
\Delta_{\mathrm{iso}}^{-1} \ll T \ll \delta E_{\mathrm{split}}^{-1}.
\label{eq:finite-time-window}
\end{equation}
The left inequality suppresses excitation out of the isolated manifold; the
right inequality prevents the weak internal splitting from overwhelming the
geometric operation.

To test finite-time Artin consistency, Fig.~\ref{fig:finite-time-dynamics}(b) compares the uninterrupted exchange words $121$ and $212$.  Their projective distance has a common low-residual interval across the sampled $\kappa$ values. This finite-time braid window balances nonadiabatic transitions against the dynamical phase accumulated from persistent splitting.

The finite-time calculation connects the geometric exchange operators to uninterrupted microscopic dynamics.

% ============================================================

\subsection{Independent two-dimensional vortex test}

\label{subsec:2d-kitaev-certificate}
% ============================================================

To test the observables in a different microscopic model, we use a four-vortex exchange protocol in the chiral Kitaev honeycomb model with fixed controls.

\begin{figure*}[t]
\centering
\includegraphics[width=0.98\textwidth]{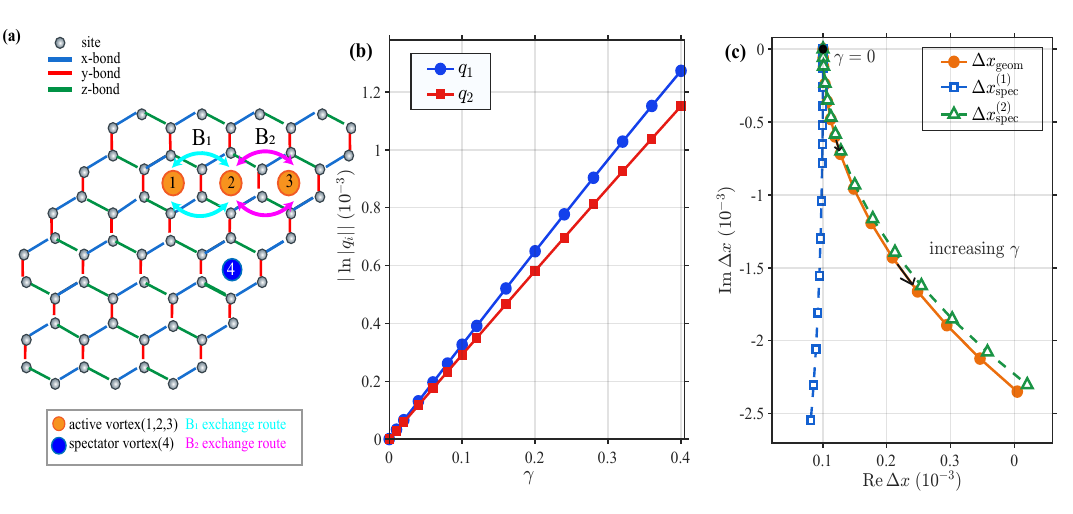}
\caption{\textbf{Independent spectral and relative-geometric responses
in a two-dimensional vortex protocol.}
(a) Schematic four-vortex configuration in the chiral Kitaev honeycomb
model, with blue, red, and green links denoting $x$, $y$, and $z$ bonds.
Plaquette vortices $1,2,3$ participate in the neighboring exchanges
$B_1:1\leftrightarrow2$ and $B_2:2\leftrightarrow3$; vortex $4$ is
a spectator. The arrows indicate the exchange pairs schematically.
Both protocols start from the same configuration and use the same
fixed non-Hermitian bond texture.
(b) Magnitudes of the single-exchange spectral responses
$|\ln|q_1||$ and $|\ln|q_2||$ versus the microscopic
non-Hermitian coupling $\gamma$. The signed values
$\ln|q_r|$ are reported in the Supplemental Material.
(c) Complex-plane trajectories of the differential relative geometry
$\Delta\mathfrak{x}_{\mathrm{geom}}$ and the two spectral benchmarks
$\Delta\mathfrak{x}_{\mathrm{spec}}^{(r)}$, where
$\Delta\mathfrak{x}(\gamma)=\mathfrak{x}(\gamma)-\mathfrak{x}(0)$ and
$\mathfrak{x}_{\mathrm{spec}}^{(r)}=q_r+q_r^{-1}-1$.
Orange circles show the relative geometry reconstructed from both exchanges; blue squares and green triangles show the spectral benchmarks for $r=1$ and $r=2$, respectively.
The trajectories begin at the Hermitian reference $\gamma=0$.
The same lattice deformation produces distinct spectral and relative-geometric responses.}
\label{fig:2d-kitaev}
\end{figure*}

In the Majorana matter representation,
\begin{equation}
H_{\mathrm{2D}}(\gamma )=\frac{i}{4}\mathbf{c}^{T}A[u;\gamma ]\mathbf{c}%
,\qquad J_{e}(\gamma )=J(1+i\gamma \xi _{e}),
\label{eq:2d-kitaev-hamiltonian}
\end{equation}%
where the spatial bond texture $\xi _{e}$ is fixed once and shared by the
two neighboring exchange protocols. Vortices $1,2,3$ define the adjacent
exchanges $B_{1}$ and $B_{2}$, while vortex $4$ fixes the spectator parity
sector, as shown in Fig.~\ref{fig:2d-kitaev}(a).

In the honeycomb model, $\gamma$ controls the microscopic bond amplitudes,
while $q_r(\gamma)$ is extracted from the resulting Riesz--Kato holonomies.
The effective spectral coordinates are
\begin{equation}
\kappa _{\mathrm{eff},r}(\gamma )=\frac{\log |q_{r}(\gamma )|}{2\pi },\qquad
\phi _{r}(\gamma )=\arg q_{r}(\gamma ),  \label{eq:add-2d-effective-kappa}
\end{equation}%

With the bond texture and exchange paths fixed, the lattice Hamiltonian
generates $B_1$ and $B_2$; their eigenvalues and eigenlines then
independently determine $q_1$, $q_2$, and $\mathfrak{x}_{\mathrm{geom}}$,
together with the two spectral predictions
$\mathfrak{x}_{\mathrm{spec}}^{(r)}=q_r+q_r^{-1}-1$.

Figure~\ref{fig:2d-kitaev}(b) shows that increasing $\gamma$ drives the
two single-exchange spectra away from the unit circle by unequal amounts.
We reconstruct the relative eigenline geometry from both exchange matrices.

To isolate the response induced by non-Hermiticity from the finite-size
Hermitian offset, we use
\begin{equation}
\Delta \mathfrak{x}(\gamma) = \mathfrak{x}(\gamma)-\mathfrak{x}(0).
\label{eq:2d-differential-x}
\end{equation}
Figure~\ref{fig:2d-kitaev}(c) compares the independently reconstructed $%
\Delta \mathfrak{x}_{\mathrm{geom}}$ with the two spectral responses $\Delta%
\mathfrak{x}_{\mathrm{spec}}^{(1,2)}$ in the complex plane.

The frozen two-dimensional protocol yields distinct single-exchange
spectra and a neighboring-eigenframe response on comparable complex scales.
Finite vortex separation produces differences between the relative-geometric response and the two spectral SGL predictions. 
The single-exchange spectra and relative eigenline geometry have distinct microscopic responses.
SGL provides the algebraic benchmark; finite separation and lattice geometry control the approach to that branch.

% ============================================================

\subsection{Microscopic fusion coherence}

\label{subsec:micro-fusion}
% ============================================================

The previous subsections concern neighboring exchanges represented in one
common fusion description. We now ask whether independently prepared fusion
charts of the same logical sector can be sewn consistently.

A weak channel resolver $J_{\mathrm{split}}$ prepares the left and right
fusion descriptions through physical $12$ and $23$ splittings, respectively,
while the low-energy manifold remains isolated from the bulk. The resolver
selects the fusion chart and is switched off during exchange; it is distinct
from persistent splitting, which remains active during the finite-time braid.

An open full-Hamiltonian Kato reassociation from the left chart to the right
chart reconstructs
\begin{equation}
F_{\mathrm{micro}} \simeq
\begin{pmatrix}
0.5+0.5i & 0.5+0.5i \\
-0.5+0.5i & 0.5-0.5i%
\end{pmatrix}
\label{eq:Fmicro-main}
\end{equation}
in the working gauge. Its rescaling-invariant cross ratio is
\begin{equation}
\Xi_F = \frac{ (F_{\mathrm{micro}})_{12}(F_{\mathrm{micro}})_{21} }{ (F_{%
\mathrm{micro}})_{11}(F_{\mathrm{micro}})_{22} } \simeq -1.
\label{eq:XiF-main}
\end{equation}
The left- and right-chart braid probes are reconstructed independently and
obey
\begin{equation}
B_i^{(R)} \simeq F_{\mathrm{micro}} B_i^{(L)} F_{\mathrm{micro}}^{-1}.
\label{eq:micro-fusion-covariance}
\end{equation}

\begin{figure*}[t]
\centering
\includegraphics[width=0.99\textwidth]{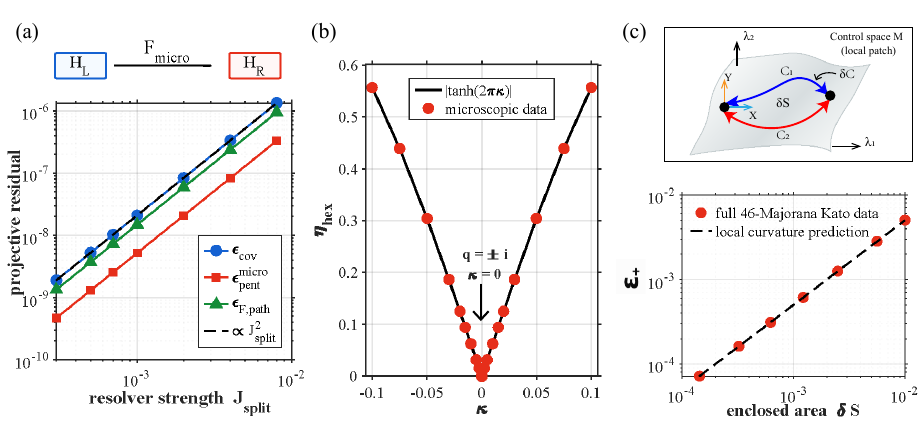}
\caption{\textbf{Fusion sewing, categorical selection, and control-path curvature.}
(a) The microscopic map $F_{\mathrm{micro}}$ connects left and right fusion
charts. The covariance residual $\epsilon_{\mathrm{cov}}$, microscopic
Pentagon residual $\epsilon_{\mathrm{pent}}^{\mathrm{micro}}$, and
reassociation-path residual $\epsilon_{F,\mathrm{path}}$ exhibit quadratic
scaling with the chart-resolver strength $J_{\mathrm{split}}$; the dashed
guide is proportional to $J_{\mathrm{split}}^2$.
Residual definitions and common-basis conventions are given in the Supplemental Material. The resolver prepares the fusion charts and is
switched off during exchange, distinguishing it from the persistent
splitting in Fig.~\ref{fig:finite-time-dynamics}.
(b) Fixed-Ising Hexagon obstruction
$\eta_{\mathrm{hex}}=|q_{\mathrm{micro}}^2+1|/(1+|q_{\mathrm{micro}}|^2)$.
Red markers show microscopic values and the black curve gives
$|\tanh(2\pi\kappa)|$. The fixed Ising fusion structure selects
$q=\pm i$; the displayed branch $q=ie^{2\pi\kappa}$ intersects this
set at $q=i$, $\kappa=0$.
(c) Two control paths with common endpoints bound a small loop $\delta C$
of area $\delta S$. The lower plot compares the full $46$-Majorana
Kato result for the normalized even-parity response
$\epsilon_+=\|W_{\delta C}^{(+)}-I_2\|_F/\sqrt{2}$
with the local-curvature prediction
$\epsilon_+/\delta S\to1/2$ in the small-area limit.
The linear area dependence identifies nonzero projective curvature.
Fusion sewing, categorical compatibility, and control-path curvature are tested separately.}
\label{fig:coherence}
\end{figure*}

We test path independence in fusion-tree space.
Different microscopic reassociation trajectories approach the same
projective sewing map, and the corresponding fusion covariance, microscopic
Pentagon, and reassociation-path residuals all vanish quadratically as the
resolver is removed,
\begin{equation}
\epsilon_{\mathrm{cov}} \propto J_{\mathrm{split}}^2, \qquad \epsilon_{%
\mathrm{pent}}^{\mathrm{micro}} \propto J_{\mathrm{split}}^2, \qquad
\epsilon_{F,\mathrm{path}} \propto J_{\mathrm{split}}^2.
\label{eq:resolver-scaling-main}
\end{equation}
Figure~\ref{fig:coherence}(a) shows this common scaling.

The quadratic law has a direct geometric interpretation. The physical
resolver deforms the fusion charts over an area of order $J_{\mathrm{split}%
}^2$ in resolver-control space. As the resolver vanishes, the finite chart-separation correction vanishes and the projective transition map remains well defined. Kato reassociation therefore realizes the fusion map as a microscopic operation.

Compatibility with fixed categorical data imposes the additional Hexagon condition. For the fixed planar Ising fusion data, the Hexagon equations impose
\begin{equation}
 q^2+1=0, \qquad q=\pm i .   \label{eq:micro-hex-selection-main}
\end{equation}
Evaluating this condition on the microscopic radial family gives
\begin{equation}
\eta_{\mathrm{hex}} = \frac{|q_{\mathrm{micro}}^2+1|} {1+|q_{\mathrm{micro}%
}|^2} = |\tanh(2\pi \kappa)|.  \label{eq:micro-hex-obstruction-main}
\end{equation}
As shown in Fig.~\ref{fig:coherence}(b), the obstruction vanishes only at
the unitary Ising point on the displayed radial branch.

 Artin consistency permits a continuous
non-Hermitian statistical geometry. Fusion sewing can be implemented
projectively over that geometry. Imposing the fixed conventional Ising
associator then selects the discrete braid--fusion-compatible point. 

% ============================================================
% ============================================================

% ============================================================

\subsection{Projective curvature and the boundary between geometry and
topology}

\label{subsec:flatness-obstruction}
% ============================================================

We finally test whether smooth control-path deformations that preserve the braid class also preserve the projective exchange operation.

For the dark subspace of the identical-arm junction, the projector is
\begin{equation}
P_{0}(n)=I-nn^{T},\qquad n^{T}n=1.  \label{eq:quadric-projector}
\end{equation}%
The control vectors lie on the complex quadric
\begin{equation}
n\in Q_{m-1}=\{n\in \mathbb{C}^{m}:n^{T}n=1\}.
\end{equation}%
For tangent vectors $X,Y\in T_nQ_{m-1}$, the Kato curvature is
\begin{equation}
\mathcal{F}_{K}(X,Y)=XY^{T}-YX^{T}.  \label{eq:quadric-curvature}
\end{equation}%
It is traceless, and hence
\begin{equation}
 \mathcal F_{\mathrm{proj}}(X,Y) = XY^{T}-YX^{T}.
\label{eq:quadric-projective-curvature}
\end{equation}%
For $m\geq 3$ and generic independent tangent directions, we have
\begin{equation}
\mathcal{F}_{\mathrm{proj}}(X,Y)\neq 0.
\end{equation}%
No full-dimensional regular open patch of this control manifold is therefore
projectively flat.

A small contractible control-path detour $\delta C$ spanning area $\delta S$
changes the projective holonomy by the curvature term
$-\mathcal{F}_{\mathrm{proj}}(X,Y)\delta S$ at leading order:
\begin{equation}
W_{\delta C}=I-\mathcal{F}_{\mathrm{proj}}(X,Y)\delta S+O(\delta S^{3/2}).
\label{eq:detour-holonomy}
\end{equation}%
The original and deformed exchange paths can have the same endpoints and the
same braid class while producing different projective Kato holonomies:
\begin{equation}
\lbrack \gamma ^{\prime }]=[\gamma ]\qquad \not \Rightarrow \qquad \lbrack
W_{\gamma ^{\prime }}]=[W_{\gamma }].  \label{eq:same-braid-no-flatness}
\end{equation}

Figure~\ref{fig:coherence}(c) tests this curvature through microscopic loop transport. The change in projective holonomy is linear in
the enclosed control-space area, with the coefficient fixed by the exact
curvature. The area-dependent response probes microscopic control-path
geometry, separately from Artin consistency and fusion-tree coherence.

Braid and fusion coherence can coexist with control-path dependence.
 SGL constrains neighboring exchanges, fusion sewing and Hexagon equations constrain their composition, and projective curvature determines their control-path dependence.

Intrinsic topological statistics requires projectively flat transport on the admissible control domain, together with locality and spectral protection.
Non-Hermitian statistical geometry therefore occupies the intermediate layer
exposed when that compression is incomplete.

% ============================================================
% ============================================================

\section{Unified Interpretation: From Statistical Geometry to Topological
Statistics}

\label{sec:interpretation}
% ============================================================

Exchange-spectrum softening changes relative channel amplification. Artin consistency links the spectrum to relative eigenline geometry; fusion and curvature impose further conditions.

 Holonomy specifies an exchange operation, braid--fusion coherence specifies its consistent composition, and projective path independence, locality, and protection give its topological meaning.

% ============================================================

\subsection{From geometric holonomy to topological statistics}

\label{subsec:holonomy-to-topology}
% ============================================================

Consistent exchange holonomies on a common protected space define the multigenerator structure of quantum statistics.

The projective spectrum and relative eigenline geometry of neighboring exchanges must satisfy Artin consistency. Different fusion-tree descriptions
must be sewn consistently by $F$ moves, while the Pentagon and Hexagon
equations impose global reassociation and braid--fusion compatibility.
Topological statistics additionally requires invariance under admissible control-path deformations.

Geometric holonomies define controlled operations. Braid and fusion constraints make their composition coherent, and projective path independence, locality, and protection give topological statistics.

 In a unitary topological realization, exchange operators preserve a common positive-definite Hermitian metric and satisfy braid--fusion coherence. Their projective structure group is compact, and projective flatness removes control-path dependence.

The conventional $F$ and $R$ symbols encode the resulting fusion and exchange data, with locality and protection supplying the physical topological realization.

For the two-channel Ising sector, local braid consistency selects the SGL locus
\begin{equation}
\mathcal{M}_{\mathrm{stat}}^{\mathrm{cand}}\supset \Gamma _{\mathrm{SGL}},
\end{equation}%
whereas compatibility with the fixed conventional Ising fusion skeleton
selects $(q,\mathfrak{x})=(\pm i,-1).$  The five-arm Kato holonomies realize SGL. Microscopic $F$ moves connect the fusion charts, and the five-chart Pentagon residual vanishes as $J_{\mathrm{split}}^2$. The planar Ising Hexagon
subsequently selects $q=\pm i$ from the continuous radial family. The fixed two-dimensional Kitaev protocol independently tests the spectral and relative-geometric observables entering the SGL/Artin comparison. The five-arm curvature calculation shows that braid and fusion coherence can coexist with control-path dependence.

$R$ specifies exchange and $F$ changes the fusion description. The Pentagon equation makes reassociation consistent, and the Hexagon equations make exchange compatible with fusion.

% ============================================================

\subsection{Relation to tunable non-Hermitian Majorana braiding}

\label{subsec:zhao}
% ============================================================

Zhao \emph{et al.}~\cite{Zhao2021} provide a tunable single-exchange spectrum to which we can apply the SGL prediction.

For their non-Hermitian Majorana qubit, the reported geometric phases are
\begin{equation}
\Delta \phi _{0}^{\beta }=-\frac{\pi }{2(1+e^{-4\beta })},\qquad \Delta \phi
_{1}^{\beta }=\frac{\pi }{4},
\end{equation}%
which give the exchange operation $\mathcal{R}_{M}^{\beta }=e^{-i\Delta \Phi
_{\beta }\tau _{z}},$ with
\begin{equation}
\Delta \Phi _{\beta }=\frac{\pi }{8}+\frac{\pi }{4(1+e^{-4\beta })}.
\label{eq:zhao-phase}
\end{equation}

In our convention, their single-exchange result gives $q_\beta=e^{2i\Delta\Phi_\beta}$. Artin consistency then constrains the relative eigenline geometry of neighboring exchanges.

If two neighboring exchanges act on a common protected fusion space and lie
on the generic irreducible rank-two braid-consistent branch, SGL predicts
\begin{equation}
\mathfrak{x}_{\beta }=q_{\beta }+q_{\beta }^{-1}-1=2\cos (2\Delta \Phi
_{\beta })-1.  \label{eq:zhao-x-general}
\end{equation}%
Using Eq.~\eqref{eq:zhao-phase}, we have
\begin{equation}
\mathfrak{x}_{\beta }=2\cos \! \left[ \frac{\pi }{4}+\frac{\pi }{%
2(1+e^{-4\beta })}\right] -1.  \label{eq:zhao-SGL-prediction}
\end{equation}%
At $\beta =0$, $(q,\mathfrak{x})=(i,-1)$ recovers the conventional Ising
point.

Zhao \emph{et al.}~\cite{Zhao2021} determine the tunable
single-exchange spectrum
$q_\beta$. If neighboring exchanges lie on the generic irreducible
rank-two braid-consistent branch, SGL predicts their relative eigenline
coordinate $\mathfrak{x}_\beta$.

The neighboring microscopic holonomies can test this conditional prediction through their eigenlines and Artin residual; projective curvature separately tests control-path dependence.

 Zhao \emph{et al.} use normalized right states to define Berry transport; we use Riesz projectors to define Kato transport. Their relation depends on the connection term generated by a parameter-dependent similarity transformation, as in Eq.~\eqref{eq:variable-similarity}.

% ============================================================

\subsection{Geometric and Galois routes to nonunitary statistics}

\label{subsec:galois-NH}
% ============================================================

Non-Hermitian deformation and Galois conjugation can reach nonunitary statistics through different mechanisms.

A continuous non-Hermitian Hamiltonian deformation allows the exchange spectrum and relative eigenline geometry to vary.

Galois conjugation changes the algebraic embedding of the fusion and braiding data while preserving their polynomial consistency relations~\cite{Freedman2012,Chen2022}.

At the level of the local rank-two projective coordinates, the unitary $%
SU(2)_{k}$ sequence may be represented, up to inversion, by $q_{k,1}=-\exp
\! \left( \frac{2\pi i}{k+2}\right) $ with corresponding SGL coordinate
\begin{equation}
\mathfrak{x}_{k,1}=-1-2\cos \! \left( \frac{2\pi }{k+2}\right) .
\label{eq:unitary-SU2-points}
\end{equation}

Let $N$ be the conductor of the cyclotomic field containing the full $F,R$ data, so the field is generated by an $N$th root of unity. An admissible Galois embedding with exponent $p$ coprime to $N$ acts locally as
\begin{equation}
q_{k,p}=-\exp \! \left( \frac{2\pi ip}{k+2}\right) ,
\end{equation}%
and therefore generates the arithmetic SGL orbit
\begin{equation}
\mathfrak{x}_{k,p}=-1-2\cos \! \left( \frac{2\pi p}{k+2}\right) .
\label{eq:galois-SGL-orbit}
\end{equation}%

Varying $k$ changes the unitary fusion theory; varying the admissible embedding $p$ at fixed $k$ gives Galois conjugates.

The Fibonacci/Yang--Lee pair gives the simplest example. For the Fibonacci sector,
\begin{equation}
q_{\mathrm{Fib}}=-e^{2\pi i/5},\qquad \mathfrak{x}_{\mathrm{Fib}}=-\varphi ,
\end{equation}%
where $\varphi =\frac{1+\sqrt{5}}{2}.$ Its nonunitary Galois conjugate gives
\begin{equation}
q_{\mathrm{YL}}=-e^{4\pi i/5},\qquad \mathfrak{x}_{\mathrm{YL}}=\varphi
^{-1}.  \label{eq:fib-YL-SGL}
\end{equation}

Both lie on the unit-modulus spectral slice $|q|=1$, but they differ in positive-metric unitarizability. 
A unit-modulus exchange spectrum can occur without a common positive-definite metric for the braid generators.

In the same projective coordinates, non-Hermitian deformation traces a continuous geometric path, whereas Galois conjugation gives a discrete algebraic orbit.

 A microscopic realization of Galois-conjugate data also requires fusion sewing, a specified metric and transport prescription, and the projective flatness, locality, and protection needed for topological exchange.

% ============================================================
% Closing paragraph
% ============================================================

% ============================================================

% ============================================================

\section{Conclusion and outlook}

\label{sec:outlook}
% ============================================================

We formulate non-Hermitian statistics through the exchange spectrum and relative eigenline geometry of neighboring generators. In the two-channel setting, $q$ describes the projective exchange spectrum and $\mathfrak{x}$ describes the gauge-invariant relative eigenline geometry.  On the generic irreducible branch, the
Artin relation re-locks them exactly through
\begin{equation}
\mathfrak{x}=q+q^{-1}-1 .  \label{eq:conclusion-SGL}
\end{equation}
Non-Hermiticity exposes how neighboring exchanges are arranged in a common protected space.

 In the five-arm network, the complex solid angle of the control path generates the radial Ising spectrum, and independently reconstructed eigenlines agree with Eq.~\eqref{eq:conclusion-SGL}. 

Geometry-detuned and fixed-axis controls produce different braid actions at matched exchange spectra and isolation gaps.%
Relative eigenline geometry is a multigenerator observable fixed by Artin consistency on the SGL locus.

The microscopic picture remains operational away from the ideal geometric
limit. Full finite-time propagation approaches the Kato exchange as nonadiabatic transitions are suppressed; persistent splitting adds an internal dynamical time scale. The window $\Delta_{\mathrm{iso}}^{-1}\ll T\ll\delta E_{\mathrm{split}}^{-1}$ realizes this exchange structure through uninterrupted microscopic dynamics. A fixed two-dimensional four-vortex Kitaev protocol provides an independent test. The same lattice deformation produces distinct single-exchange spectral and relative-geometric responses.

 Microscopic
reassociation maps sew independently prepared fusion charts, while fusion
covariance and Pentagon closure establish their consistent composition. The
Hexagon condition then asks whether the locally braid-consistent exchange
data are compatible with a specified fusion theory. For the fixed Ising fusion data, they select $q=\pm i$ and $\mathfrak{x}=-1$ from the continuous SGL family. 

 The five-arm dark subspace has generically nonzero projective Kato curvature, $\mathcal F_{\mathrm{proj}}(X,Y)=XY^T-YX^T$. A small control loop therefore gives $W_{\delta C}=I-\mathcal F_{\mathrm{proj}}(X,Y)\delta S+\cdots$. Consequently, exchange protocols with the same endpoints and
braid class can retain different projective holonomies. 
Coherent exchange data can therefore remain control-path dependent.%
Topological exchange additionally requires projective flatness, locality, and protection.

Artin consistency locks softened spectra to relative eigenline geometry; fusion coherence constrains their many-anyon composition; projective flatness, locality, and protection give intrinsic topological statistics.

Future directions include projectively flat non-Hermitian realizations, higher-rank relative geometry at fixed spectra, and local realizations of nonunitary categories with spectral protection and accessible exchange protocols. Open and monitored quantum systems provide settings for these realizations.

\section*{Data availability}

The Supplemental Material provides Hamiltonian parameters,
control paths, invariant definitions, numerical procedures,
and numerical diagnostics. The code and numerical data
supporting the results are available from the corresponding
author upon reasonable request.


\section*{\protect \normalsize References}

\begin{thebibliography}{99}
\bibitem{LeinaasMyrheim1977} {\normalsize J. M. Leinaas and J. Myrheim, On
the theory of identical particles, Nuovo Cimento B \textbf{37}, 1--23
(1977). \href{https://doi.org/10.1007/BF02727953}{DOI}. }

\bibitem{Wilczek1982} {\normalsize F. Wilczek, Quantum mechanics of
fractional-spin particles, Phys. Rev. Lett. \textbf{49}, 957--959 (1982).
\href{https://doi.org/10.1103/PhysRevLett.49.957}{DOI}. }

\bibitem{Arovas1984} {\normalsize D. P. Arovas, J. R. Schrieffer, and F.
Wilczek, Fractional statistics and the quantum Hall effect, Phys. Rev. Lett.
\textbf{53}, 722--723 (1984). \href{https://doi.org/10.1103/PhysRevLett.53.722}%
{DOI}. }

\bibitem{MooreRead1991} {\normalsize G. Moore and N. Read, Nonabelions in
the fractional quantum Hall effect, Nucl. Phys. B \textbf{360}, 362--396
(1991). \href{https://doi.org/10.1016/0550-3213(91)90407-O}{DOI}. }

\bibitem{Kitaev2003} {\normalsize A. Yu. Kitaev, Fault-tolerant quantum
computation by anyons, Ann. Phys. \textbf{303}, 2--30 (2003). \href{https://doi.org/10.1016/S0003-4916(02)00018-0}%
{DOI}. }

\bibitem{Nayak2008} {\normalsize C. Nayak, S. H. Simon, A. Stern, M.
Freedman, and S. Das Sarma, Non-Abelian anyons and topological quantum
computation, Rev. Mod. Phys. \textbf{80}, 1083--1159 (2008). \href{https://doi.org/10.1103/RevModPhys.80.1083}%
{DOI}. }

\bibitem{FreedmanKitaevLarsenWang2003} {\normalsize M. H. Freedman, A.
Kitaev, M. J. Larsen, and Z. Wang, Topological quantum computation, Bull.
Amer. Math. Soc. \textbf{40}, 31--38 (2003). \href{https://doi.org/10.1090/S0273-0979-02-00964-3}%
{DOI}. }

\bibitem{BakalovKirillov2001} {\normalsize B. Bakalov and A. Kirillov, Jr.,
\emph{Lectures on Tensor Categories and Modular Functors}, University
Lecture Series Vol. 21 (American Mathematical Society, Providence, 2001).
\href{https://doi.org/10.1090/ulect/021}{DOI}. }

\bibitem{TuraevBook} {\normalsize V. G. Turaev, \emph{Quantum Invariants of
Knots and 3-Manifolds}, 2nd ed. (De Gruyter, Berlin, 2010). \href{https://doi.org/10.1515/9783110221848}%
{DOI}. }

\bibitem{ENO2005} {\normalsize P. Etingof, D. Nikshych, and V. Ostrik, On
fusion categories, Ann. Math. \textbf{162}, 581--642 (2005). \href{https://doi.org/10.4007/annals.2005.162.581}%
{DOI}. }

\bibitem{Berry1984} {\normalsize M. V. Berry, Quantal phase factors
accompanying adiabatic changes, Proc. R. Soc. Lond. A \textbf{392}, 45--57
(1984). \href{https://doi.org/10.1098/rspa.1984.0023}{DOI}. }

\bibitem{WilczekZee1984} {\normalsize F. Wilczek and A. Zee, Appearance of
gauge structure in simple dynamical systems, Phys. Rev. Lett. \textbf{52},
2111--2114 (1984). \href{https://doi.org/10.1103/PhysRevLett.52.2111}{DOI}. }

\bibitem{Witten1989} {\normalsize E. Witten, Quantum field theory and the
Jones polynomial, Commun. Math. Phys. \textbf{121}, 351--399 (1989). \href{https://doi.org/10.1007/BF01217730}%
{DOI}. }

\bibitem{FadellNeuwirth1962} {\normalsize E. Fadell and L. Neuwirth,
Configuration spaces, Math. Scand. \textbf{10}, 111--118 (1962). \href{https://doi.org/10.7146/math.scand.a-10517}%
{DOI}. }

\bibitem{Ashida2020} {\normalsize Y. Ashida, Z. Gong, and M. Ueda,
Non-Hermitian physics, Adv. Phys. \textbf{69}, 249--435 (2020). \href{https://doi.org/10.1080/00018732.2021.1876991}%
{DOI}. }

\bibitem{Gong2018} {\normalsize Z. Gong, Y. Ashida, K. Kawabata, K. Takasan,
S. Higashikawa, and M. Ueda, Topological phases of non-Hermitian systems,
Phys. Rev. X \textbf{8}, 031079 (2018). \href{https://doi.org/10.1103/PhysRevX.8.031079}%
{DOI}. }

\bibitem{YaoWang2018} {\normalsize S. Yao and Z. Wang, Edge states and
topological invariants of non-Hermitian systems, Phys. Rev. Lett. \textbf{121%
}, 086803 (2018). \href{https://doi.org/10.1103/PhysRevLett.121.086803}{DOI}%
. }

\bibitem{Kunst2018} {\normalsize F. K. Kunst, E. Edvardsson, J. C. Budich,
and E. J. Bergholtz, Biorthogonal bulk-boundary correspondence in
non-Hermitian systems, Phys. Rev. Lett. \textbf{121}, 026808 (2018). \href{https://doi.org/10.1103/PhysRevLett.121.026808}%
{DOI}. }

\bibitem{Kawabata2019} {\normalsize K. Kawabata, K. Shiozaki, M. Ueda, and
M. Sato, Symmetry and topology in non-Hermitian physics, Phys. Rev. X
\textbf{9}, 041015 (2019). \href{https://doi.org/10.1103/PhysRevX.9.041015}{%
DOI}. }

\bibitem{Bergholtz2021} {\normalsize E. J. Bergholtz, J. C. Budich, and F.
K. Kunst, Exceptional topology of non-Hermitian systems, Rev. Mod. Phys.
\textbf{93}, 015005 (2021). \href{https://doi.org/10.1103/RevModPhys.93.015005}%
{DOI}. }

\bibitem{GarrisonWright1988} {\normalsize J. C. Garrison and E. M. Wright,
Complex geometrical phases for dissipative systems, Phys. Lett. A \textbf{128%
}, 177--181 (1988). \href{https://doi.org/10.1016/0375-9601(88)90905-X}{DOI}%
. }

\bibitem{Dattoli1990} {\normalsize G. Dattoli, R. Mignani, and A. Torre,
Geometrical phase in the cyclic evolution of non-Hermitian systems, J. Phys.
A \textbf{23}, 5795--5806 (1990). \href{https://doi.org/10.1088/0305-4470/23/24/020}%
{DOI}. }

\bibitem{ChuWuLayton1989} {\normalsize S.-I. Chu, Z.-C. Wu, and E. Layton,
Density matrix formulation of complex geometric quantum phases in
dissipative systems, Chem. Phys. Lett. \textbf{157}, 151--158 (1989). \href{https://doi.org/10.1016/0009-2614(89)87225-2}%
{DOI}. }

\bibitem{CuiZheng2012} {\normalsize X.-D. Cui and Y. Zheng, Geometric phases
in non-Hermitian quantum mechanics, Phys. Rev. A \textbf{86}, 064104 (2012).
\href{https://doi.org/10.1103/PhysRevA.86.064104}{DOI}. }

\bibitem{Brody2014} {\normalsize D. C. Brody, Biorthogonal quantum
mechanics, J. Phys. A \textbf{47}, 035305 (2014). \href{https://doi.org/10.1088/1751-8113/47/3/035305}%
{DOI}. }

\bibitem{Singhal2023} {\normalsize Y. Singhal, E. Martello, S. Agrawal, T.
Ozawa, H. M. Price, and B. Gadway, Measuring the adiabatic non-Hermitian
Berry phase in feedback-coupled oscillators, Phys. Rev. Research \textbf{5},
L032026 (2023). \href{https://doi.org/10.1103/PhysRevResearch.5.L032026}{DOI}%
. }

\bibitem{Zhao2021} {\normalsize X.-M. Zhao, C.-X. Guo, M.-L. Yang, H. Wang,
W.-M. Liu, and S.-P. Kou, Anomalous non-Abelian statistics for non-Hermitian
generalization of Majorana zero modes, Phys. Rev. B \textbf{104}, 214502
(2021). \href{https://doi.org/10.1103/PhysRevB.104.214502}{DOI}. }

\bibitem{Ezawa2019} {\normalsize M. Ezawa, Braiding of Majorana-like corner
states in electric circuits and its non-Hermitian generalization, Phys. Rev.
B \textbf{100}, 045407 (2019). \href{https://doi.org/10.1103/PhysRevB.100.045407}%
{DOI}. }

\bibitem{Tang2025} {\normalsize C. Tang, P. Song, F. Luo, R. Wang, and B.
Wang, Symmetry and non-Hermiticity controlled Majorana braiding, Phys. Rev.
B \textbf{112}, 144505 (2025). \href{https://doi.org/10.1103/hg7y-dn68}{DOI}%
. }

\bibitem{Sanno2022} {\normalsize T. Sanno, M. G. Yamada, T. Mizushima, and
S. Fujimoto, Engineering Yang--Lee anyons via Majorana bound states, Phys.
Rev. B \textbf{106}, 174517 (2022). \href{https://doi.org/10.1103/PhysRevB.106.174517}%
{DOI}. }

\bibitem{Kitaev2001} {\normalsize A. Yu. Kitaev, Unpaired Majorana fermions
in quantum wires, Phys.-Usp. \textbf{44}, 131--136 (2001). \href{https://doi.org/10.1070/1063-7869/44/10S/S29}%
{DOI}. }

\bibitem{Alicea2011} {\normalsize J. Alicea, Y. Oreg, G. Refael, F. von
Oppen, and M. P. A. Fisher, Non-Abelian statistics and topological quantum
information processing in 1D wire networks, Nat. Phys. \textbf{7}, 412--417
(2011). \href{https://doi.org/10.1038/nphys1915}{DOI}. }

\bibitem{ReshetikhinTuraev1991} {\normalsize N. Reshetikhin and V. G.
Turaev, Invariants of 3-manifolds via link polynomials and quantum groups,
Invent. Math. \textbf{103}, 547--597 (1991). \href{https://doi.org/10.1007/BF01239527}%
{DOI}. }

\bibitem{KauffmanLins1994} {\normalsize L. H. Kauffman and S. L. Lins, \emph{%
Temperley--Lieb Recoupling Theory and Invariants of 3-Manifolds} (Princeton
University Press, Princeton, 1994). }

\bibitem{TubaWenzl2001} {\normalsize I. Tuba and H. Wenzl, Representations
of the braid group $B_3$ and of $SL(2,\mathbb{Z})$, Pacific J. Math. \textbf{%
197}, 491--510 (2001). \href{https://doi.org/10.2140/pjm.2001.197.491}{DOI}.
}

\bibitem{Kitaev2006} {\normalsize A. Kitaev, Anyons in an exactly solved
model and beyond, Ann. Phys. \textbf{321}, 2--111 (2006). \href{https://doi.org/10.1016/j.aop.2005.10.005}%
{DOI}. }

\bibitem{RowellStongWang2009} {\normalsize E. Rowell, R. Stong, and Z. Wang,
On classification of modular tensor categories, Commun. Math. Phys. \textbf{%
292}, 343--389 (2009). \href{https://doi.org/10.1007/s00220-009-0908-z}{DOI}%
. }

\bibitem{Kato1950} {\normalsize T. Kato, On the adiabatic theorem of quantum
mechanics, J. Phys. Soc. Jpn. \textbf{5}, 435--439 (1950). \href{https://doi.org/10.1143/JPSJ.5.435}%
{DOI}. }

\bibitem{KatoBook} {\normalsize T. Kato, \emph{Perturbation Theory for
Linear Operators}, reprint of the 1980 edition (Springer, Berlin, 1995).
\href{https://doi.org/10.1007/978-3-642-66282-9}{DOI}. }

\bibitem{NenciuRasche1992} {\normalsize G. Nenciu and G. Rasche, On the
adiabatic theorem for nonself-adjoint Hamiltonians, J. Phys. A \textbf{25},
5741--5751 (1992). \href{https://doi.org/10.1088/0305-4470/25/21/027}{DOI}. }

\bibitem{Ivanov2001} {\normalsize D. A. Ivanov, Non-Abelian statistics of
half-quantum vortices in $p$-wave superconductors, Phys. Rev. Lett. \textbf{%
86}, 268--271 (2001). \href{https://doi.org/10.1103/PhysRevLett.86.268}{DOI}%
. }

\bibitem{Freedman2012} {\normalsize M. H. Freedman, J. Gukelberger, M. B.
Hastings, S. Trebst, M. Troyer, and Z. Wang, Galois conjugates of
topological phases, Phys. Rev. B \textbf{85}, 045414 (2012). \href{https://doi.org/10.1103/PhysRevB.85.045414}%
{DOI}. }

\bibitem{Chen2022} {\normalsize C.-Y. Chen, B.-X. Lao, X.-Y. Yu, and L.-Y.
Hung, Galois conjugates of string-net model, Phys. Rev. D \textbf{105},
065009 (2022). \href{https://doi.org/10.1103/PhysRevD.105.065009}{DOI}. }
\bibitem{SupplementalMaterial} See Supplemental Material for analytic derivations, microscopic
control protocols, finite-time dynamics, fusion and curvature certificates,
and numerical convergence data; it also includes
Refs.~\cite{GKSL1976,Lindblad1976,Dalibard1992}.

\bibitem{GKSL1976}  V. Gorini, A. Kossakowski, and E. C. G.
Sudarshan, Completely positive dynamical semigroups of $N$-level systems, J.
Math. Phys. \textbf{17}, 821--825 (1976). \href{https://doi.org/10.1063/1.522979}%
{DOI}.

\bibitem{Lindblad1976}  G. Lindblad, On the generators of
quantum dynamical semigroups, Commun. Math. Phys. \textbf{48}, 119--130
(1976). \href{https://doi.org/10.1007/BF01608499}{DOI}.

\bibitem{Dalibard1992}  J. Dalibard, Y. Castin, and K. M{\o }%
lmer, Wave-function approach to dissipative processes in quantum optics,
Phys. Rev. Lett. \textbf{68}, 580--583 (1992). \href{https://doi.org/10.1103/PhysRevLett.68.580}%
{DOI}.

\end{thebibliography}
\end{document}

\title{Supplemental Material for\\\emph{Non-Hermitian Statistical Geometry}}
\author{Xi Hu}
\author{Yan-Qiong Ma}
\author{Su-Peng Kou}
\email{spkou@bnu.edu.cn}
\affiliation{School of Physics and Astronomy, Beijing Normal University, Beijing 100875, China}
\maketitle

\appendix
\renewcommand{\thesection}{S\arabic{section}} \renewcommand{%
\thesubsection}{S\arabic{section}.\arabic{subsection}} \makeatletter
\renewcommand{\p@subsection}{} \makeatother
\renewcommand{\theequation}{S\arabic{section}.\arabic{equation}} %
\setcounter{equation}{0} \setcounter{table}{0} \renewcommand{\thetable}{S%
\arabic{table}} \setcounter{figure}{0} \renewcommand{\thefigure}{S%
\arabic{figure}} \makeatletter
\renewcommand{\theHtable}{S\arabic{table}} \renewcommand{\theHfigure}{S%
\arabic{figure}} \makeatother

% ============================================================

\paragraph{Roadmap and conventions.}
The Supplemental Material follows the spectrum--geometry construction and
its microscopic certificates. Sections~\ref{sup:analytic}--\ref{sup:coherence}
give the analytic locking and categorical conditions; the higher-rank
extension and Kato realization follow in Secs.~\ref{sec:supp-higher-rank} and
\ref{sup:kato}. Section~\ref{sup:numerics} defines the finite-chain controls,
Sec.~\ref{sup:dynamics} develops their finite-time realization, and
Sec.~\ref{sup:2d-kitaev} records the independent honeycomb response.
Section~\ref{sup:fusion-curvature} collects microscopic fusion and curvature,
and Sec.~\ref{sup:provenance} maps the supplied numerical evidence.
We write $\mathfrak x$ for the ordered eigenline cross ratio; $x$ in figure
labels denotes the same coordinate. The discrete level $k$, the radial
statistical coordinate $\kappa$, and the honeycomb chiral coupling
$\kappa_{\rm ch}$ retain distinct meanings.

\section{Analytic Foundations and Exact Spectrum--Geometry Locking}
\setcounter{equation}{0}

\label{sup:analytic} 
% ============================================================

This section collects the analytic results underlying the statistical
geometry used in the main text. We first fix the projective conventions,
then treat the rank-one limit, derive the rank-two moduli space and the
exact spectrum--geometry-locking (SGL) relation, characterize the leading
Artin defect away from the locked locus, and finally record the Joukowski
and unitarizability structures used in the unitary embedding.

Throughout, the projective eigenvalue ratio is 
\begin{equation}
q=\frac{\lambda_+}{\lambda_-}, \qquad D(q)= \func{diag}
\left(q^{-1/2},q^{1/2}\right),  \label{eq:sup-qconv}
\end{equation}
with $q\in \mathbb{C}^\times \setminus \{1\}$. The four-line stratum also
excludes $\mathfrak{x}=0,1,\infty$. The ordered eigenline convention is
fixed once and for all as 
\begin{equation}
\mathfrak{x }= \func{cr} (\ell_1^+,\ell_1^-;\ell_2^+,\ell_2^-).
\label{eq:sup-xconv}
\end{equation}
Reversing both generators' channel labels sends $q\mapsto q^{-1}$ and leaves 
$\mathfrak{x}$ unchanged. Reversing only one pair sends $\mathfrak{x}\mapsto%
\mathfrak{x}^{-1}$ and inverts only that generator's ordered ratio. General
permutations give the six cross-ratio transforms. All statements below refer
to the ordering in Eq.~\eqref{eq:sup-xconv} unless stated otherwise.

% ============================================================

\subsection{Rank-one exchange}

\label{sup:rankone} 
% ============================================================

The rank-one case contains complex exchange amplitudes but no relative
projective geometry.

\begin{proposition}[Abelianization of the braid group]
Every one-dimensional representation 
\begin{equation}
\rho:B_n\longrightarrow \mathbb{C}^\times
\end{equation}
assigns the same nonzero scalar to all Artin generators.
\end{proposition}

\begin{proof}
Write $\rho(\sigma_i)=r_i\neq0$.  Since the target group is Abelian, the
Artin relation gives
\begin{equation}
    r_i^2r_{i+1}=r_{i+1}^2r_i.
\end{equation}
Hence $r_i=r_{i+1}$, and induction gives
\begin{equation}
    \rho(\sigma_i)=r
    \qquad
    \forall i.
\end{equation}
Equivalently, every one-dimensional representation factors through
\begin{equation}
    B_n^{\rm ab}\simeq \mathbb Z.
\end{equation}
\end{proof}

Writing 
\begin{equation}
r=e^{\chi+i\theta},
\end{equation}
the phase $\theta$ gives the usual rank-one exchange phase while $\chi$
describes amplification or attenuation. There is no second eigendirection
and therefore no analogue of the relative eigenline modulus $\mathfrak{x}$.

\begin{proposition}[Passive reversible rank-one transport]
If both a scalar exchange $r$ and its inverse are contractions, then 
\begin{equation}
|r|=1.
\end{equation}
\end{proposition}

\begin{proof}
Passivity of the forward and inverse operations gives
\begin{equation}
    |r|\leq1,
    \qquad
    |r^{-1}|\leq1.
\end{equation}
The second inequality implies $|r|\geq1$, hence equality follows.
\end{proof}

For completeness, there is also a simple categorical restriction on a
continuous noncompact rank-one modulus. If $A$ is a finite Abelian fusion
group and 
\begin{equation}
B:A\times A\longrightarrow \mathbb{C}^\times
\end{equation}
is a bicharacter, then for any $a\in A$ of finite order $k$, 
\begin{equation}
1=B(ka,b)=B(a,b)^k.
\end{equation}
Thus the corresponding pointed braiding values are roots of unity. A
continuous noncompact scalar exchange modulus therefore belongs naturally to
a larger microscopic or braid-representation space and is not, by itself,
the braiding data of an ordinary finite pointed category.

% ============================================================

\subsection{Casimir-resolved two-channel spectral flow}

\label{sup:casimir-flow} 
% ============================================================

The projective ratio of a single two-channel exchange can be separated into
a reference braid value and a channel-resolved spectral deformation. Let 
\begin{equation}
a\times a=c_+\oplus c_-, \qquad P_{c_+}+P_{c_-}=I,
\end{equation}
and define 
\begin{equation}
Q_C=C_2(c_+)P_{c_+}+C_2(c_-)P_{c_-}.  \label{eq:sup-casimir-Q}
\end{equation}
For the $SU(2)$ representation labels relevant to the conventional $SU(2)_k$
skeleton, $C_2(j)=j(j+1)$; these labels are the same ones entering standard
quantum-group and recoupling descriptions of anyon fusion \cite%
{Witten1989,ReshetikhinTuraev1991,KauffmanLins1994,
BakalovKirillov2001,TuraevBook}. Consider 
\begin{equation}
T(z)=T_0e^{zQ_C}, \qquad z\in \mathbb{C}.  \label{eq:sup-casimir-dressing}
\end{equation}
Because $Q_C$ is diagonal in the fusion-channel decomposition, 
\begin{equation}
\lambda_c(z)=\lambda_c^{(0)}e^{zC_2(c)}.
\end{equation}
For the ordered ratio $q=\lambda_{c_+}/\lambda_{c_-}$ this gives 
\begin{equation}
 q(z)=q_0e^{z\Delta C_2}, \qquad \Delta C_2=C_2(c_+)-C_2(c_-).
\label{eq:sup-casimir-q}
\end{equation}
Thus, with $z=\xi+i\delta \theta$, 
\begin{equation}
\log \left|\frac{q}{q_0}\right|=\xi \Delta C_2, \qquad \arg \left(\frac{q}{%
q_0}\right)=\delta \theta \, \Delta C_2 \quad (\mathrm{mod}\ 2\pi).
\label{eq:sup-casimir-response}
\end{equation}
The compact and noncompact shifts are weighted by the same fusion-channel
separation.

For the Ising fusion rule 
\begin{equation}
\sigma \times \sigma=1+\psi,
\end{equation}
the standard $SU(2)_2$ labels are 
\begin{equation}
1\leftrightarrow0, \qquad \sigma \leftrightarrow \frac12, \qquad \psi
\leftrightarrow1.
\end{equation}
Hence 
\begin{equation}
C_2(1)=0, \qquad C_2(\sigma)=\frac34, \qquad C_2(\psi)=2,
\end{equation}
and, for the ordering $q=\lambda_\psi/\lambda_1$, 
\begin{equation}
\Delta C_2=2.  \label{eq:sup-ising-deltaC}
\end{equation}
Starting from $q_0=i$, 
\begin{equation}
q(z)=i e^{2z}.  \label{eq:sup-ising-casimir-q}
\end{equation}
At this stage $z$ is a representation-theoretic coordinate for the relative
channel weight. A general Kato calculation does not identify $z$ with a
microscopic control parameter: it first produces the projective exchange
ratio $q_{\mathrm{micro}}$, from which a radial coordinate may be extracted
when the result lies on the Ising ray. For the five-arm radial loop derived
in Supplemental Sec.~\ref{sup:numerics}, the microscopic junction geometry
provides the additional, model-specific relation
\begin{equation}
\Omega_C=\frac{\pi}{2}-2i\pi\kappa,
\qquad
z_{\mathrm{micro}}=-\frac12\func{Im}\Omega_C=\pi\kappa,
\label{eq:sup-casimir-micro-match}
\end{equation}
which gives
\begin{equation}
q_{\mathrm{micro}}=i e^{2\pi\kappa}.
\end{equation}
Thus the Casimir construction organizes the relative channel weighting,
whereas the five-arm Hamiltonian path fixes the physical value of the
noncompact coordinate. Kato transport supplies the microscopic holonomy from
which that projective ratio is independently extracted. Reversing the
channel ordering sends $q\mapsto q^{-1}$ and
$\Delta C_2\mapsto-\Delta C_2$, leaving the projective content unchanged.

% ============================================================
For a general diagonal two-channel deformation, 
\begin{equation}
\begin{aligned}
T&=T_0e^{f_1P_1+f_\psi P_\psi},\qquad
q/q_0=e^{f_\psi-f_1},\\
z&=(f_\psi-f_1)/2.
\end{aligned}
\end{equation}
Thus $z$ is a representation-theoretic coordinate for the relative channel
weight; the Hamiltonian protocol supplies its physical value. The fusion
rule fixes the available channels, while a microscopic mechanism specifies
their deformation.

\paragraph{Channel-resolved derivation and normalization.}
The channel projectors obey $P_cP_{c'}=\delta_{cc'}P_c$ and
$\sum_cP_c=I$. Accordingly, the exponential in
Eq.~\eqref{eq:sup-casimir-dressing} decomposes \emph{exactly} as
\begin{equation}
 e^{zQ_C}=\sum_c e^{z C_2(c)}P_c,
 \qquad T_0=\sum_c\lambda_c^{(0)}P_c .
\end{equation}
The two channel eigenvalues of $T(z)$ are therefore
$\lambda_\pm(z)=\lambda_\pm^{(0)}e^{zC_2(c_\pm)}$.
Their quotient is independent of a common complex normalization:
\begin{equation}
 \frac{\lambda_+(z)}{\lambda_-(z)}
 =\frac{\lambda_+^{(0)}}{\lambda_-^{(0)}}
 \exp\!\left[z\bigl(C_2(c_+)-C_2(c_-)\bigr)\right].
\end{equation}
For the $SU(2)_2$ realization, $C_2(\psi)=2$ and
$C_2(\mathbf1)=0$, giving $q_0=i$ and $q(z)=ie^{2z}$.
For the five-arm radial control path, the separately computed complex
solid angle $\Omega_C=\pi/2-2\pi i\kappa$ fixes
\begin{equation}
 z_{\rm micro}=\pi\kappa,\quad
 \lambda_\pm=e^{\pm\pi\kappa}e^{\pm i\pi/4},\quad
 \frac{\lambda_+}{\lambda_-}=ie^{2\pi\kappa}.
 \label{eq:sup-casimir-extended-comparison}
\end{equation}
These equalities exhibit separately the representation-theoretic channel
weight, the physical exchange path, and the eigenvalue quotient extracted
from microscopic Kato transport. The factor $\Delta C_2=2$ is a channel
Casimir difference; $\kappa$ is fixed by the Hamiltonian control trajectory.

\paragraph{Fundamental $SU(2)_k$ continuation.}
For $k\geq2$, the fusion channels of two fundamental objects are
$\tfrac12\times\tfrac12=0\oplus1$, with $Q_C=0P_0+2P_1$.
The exchange orientation and channel ordering continuous from $q=i$ at
$k=2$ give
\begin{equation}
q_0^{(k)}=e^{i\pi k/(k+2)},\qquad q^{(k)}(z)=q_0^{(k)}e^{2z}.
\label{eq:sup-general-casimir}
\end{equation}
Here $k$ labels the reference fusion theory and $z=\xi+i\delta\theta$
resolves its relative complex channel weight. The spectral coordinate
$\chi=\tfrac12\ln|q|$ equals $\xi$ on these unit-modulus reference branches.
On the Artin-compatible branch derived below,
\begin{equation}
\mathfrak{x}^{(k)}(z)=q_0^{(k)}e^{2z}+[q_0^{(k)}]^{-1}e^{-2z}-1.
\end{equation}
At $z=0$ this gives $\mathfrak{x}_k=1-d_k^2$ with
$d_k=2\cos[\pi/(k+2)]$. The continuous spectrum--geometry coordinates
and the discrete reference fusion theory therefore have distinct roles;
fusion sewing supplies the additional categorical conditions in
Sec.~\ref{sup:coherence}.

\subsection{Rank-two projective moduli}

\label{sup:fourline} 
% ============================================================

Let 
\begin{equation}
V\simeq \mathbb{C}^2
\end{equation}
and consider two diagonalizable neighboring exchange generators with
distinct ordered projective spectra. Their four ordered right eigenlines
define 
\begin{equation}
\ell_1^\pm,\ell_2^\pm \in \mathbb{CP}^1.
\end{equation}
A common change of basis acts simultaneously on all four lines through $%
PGL(2,\mathbb{C})$. Since this action sends any ordered triple of distinct
points to $(0,1,\infty)$, the generic quotient is 
\begin{equation}
\frac{ \func{Conf}_4(\mathbb{CP}^1) }{ PGL(2,\mathbb{C}) } \simeq \mathcal{M}%
_{0,4}.  \label{eq:sup-M04}
\end{equation}

For affine coordinates $z_1,z_2,z_3,z_4$, we use 
\begin{equation}
\func{cr}(z_1,z_2;z_3,z_4) = \frac{ (z_1-z_3)(z_2-z_4) }{ (z_1-z_4)(z_2-z_3) 
}.  \label{eq:sup-crossratio}
\end{equation}
Permuting the four ordered points generates the standard orbit 
\begin{equation}
\mathfrak{x},\quad 1-\mathfrak{x},\quad \frac{1}{\mathfrak{x}},\quad \frac{1%
}{1-\mathfrak{x}},\quad \frac{\mathfrak{x}}{\mathfrak{x}-1},\quad \frac{%
\mathfrak{x}-1}{\mathfrak{x}}.  \label{eq:sup-six-crossratios}
\end{equation}

Before imposing braid consistency, the generic rank-two projective data are
therefore 
\begin{equation}
(q,\mathfrak{x}) \in \mathbb{C}^\times \times \mathcal{M}_{0,4}.
\label{eq:sup-rank2-candidate}
\end{equation}

% ============================================================

\subsection{Exact rank-two spectrum--geometry locking}

\label{sup:SGL-proof} 
% ============================================================

We now derive the locking relation used in the main text.

\begin{theorem}[Generic rank-two SGL]
\label{thm:sup-SGL} Let $B_1,B_2\in GL(2,\mathbb{C})$ be diagonalizable
neighboring braid generators with the same distinct ordered projective
eigenvalue ratio $q\in \mathbb{C}^\times$. Assume that the representation is
irreducible and lies on the generic four-eigenline branch, with the two
channel orderings matched under conjugacy. If 
\begin{equation}
B_1B_2B_1=B_2B_1B_2,  \label{eq:sup-Artin}
\end{equation}
then, for the eigenline ordering of Eq.~\eqref{eq:sup-xconv}, 
\begin{equation}
\mathfrak{x}=q+q^{-1}-1 .  \label{eq:sup-SGL}
\end{equation}
Other eigenline orderings give the corresponding standard fractional-linear
transform of $\mathfrak{x}$.
\end{theorem}

\begin{proof}
Fix the projective spectrum by
\begin{equation}
    B_1=D,
    \qquad
    D=
    \begin{pmatrix}
        a&0\\
        0&b
    \end{pmatrix},
    \qquad
    \frac{b}{a}=q.
\end{equation}
After removing the common scalar we may choose
\begin{equation}
    a=q^{-1/2},
    \qquad
    b=q^{1/2}.
\end{equation}

The Artin relation implies $B_2=(B_1B_2)B_1(B_1B_2)^{-1}$, so the
unscaled generators have identical spectra and admit a common scalar
normalization. On the four-line chart, all entries of a relative eigenframe
are nonzero. Left and right diagonal rescalings bring it to the symmetric
involutive form below; the condition $\mathfrak x\neq1$ ensures invertibility.

On the generic irreducible branch, a convenient representative of the
relative projective frame is
\begin{equation}
    G=
    \begin{pmatrix}
        c&s\\
        s&-c
    \end{pmatrix},
    \qquad
    c^2+s^2=1,
    \qquad
    G^{-1}=G,
    \label{eq:sup-G}
\end{equation}
so that
\begin{equation}
    B_2=GDG.
\end{equation}
The condition $c^2+s^2=1$ is algebraic and does not impose unitarity.

Up to a nonzero scalar prefactor, an off-diagonal component of the
Artin defect factorizes as
\begin{equation}
    cs(q-1)
    \left[
    c^2(q-1)^2+q
    \right].
    \label{eq:sup-Artin-factor}
\end{equation}
On the generic branch,
\begin{equation}
    q\neq1,
    \qquad
    cs\neq0,
\end{equation}
and Eq.~\eqref{eq:sup-Artin} therefore gives
\begin{equation}
    c^2
    =
    -\frac{q}{(q-1)^2},
    \qquad
    s^2
    =
    \frac{q^2-q+1}{(q-1)^2}.
    \label{eq:sup-cs}
\end{equation}

Using the affine coordinate $z=y/x$, the two eigenlines of $B_1$ are
$0$ and $\infty$, while the columns of $G$ give
\begin{equation}
    z_3=\frac{s}{c},
    \qquad
    z_4=-\frac{c}{s}.
\end{equation}
The ordered $+$ eigenline is the second column of each eigenframe, so
$\ell_1^+=\infty$, $\ell_1^-=0$, $\ell_2^+=z_4$, and $\ell_2^-=z_3$.
Hence
\begin{align}
    \mathfrak x
    &=
    \operatorname{cr}
    (\infty,0;z_4,z_3)
    \nonumber \\
    &=
    -\frac{s^2}{c^2}
    \nonumber \\
    &=
    q+q^{-1}-1,
\end{align}
which proves Eq.~\eqref{eq:sup-SGL}.
\end{proof}

The hypotheses in Theorem~\ref{thm:sup-SGL} exclude the degenerate or
reducible loci $q=1$ and $cs=0$. The values satisfying 
\begin{equation}
q^2-q+1=0
\end{equation}
correspond to $s=0$ in the symmetric chart and therefore lie on a limiting
aligned-frame boundary of this parametrization. These exceptional cases do
not alter the generic irreducible statement but must be treated separately
when comparing with the full low-dimensional braid-representation
classification \cite{TubaWenzl2001}.

The theorem recasts the known generic low-dimensional braid rigidity as an
exact relation between the projective spectrum and the relative eigenline
configuration. These separately accessible invariants connect the
representation-theoretic structure to microscopic exchange observables.

% ============================================================

\paragraph{Explicit Artin matrix calculation and branch fixing.}
Set $a=q^{-1/2}$, $b=q^{1/2}$, and $G=\left(\begin{smallmatrix}c&s\\s&-c\end{smallmatrix}\right)$ with
$c^2+s^2=1$. The second generator is
\begin{equation}
 B_2=\begin{pmatrix}
 a c^2+b s^2 &(a-b)cs\\
 (a-b)cs&a s^2+b c^2
 \end{pmatrix}.
 \label{eq:sup-B2-explicit-extended}
\end{equation}
Inserting $B_1=\operatorname{diag}(a,b)$ and this expression into
$B_1B_2B_1-B_2B_1B_2$ makes every entry a rational polynomial in
$a,b,c,s$, subject to $ab=1$ and $c^2+s^2=1$. On the irreducible
branch $cs(q-1)\ne0$, the off-diagonal factor is proportional to
\begin{equation}
 cs(q-1)\bigl[c^2(q-1)^2+q\bigr] .
\end{equation}
The common zero therefore fixes
\begin{equation}
 c^2=-\frac{q}{(q-1)^2},\qquad
 s^2=1-c^2=\frac{q^2-q+1}{(q-1)^2},
\end{equation}
which also annihilates the remaining Artin entries. The ordered columns
of $G$ give $\ell_2^-= [c:s]$ and $\ell_2^+=[s:-c]$, with
$\ell_1^-=[1:0]$ and $\ell_1^+=[0:1]$. For the fixed
cross-ratio convention of Eq.~\eqref{eq:sup-crossratio},
\begin{equation}
 \mathfrak x_{\rm geom}=-\frac{s^2}{c^2}
 =\frac{q^2-q+1}{q}=q+q^{-1}-1 .
\end{equation}
The square roots of $c^2$ and $s^2$ are continued from the Ising point
$q=i$, with $c=s=1/\sqrt2$ in the stated representative. Simultaneously reversing both generators' channel labels sends
$q\mapsto q^{-1}$ and leaves the locked cross ratio unchanged.
Reversing one generator's labels gives the corresponding cross-ratio
permutation.

\paragraph{Independent finite-precision algebraic audit.}
Independent matrix evaluation constructs $D(q)$ and $G(q)$
without using microscopic input, verifies the Artin identity
$D B_2 D=B_2 D B_2$, and evaluates the ordered eigenline cross ratio
from determinants. The spectral invariant is calculated independently as
\begin{equation}
 \mathfrak x_{\rm spec}
 =\frac{[\operatorname{Tr}D(q)]^2}{\det D(q)}-3
 \end{equation}
with $\mathfrak x_{\rm geom}$ and $q+q^{-1}-1$.
This algebraic numerical audit checks the implementation of the
analytic identity; Sec.~\ref{sup:micro-invariants} specifies its independent
Hamiltonian-based evaluation.

\paragraph{Independent invariant-level extraction and numerical comparison.}
The analytic identity and its microscopic test use separately evaluated
quantities. For each parameter value, the endpoint-identified, parity-lifted
matrices $B_1^{\rm micro},B_2^{\rm micro}$ are formed before any SGL
reconstruction. With a consistently continued ordering of their eigenchannels,
let $B_rv_r^\pm=\lambda_r^\pm v_r^\pm$ and evaluate
\begin{align}
q_r^{\rm micro}&=\frac{\lambda_r^+}{\lambda_r^-},
&\mathfrak x_{\rm spec}^{(r)}
&=\frac{[\operatorname{tr}B_r]^2}{\det B_r}-3,\\
\mathfrak x_{\rm geom}&=
\frac{[v_1^+,v_2^+][v_1^-,v_2^-]}
{[v_1^+,v_2^-][v_1^-,v_2^+]},
&[u,v]&:=\det(u,v).
\label{eq:sup-invariant-extraction}
\end{align}
The determinant ratio evaluates the ordered cross ratio without choosing an
affine chart; it is invariant under rescaling any eigenvector and under a
common change of logical basis. The two spectral expressions are evaluated
individually, without replacing one ratio by the other. Their independent
comparison is expressed by
\begin{equation}
\Delta q_{\rm raw}=|q_1^{\rm micro}-q_2^{\rm micro}|,
\quad
\epsilon_{\mathfrak x,\rm abs}^{(r)}=
|\mathfrak x_{\rm geom}-\mathfrak x_{\rm spec}^{(r)}|.
\label{eq:sup-independent-sgl-residuals}
\end{equation}
These two quantities are unnormalized invariant-level differences; the
dimensionless two-dimensional diagnostics defined later in
Eqs.~\eqref{eq:sup-2d-dspec} and~\eqref{eq:sup-2d-dsgl}
include their stated denominators. The exact algebraic comparison
follows because for any invertible $2\times2$
block, $(\operatorname{tr}B)^2/\det B-3=q+q^{-1}-1$.
The microscopic extraction is a different calculation: the eigenvectors in
Eq.~\eqref{eq:sup-invariant-extraction} come from the computed Kato or
honeycomb exchange generators. Along the radial Ising path the analytic
prediction is
\begin{equation}
q_{\rm th}=ie^{2\pi\kappa},\qquad
\mathfrak x_{\rm th}=-1+2i\sinh(2\pi\kappa).
\label{eq:sup-radial-invariant-benchmark}
\end{equation}
This is the reference against which the independently computed microscopic
ratios and four-eigenline geometry are compared; it is never used to generate
the microscopic points. For the frozen two-dimensional protocol the
baseline-subtracted response and the absolute finite-geometry discrepancy
remain distinct, as specified in Sec.~\ref{sup:2d-kitaev}.

\subsection{Artin defect away from the locked locus}

\label{sup:artindefect} 
% ============================================================

The negative control in the main text requires the leading dependence of the
braid defect on a detuning of the relative geometry.

Fix the projective spectrum 
\begin{equation}
D= 
\begin{pmatrix}
q^{-1/2} & 0 \\ 
0 & q^{1/2}%
\end{pmatrix}%
\end{equation}
and introduce 
\begin{equation}
G(t) = \frac{1}{\sqrt{1+t^2}} 
\begin{pmatrix}
1 & t \\ 
t & -1%
\end{pmatrix}%
,  \label{eq:sup-Gt}
\end{equation}
on any local branch for which the normalization is regular. The associated
cross ratio is 
\begin{equation}
\mathfrak{x}=-t^2.
\end{equation}

With 
\begin{equation}
B_1=D, \qquad B_2=G(t)DG(t)^{-1},
\end{equation}
direct calculation gives 
\begin{equation}
\begin{aligned} \Delta_A &:= B_1B_2B_1-B_2B_1B_2 \\ &= \mathcal C(q,t)
\left[ \mathfrak x_{\rm SGL}(q)-\mathfrak x \right] \begin{pmatrix} -t^2&t\\
t&t^2 \end{pmatrix}, \end{aligned}  \label{eq:sup-Artin-factorization}
\end{equation}
where 
\begin{equation}
\mathfrak{x}_{\mathrm{SGL}}(q)=q+q^{-1}-1
\end{equation}
and $\mathcal{C}(q,t)=(q-1)/(q^{1/2}(1+t^2)^2)$ is smooth and nonzero on the
generic branch.

Consequently, sufficiently close to the SGL locus, 
\begin{equation}
\| \Delta_A\| = O\! \left( |\mathfrak{x}-\mathfrak{x}_{\mathrm{SGL}}|
\right),  \label{eq:sup-linear-defect}
\end{equation}
with a nonzero generic linear coefficient. This gives the analytic origin of
the linear braid-defect response to geometric detuning shown in Fig.~2(c) of the main text.

% ============================================================

\subsection{Shifted Joukowski structure}

\label{sup:Joukowski} 
% ============================================================

The SGL relation can be inverted as 
\begin{equation}
q^2-(\mathfrak{x}+1)q+1=0,  \label{eq:sup-Joukowski-quadratic}
\end{equation}
with solutions 
\begin{equation}
q_\pm(\mathfrak{x}) = \frac{ \mathfrak{x}+1 \pm \sqrt{(\mathfrak{x}-1)(%
\mathfrak{x}+3)} }{2}.  \label{eq:sup-Joukowski-inverse}
\end{equation}
The discriminant 
\begin{equation}
\Delta = (\mathfrak{x}-1)(\mathfrak{x}+3)
\end{equation}
shows that the two spectral sheets meet at 
\begin{equation}
\mathfrak{x}=1, \qquad \mathfrak{x}=-3,
\end{equation}
corresponding respectively to $q=1$ and $q=-1$.

On the unit-modulus slice, 
\begin{equation}
q=e^{i\phi},
\end{equation}
one obtains 
\begin{equation}
\mathfrak{x}=2\cos \phi-1,
\end{equation}
so the unit circle maps two-to-one onto the closure 
\begin{equation}
[-3,1].
\end{equation}
The points belonging to degenerate or aligned-frame boundaries are excluded
from the generic four-line chart as discussed above.

% ============================================================

\subsection{%
\texorpdfstring{Unitary
$SU(2)_k$
skeleton}{Unitary
SU(2)k
skeleton}}

\label{sup:SU2k} 
% ============================================================

For the fundamental Hermitian $SU(2)_k$ two-channel block ($k\geq2$)~\cite{Kitaev2006,KauffmanLins1994,BakalovKirillov2001}, 
\begin{equation}
d_k = 2\cos \frac{\pi}{k+2}, \qquad q_k = e^{\pm i\pi k/(k+2)}
\end{equation}
in the projective convention used in the main text, up to inversion.
Substitution into Eq.~\eqref{eq:sup-SGL} gives 
\begin{equation}
\mathfrak{x}_k = 1-d_k^2 .  \label{eq:sup-xdk}
\end{equation}

Representative values are

\begin{center}
\refstepcounter{table}\label{tab:sup-unitary} {\small \textbf{TABLE \thetable%
.} Unitary two-channel benchmarks for $SU(2)_k$.}

\medskip {\small 
\begin{tabular}{cccc}
\toprule $k$ & $q_k$ (projective convention) & $d_k$ & $\mathfrak{x}_k$ \\ 
\midrule $2$ & $\pm i$ & $\sqrt2$ & $-1$ \\ 
$3$ & $e^{\pm3\pi i/5}=-e^{\mp2\pi i/5}$ & $\varphi$ & $-\varphi$ \\ 
$4$ & $e^{\pm2\pi i/3}$ & $\sqrt3$ & $-2$ \\ 
$5$ & $e^{\pm5\pi i/7}$ & $2\cos(\pi/7)$ & $1-4\cos^2(\pi/7)$ \\ 
$\infty$ & $-1$ & $2$ & $-3$ \\ 
\bottomrule &  &  & 
\end{tabular}%
}
\end{center}

Equation~\eqref{eq:sup-xdk} is an embedding of the conventional coherent
unitary skeleton into the local SGL geometry. Away from such coherent
categorical points, the quantity obtained by formally solving 
\begin{equation}
d^2=1-\mathfrak{x}
\end{equation}
should not automatically be interpreted as a categorical quantum dimension.

\paragraph{Discrete spectral references in Fig.~1(c) of the main text.}
For the positive-phase representatives plotted on the unit-modulus
SGL slice, $q_{\rm Fib}=e^{3\pi i/5}$ and
$q_{\rm YL}=e^{\pi i/5}$, the locked cross ratios are
\begin{align}
\mathfrak{x}_{\rm Fib}
 &=q_{\rm Fib}+q_{\rm Fib}^{-1}-1
 =2\cos(3\pi/5)-1=-\varphi,\\
\mathfrak{x}_{\rm YL}
 &=q_{\rm YL}+q_{\rm YL}^{-1}-1
 =2\cos(\pi/5)-1=\varphi^{-1},
\end{align}
where $\varphi=(1+\sqrt5)/2$. These values specify the
SGL images of the displayed single-exchange spectral references.
Cyclotomic conjugation acts discretely on these reference phases;
identification of a full braided fusion theory also requires its
fusion rules, associators and $R$ symbols.

% ============================================================
\subsection{Simultaneous unitarizability and the positive metric}

\label{sup:unitarizability} 
% ============================================================

Simultaneous unitarizability combines a unit-modulus projective spectrum
with a common positive Hermitian form for all generators.

\begin{definition}[Simultaneous projective unitarizability]
A projective braid representation is simultaneously unitarizable if one can
choose representatives $\widetilde \rho(\beta)$ and a single positive
Hermitian form $h>0$ such that 
\begin{equation}
\widetilde \rho(\beta)^\dagger h \widetilde \rho(\beta) = h \qquad \forall
\beta \in B_n.  \label{eq:sup-h-unitarity}
\end{equation}
\end{definition}

Before removing central scalars, the equivalent condition is 
\begin{equation}
\rho(\beta)^\dagger h \rho(\beta) = c_\beta h, \qquad c_\beta>0.
\label{eq:sup-projective-unitarity}
\end{equation}
It is sufficient to impose this condition on a generating set.

For $r=2$, write 
\begin{equation}
h= 
\begin{pmatrix}
u & v \\ 
v^* & w%
\end{pmatrix}%
, \qquad u>0, \quad w>0, \quad uw-|v|^2>0.  \label{eq:sup-h2}
\end{equation}
For fixed exchange generators, the equations 
\begin{equation}
\widetilde B_i^\dagger h \widetilde B_i = h
\end{equation}
are linear in the entries of $h$; positivity is the additional nonlinear
constraint.

If such an $h$ exists, a common frame transformation sending $h$ to the
identity reduces the projective structure group, 
\begin{equation}
PGL(r,\mathbb{C}) \longrightarrow PU(r).  \label{eq:sup-HermitianReduction}
\end{equation}

This criterion makes precise why 
\begin{equation}
|q|=1
\end{equation}
constrains the eigenvalue moduli, while simultaneous unitarizability
additionally constrains the relative eigenframes.

\paragraph{Exact positive-metric locus on the SGL branch.}
For $q=e^{i\varphi}\ne1$, Eq.~\eqref{eq:sup-qconv} has distinct
unit-modulus eigenvalues. Its invariant positive Hermitian form is diagonal,
$h=\operatorname{diag}(u,w)$ with $u,w>0$. The symmetric relative frame gives
\begin{equation}
c^2=\frac{1}{4\sin^2(\varphi/2)},\qquad s^2=1-c^2,
\qquad \mathfrak{x}=-s^2/c^2.
\end{equation}
When $\mathfrak{x}<0$, real $c,s$ make both normalized generators unitary
with $h=I$. When $\mathfrak{x}>0$, choose $c$ real and $s=i\sigma$ with
$\sigma$ real and nonzero. Orthogonality of the distinct eigenvectors of
$B_2$ in the positive metric would require
$ic\sigma(u+w)=0$, which contradicts $u,w>0$.
Thus the generic SGL branch admits one common positive metric precisely for
\begin{equation}
|q|=1,\qquad -3\leq\mathfrak{x}<0.
\label{eq:sup-positive-locus}
\end{equation}
The aligned-frame boundary $\mathfrak{x}=0$ is treated separately from the
four-line stratum. This derivation identifies the relative-geometric
condition that completes the unit-modulus spectral criterion.

Finally, unitarizability and projective flatness are logically independent. A
unitary Berry or Wilczek--Zee connection may have nonzero
curvature, while a non-Hermitian Kato connection can in principle be
projectively flat. The following properties must therefore be kept distinct: 
\begin{equation}
\begin{gathered} \text{unit-modulus projective spectrum},\\
\text{simultaneous unitarizability},\\ \text{braid--fusion coherence},\\
\text{projective flatness}. \end{gathered}  \label{eq:sup-four-properties}
\end{equation}

% ============================================================

\section{Fusion Coherence and Fixed-Ising Categorical Selection}
\setcounter{equation}{0}

\label{sup:coherence} 
% ============================================================

This section collects the coherence conditions that supplement the local
rank-two braid geometry derived in Sec.~\ref{sup:analytic}. The discussion
is organized at two normalization levels. At the projective level, overall
nonzero complex scalars are removed; once a categorical normalization of the
relevant one-dimensional central lines has been fixed, the corresponding
relations become ordinary matrix equalities.

Throughout this section we write 
\begin{equation}
A\doteq B \qquad \Longleftrightarrow \qquad A=\zeta B \quad \text{for some}%
\quad \zeta \in \mathbb{C}^\times ,  \label{eq:sup-projective-equivalence}
\end{equation}
for projective equivalence of two nonzero linear maps with the same domain
and codomain. When the central normalization is fixed, $\doteq$ is replaced
by equality.

The role of the three coherence layers is then: 
\begin{equation}
\begin{aligned}
\text{fusion sewing}&:\quad \text{covariance between fusion charts},\\
\text{Pentagon}&:\quad \text{consistency of iterated reassociation},\\
\text{Hexagon}&:\quad \text{braid--fusion compatibility}.
\end{aligned}
\label{eq:sup-coherence-levels}
\end{equation}

% ============================================================

\subsection{Fusion-chart covariance and projective sewing}

\label{sup:F-sewing} 
% ============================================================

Let $\mathcal{H}_L$ and $\mathcal{H}_R$ denote two fusion-tree resolutions
of the same isolated statistical sector, and let 
\begin{equation}
F_{\mathrm{micro}}: \mathcal{H}_L \longrightarrow \mathcal{H}_R
\end{equation}
be the corresponding microscopic sewing map. If $B_i^{(L)}$ and $B_i^{(R)}$
represent the same elementary braid operation in the two fusion frames,
covariance requires 
\begin{equation}
B_i^{(R)} \doteq F_{\mathrm{micro}} B_i^{(L)} F_{\mathrm{micro}}^{-1}.
\label{eq:sup-Fcov-projective}
\end{equation}

A projective-safe residual is therefore 
\begin{equation}
\epsilon_F = \inf_{\zeta \in \mathbb{C}^\times} \frac{ \left \| B_i^{(R)} -
\zeta F_{\mathrm{micro}} B_i^{(L)} F_{\mathrm{micro}}^{-1} \right \| }{
\|B_i^{(R)}\| }.  \label{eq:sup-Fres}
\end{equation}
Here and below the same matrix norm is used in all numerical evaluations. If
the categorical normalization fixes the central scalar unambiguously, the
minimization over $\zeta$ is omitted and Eq.~\eqref{eq:sup-Fcov-projective}
becomes the ordinary covariance relation 
\begin{equation}
B_i^{(R)} = F_{\mathrm{micro}} B_i^{(L)} F_{\mathrm{micro}}^{-1}.
\label{eq:sup-Fcov-exact}
\end{equation}

The distinction between projective and normalized sewing is essential.
Equation~\eqref{eq:sup-Fcov-projective} tests whether the two matrices
define the same projective braid action, whereas Eq.~%
\eqref{eq:sup-Fcov-exact} additionally requires agreement of the chosen
central normalization.

% ============================================================

\subsection{Pentagon coherence and projective discrete flatness}

\label{sup:Pentagon} 
% ============================================================

For four anyons, the five fusion-tree resolutions form the vertices of the
associahedron $K_4$. Each oriented edge is assigned an elementary
reassociation map 
\begin{equation}
F_{\beta \alpha}: \mathcal{H}_\alpha \longrightarrow \mathcal{H}_\beta .
\end{equation}
For an oriented path 
\begin{equation}
\gamma: \alpha_0\rightarrow \alpha_1\rightarrow \cdots \rightarrow \alpha_m,
\end{equation}
define the ordered reassociation transport 
\begin{equation}
\mathcal{P}_\gamma = F_{\alpha_m\alpha_{m-1}} \cdots F_{\alpha_2\alpha_1}
F_{\alpha_1\alpha_0}.  \label{eq:sup-reassociation-transport}
\end{equation}

Let $\gamma_1$ and $\gamma_2$ denote the two elementary paths in the
pentagon connecting the same initial and final fusion trees, and set 
\begin{equation}
\mathcal{P}_1=\mathcal{P}_{\gamma_1}, \qquad \mathcal{P}_2=\mathcal{P}%
_{\gamma_2}.
\end{equation}
Projective Pentagon coherence requires 
\begin{equation}
\mathcal{P}_1\doteq \mathcal{P}_2.  \label{eq:sup-projective-pentagon}
\end{equation}
A corresponding projective-safe residual is 
\begin{equation}
\epsilon_{\mathrm{pent}} = \inf_{\zeta \in \mathbb{C}^\times} \frac{ \|%
\mathcal{P}_1-\zeta \mathcal{P}_2\| }{ \| \mathcal{P}_1\|+\| \mathcal{P}_2\| 
}.  \label{eq:sup-Pentres}
\end{equation}

Equivalently, whenever $\mathcal{P}_2$ is invertible, define the closed
reassociation holonomy 
\begin{equation}
W_{\mathrm{assoc}} = \mathcal{P}_2^{-1}\mathcal{P}_1.  \label{eq:sup-Wassoc}
\end{equation}
Then projective Pentagon coherence is 
\begin{equation}
[W_{\mathrm{assoc}}] = [I] \qquad \text{in } PGL(\mathcal{H}_{\alpha_0}).
\label{eq:sup-projective-Wassoc}
\end{equation}
An ordinary fusion category must additionally admit compatible central
normalizations for all reassociations, so that 
\begin{equation}
\mathcal{P}_1=\mathcal{P}_2, \qquad W_{\mathrm{assoc}}=I.
\label{eq:sup-exact-pentagon}
\end{equation}

Vanishing projective residual is only a necessary test for the ordinary
Pentagon identity; a global central coherence obstruction must also be
excluded. In particular, ``discrete flatness'' here means trivial holonomy
in fusion-tree space; it should not be identified with the continuous Kato
flatness of Sec.~\ref{sup:kato}.

% ============================================================

\subsection{Fixed-Ising Hexagon selection}

\label{sup:Hexagon} 
% ============================================================

We now specialize to the conventional Ising fusion skeleton 
\begin{equation}
\sigma \times \sigma = \mathbf{1}+\psi, \qquad \psi \times \sigma=\sigma,
\qquad \psi \times \psi=\mathbf{1},  \label{eq:sup-Ising-fusion}
\end{equation}
with the standard two-channel associator 
\begin{equation}
F^{\sigma \sigma \sigma}_{\sigma} = \frac{1}{\sqrt2} 
\begin{pmatrix}
1 & 1 \\ 
1 & -1%
\end{pmatrix}%
.  \label{eq:sup-Ising-F}
\end{equation}

Let the two $\sigma \times \sigma$ fusion channels be ordered according to
the projective convention fixed in Sec.~\ref{sup:analytic}, and let 
\begin{equation}
q = \frac{\lambda_+}{\lambda_-}
\end{equation}
denote the corresponding projective braid-eigenvalue ratio. A necessary
projective selection rule for a braided extension is the following; its
roots are realized by the conventional Ising extensions.

\begin{proposition}[Fixed-Ising projective Hexagon selection]
\label{prop:sup-Ising-Hexagon} For the fusion rules Eq.~%
\eqref{eq:sup-Ising-fusion}, the associator Eq.~\eqref{eq:sup-Ising-F}, and
the projective braid convention used throughout this work, braid--fusion
compatibility requires 
\begin{equation}
q+q^{-1}=0.  \label{eq:sup-IsingHex-linear}
\end{equation}
Equivalently, 
\begin{equation}
q^2+1=0 .  \label{eq:sup-IsingHex}
\end{equation}
Hence 
\begin{equation}
q=\pm i.
\end{equation}
Combining this with the rank-two SGL relation gives 
\begin{equation}
\mathfrak{x}=-1 .  \label{eq:sup-Ising-x}
\end{equation}
\end{proposition}

\begin{proof}
A braided extension must represent adjacent exchanges in the three-$\sigma$
sector by $D(q)$ and $FD(q)F^{-1}$. Their Artin relation is a necessary
consequence of braided coherence. Here $c^2=s^2=1/2$ in
Eq.~\eqref{eq:sup-G}, so Eq.~\eqref{eq:sup-cs} gives
\begin{equation}
 \frac12=-\frac{q}{(q-1)^2}
 \quad \Longleftrightarrow \quad q^2+1=0.
\end{equation}
The conventional Ising braided extensions realize both ratios $q=\pm i$
with compatible scalar $R$ symbols~\cite{Kitaev2006}. Thus these are exactly
the allowed ratios for those extensions. This argument selects the projective ratio that enters SGL; the scalar
$R$-symbol sector is central to this projective coordinate. Substitution into
SGL gives $\mathfrak x=-1$.
\end{proof}

Proposition~\ref{prop:sup-Ising-Hexagon} defines the conventional planar
Ising categorical slice: it combines the Ising fusion rules, the associator
Eq.~\eqref{eq:sup-Ising-F}, and the projective channel convention fixed
above. Equation~\eqref{eq:sup-IsingHex} is the corresponding exact Hexagon
selection in these statistical-geometric coordinates.

% ============================================================

\paragraph{Polynomial elimination and evidence levels.}
The fixed-Ising categorical calculation in
Sec.~\ref{sup:micro-coherence} introduces nonzero $R$-symbol variables
$r_1,r_\psi,r_{\psi\sigma}$ and solves the clockwise and anticlockwise
Hexagon constraints in that normalization. Three of their scalar
consequences are
\begin{equation}
 r_{\psi\sigma}^2+1=0,\qquad
 r_1=r_\psi r_{\psi\sigma},\qquad
 q r_1=r_\psi .
\end{equation}
Eliminating $r_1,r_\psi$ gives $q r_{\psi\sigma}=1$ and hence
$q^2+1=0$. The ordinary categorical Hexagon identities use the
specified scalar $R$ symbols and associator; their exact algebraic
checks are distinct from the finite-chain quantity $q_{\rm micro}$.
At the microscopic level, the measured object is the ratio of the two
fusion-channel exchange multipliers, and its normalized polynomial
obstruction is $\eta_{\rm hex}(q_{\rm micro})$. Its numerical agreement
with $|\tanh(2\pi\kappa)|$ checks the microscopic radial spectral
coordinate against the fixed-Ising selector. Microscopic reassociation
and Pentagon residuals measure separate closed-path operations in
fusion-tree space, while the finite-time $B_{\rm dyn}(T)$ tests physical
time propagation. None of these observables is substituted for another.

\paragraph{Complete categorical benchmark convention.}
For the exact Ising reference, $[F^{abc}_d]_{fe}$ maps
$((ab)_ec)_d$ into $(a(bc)_f)_d$. In the gauge used by the supplied
symbolic audit, the nontrivial associators are
\begin{equation}
\begin{aligned}
 F^{\sigma\sigma\sigma}_{\sigma}&=\frac1{\sqrt2}
 \begin{pmatrix}1&1\\1&-1\end{pmatrix},\\
 F^{\psi\sigma\psi}_{\sigma}&=-1,\qquad
 F^{\sigma\psi\sigma}_{\psi}=-1.
\end{aligned}
\label{eq:sup-complete-Ising-F}
\end{equation}
Every other admissible scalar associator equals one in this gauge.
For the $q=i$ orientation, a scalar completion is
\begin{equation}
\begin{aligned}
 R^{\sigma\sigma}_1&=e^{-i\pi/8},&
 R^{\sigma\sigma}_\psi&=e^{3i\pi/8},\\
 R^{\sigma\psi}_\sigma&=-i,&R^{\psi\sigma}_\sigma&=-i,\\
 R^{\psi\psi}_1&=-1,&R^{1a}_a=R^{a1}_a&=1.
\end{aligned}
\label{eq:sup-complete-Ising-R}
\end{equation}
Complex conjugation supplies the reverse orientation. These exact
categorical data specify a reference associator and its scalar braiding
completion; the microscopic sewing matrix is independently transported.

Let $R_{ab}$ denote the diagonal operator with entries $R^{ab}_e$ on
the allowed $ab$ channels, restricted by the adjacent total-charge
conditions. The two Hexagons in this convention are
\begin{equation}
\begin{aligned}
 \operatorname{diag}_{f\in b\times c}R^{af}_d
 &=(F^{bca}_d)^{-1}R_{ac}F^{bac}_dR_{ab}(F^{abc}_d)^{-1},\\
 \operatorname{diag}_{e\in a\times b}R^{ec}_d
 &=F^{cab}_dR_{ac}(F^{acb}_d)^{-1}R_{bc}F^{abc}_d.
\end{aligned}
\label{eq:sup-benchmark-two-Hexagons}
\end{equation}
Each diagonal factor acts in the intermediate fusion space to its right.
The symbolic audit builds all admissible channel spaces from the stated
fusion rules, checks the five-tree Pentagon edge products, and evaluates
both Hexagons in each allowed charge sector. Exact symbolic equalities
vanish identically. The machine-precision residuals reported in
Sec.~\ref{sup:micro-coherence} are floating-point evaluations of
categorical identities, whereas the full-chain sewing and finite-time
propagation retain their separately defined diagnostics.

\subsection{Normalized Hexagon obstruction}

\label{sup:Hexagon-obstruction} 
% ============================================================

For quantitative comparison away from the fixed-Ising locus, define 
\begin{equation}
\eta_{\mathrm{hex}}(q) = \frac{|q^2+1|} {1+|q|^2}.  \label{eq:sup-etahex}
\end{equation}
This normalization is a bounded spectral obstruction proxy, not a normalized
norm of every Hexagon matrix defect. It has the following properties: 
\begin{equation}
0\leq \eta_{\mathrm{hex}}(q)\leq1,
\end{equation}
and 
\begin{equation}
\eta_{\mathrm{hex}}(q^{-1}) = \eta_{\mathrm{hex}}(q),
\label{eq:sup-eta-inversion}
\end{equation}
so the obstruction is invariant under reversal of the ordered projective
spectral ratio. Moreover, 
\begin{equation}
\eta_{\mathrm{hex}}(q)=0 \quad \Longleftrightarrow \quad q=\pm i,
\end{equation}
within the nonzero projective spectral domain.

For the radial family 
\begin{equation}
q(\kappa) = i e^{2\pi \kappa},
\end{equation}
one finds 
\begin{align}
\eta_{\mathrm{hex}}(\kappa) &= \frac{ |1-e^{4\pi \kappa}| }{ 1+e^{4\pi
\kappa} }  \notag \\
&= |\tanh(2\pi \kappa)|.  \label{eq:sup-eta-radial}
\end{align}
Near the conventional Ising point, 
\begin{equation}
\begin{aligned} \eta_{\rm hex}(\kappa) &= 2\pi|\kappa| - \frac{8\pi^3}{3}
|\kappa|^3 + O(|\kappa|^5). \end{aligned}  \label{eq:sup-eta-small}
\end{equation}
Thus the fixed-Ising Hexagon obstruction opens linearly under the radial
non-Hermitian deformation, even though the latter remains on the
braid-consistent SGL locus.

% ============================================================

\subsection{Nested coherence loci and scope}

\label{sup:coherence-scope} 
% ============================================================

It is useful to formulate the logical structure as a sequence of nested
constraint loci rather than as implications between equations. Let $\mathcal{%
C}_{\mathrm{SGL}}$ denote the braid-consistent spectrum--geometry locus of
Sec.~\ref{sup:SGL-proof}. Define 
\begin{align}
\mathcal{C}_{\mathrm{sew}} &= \mathcal{C}_{\mathrm{SGL}} \cap \mathcal{C}_F,
\\
\mathcal{C}_{\mathrm{pent}} &= \mathcal{C}_{\mathrm{sew}} \cap \mathcal{C}_{%
\mathrm{Pentagon}}, \\
\mathcal{C}_{\mathrm{coh}} &= \mathcal{C}_{\mathrm{pent}} \cap \mathcal{C}_{%
\mathrm{Hexagon}}.  \label{eq:sup-nested-coherence}
\end{align}
Hence 
\begin{equation}
\mathcal{C}_{\mathrm{coh}} \subseteq \mathcal{C}_{\mathrm{pent}} \subseteq 
\mathcal{C}_{\mathrm{sew}} \subseteq \mathcal{C}_{\mathrm{SGL}}.
\label{eq:sup-coherence-inclusions}
\end{equation}

For the fixed conventional Ising skeleton, the rank-two SGL locus contains a
continuous complex family, whereas the Hexagon condition selects the
discrete projective points 
\begin{equation}
(q,\mathfrak{x}) = (\pm i,-1).  \label{eq:sup-Ising-selection-final}
\end{equation}
This is precisely the distinction used in the main text between local braid
consistency and compatibility with a specified fusion skeleton.

The present section concerns algebraic and projective coherence only.
Continuous microscopic projective flatness is a separate condition on the
Kato connection and is treated in Sec.~\ref{sup:kato}. Likewise, the
numerical realization of the sewing, Pentagon, and Hexagon tests is
collected in Sec.~\ref{sup:fusion-curvature}. Keeping these layers separate prevents
a projectively consistent braid representation, a coherent fusion skeleton,
and a path-independent microscopic realization from being identified with
one another.

%======================================================================

%======================================================================

\section{Higher-Rank Statistical Geometry: Fiber Structure and Low-Rank
Rigidity}
\setcounter{equation}{0}

\label{sec:supp-higher-rank} 
%======================================================================

The rank-two relation 
\begin{equation}
\mathfrak{x}=q+q^{-1}-1
\end{equation}
shows that, on the generic irreducible branch, braid consistency completely
locks the relative geometry of two neighboring exchange generators to their
projective spectrum. We now formulate the higher-rank analogue.

The essential change is that relative eigenframe geometry is no longer
described by one cross ratio. On the regular-semisimple stratum it is a
double-torus quotient, 
\begin{equation}
\mathcal{M}_{\mathrm{geom}}^{(r)} = T_r\backslash PGL(r,\mathbb{C})/T_r,
\label{eq:HR-Mgeom}
\end{equation}
where $T_r\subset PGL(r,\mathbb{C})$ is the maximal diagonal torus. The
Artin relation defines a correspondence inside the product of projective
spectral and relative geometric data. The geometry of its fixed-spectrum
fiber determines whether locking is single-valued, finitely branched, or has
positive-dimensional fixed-spectrum fibers.

For $r\leq5$, the low-dimensional rigidity results of Tuba and Wenzl \cite%
{TubaWenzl2001} allow this fiber structure to be determined explicitly. Rank
three remains completely locked; ranks four and five acquire finite
algebraic branching. The double-torus formulation also supplies the invariant framework for
higher-rank fixed-spectrum fiber calculations.

%======================================================================

\subsection{Generic projective moduli and generalized cross ratios}

\label{subsec:HR-moduli} 
%======================================================================

Let 
\begin{equation}
B_1,B_2\in GL(r,\mathbb{C})
\end{equation}
be adjacent exchange generators with the same ordered regular-semisimple
spectrum 
\begin{equation}
\begin{gathered}
\func{Spec}(B_i)=(\lambda_1,\ldots,\lambda_r),\qquad \lambda_a\neq0,\\
\lambda_a\neq\lambda_b\quad(a\neq b).
\end{gathered}
\label{eq:HR-regular-spectrum}
\end{equation}
After removing a common scalar, the projective spectrum is parametrized by 
\begin{equation}
q_a=\frac{\lambda_{a+1}}{\lambda_1}, \qquad a=1,\ldots,r-1,
\end{equation}
so that 
\begin{equation}
\mathcal{M}_{\mathrm{spec}}^{(r)} \simeq (\mathbb{C}^\times)^{r-1}, \qquad
\dim_{\mathbb{C}}\mathcal{M}_{\mathrm{spec}}^{(r)}=r-1.
\label{eq:HR-spectrum-space}
\end{equation}

Let $E_i$ be ordered eigenvector matrices, 
\begin{equation}
B_iE_i = E_iD_\lambda, \qquad D_\lambda = \func{diag} (\lambda_1,\ldots,%
\lambda_r),
\end{equation}
and define the relative eigensplitting matrix 
\begin{equation}
G=E_1^{-1}E_2.
\end{equation}
Independent rescaling of the ordered eigenvectors acts as 
\begin{equation}
G\longmapsto D_L^{-1}GD_R,
\end{equation}
with $D_L,D_R$ diagonal. After projectivization this gives Eq.~%
\eqref{eq:HR-Mgeom}. On the generic locus with zero-dimensional stabilizer, 
\begin{equation}
\dim_{\mathbb{C}}\mathcal{M}_{\mathrm{geom}}^{(r)} = (r-1)^2.
\label{eq:HR-Mgeom-dim}
\end{equation}
Hence the generic kinematic statistical space has 
\begin{equation}
\begin{aligned}
\mathcal{M}_{\mathrm{stat,cand}}^{(r)}
&\sim_{\mathrm{bir}}(\mathbb{C}^\times)^{r-1}
\times\left(T_r\backslash PGL(r,\mathbb{C})/T_r\right),\\
\dim_{\mathbb C}\mathcal{M}_{\mathrm{stat,cand}}^{(r)}&=r(r-1).
\end{aligned}
\label{eq:HR-Mstat}
\end{equation}
Here $\sim_{\mathrm{bir}}$ denotes the generic regular-semisimple birational
stratum, rather than a statement about singular orbit closures.

A convenient rational coordinate system on a dense big cell is obtained from
a lift $G=(g_{ij})\in GL(r,\mathbb{C})$. Whenever the required matrix
entries are nonzero, define 
\begin{equation}
\mathfrak{x}_{ij} = \frac{g_{i1}g_{1j}} {g_{ij}g_{11}}, \qquad
i,j=2,\ldots,r.  \label{eq:HR-xij}
\end{equation}
These quantities are invariant under 
\begin{equation}
g_{ij}\mapsto a_i g_{ij}b_j
\end{equation}
and provide $(r-1)^2$ rational coordinates on the double quotient. Indeed,
the left--right torus action can be used to normalize the first row and
column of $G$ to unity, after which 
\begin{equation}
\widetilde G_{ij} = \mathfrak{x}_{ij}^{-1}, \qquad i,j\geq2.
\end{equation}
For $r=2$, $\mathfrak{x}_{22}$ reduces to the ordinary rank-two cross ratio
used in the main text.

%======================================================================

\subsection{The higher-rank spectrum--geometry correspondence}

\label{subsec:HR-SGC} 
%======================================================================

Fix 
\begin{equation}
B_1=D(\mathbf{q}), \qquad B_2=GD(\mathbf{q})G^{-1},
\end{equation}
where 
\begin{equation}
D(\mathbf{q}) = \func{diag} (1,q_1,\ldots,q_{r-1}).
\end{equation}
The matrix-valued Artin defect is 
\begin{equation}
\Delta_r(D,G) = D(GDG^{-1})D - (GDG^{-1})D(GDG^{-1}).  \label{eq:HR-defect}
\end{equation}
Braid consistency is the algebraic condition 
\begin{equation}
\Delta_r(D,G)=0.
\end{equation}

We therefore define 
\begin{equation}
\mathfrak{C}_{\mathrm{SG}}^{(r)} = \left \{ (\mathbf{q},[G]): \Delta_r(D(%
\mathbf{q}),G)=0 \right \} \subset \mathcal{M}_{\mathrm{spec}}^{(r)} \times 
\mathcal{M}_{\mathrm{geom}}^{(r)}.  \label{eq:HR-CSG}
\end{equation}
Its spectral projection 
\begin{equation}
\pi_{\mathrm{spec}}^{(r)}: \mathfrak{C}_{\mathrm{SG}}^{(r)} \longrightarrow 
\mathcal{M}_{\mathrm{spec}}^{(r)}  \label{eq:HR-pispec}
\end{equation}
is the natural higher-rank object replacing a single-valued SGL function.

For generic $\mathbf{q}$, let 
\begin{equation}
\nu_r = \dim_{\mathbb{C}} \left[ \left( \pi_{\mathrm{spec}}^{(r)}
\right)^{-1}(\mathbf{q}) \right].  \label{eq:HR-nur}
\end{equation}
We distinguish 
\begin{equation}
\begin{gathered}
\nu_r=0,\; \#\pi_{\mathrm{spec}}^{-1}(\mathbf q)=1\\[-1mm]
\Longrightarrow \text{single-valued complete locking},\\[1mm]
\nu_r=0,\; 1<\#\pi_{\mathrm{spec}}^{-1}(\mathbf q)<\infty\\[-1mm]
\Longrightarrow \text{finite-branched locking},\\[1mm]
\nu_r>0\Longrightarrow \text{partial locking}.
\end{gathered}
\label{eq:HR-locking-types}
\end{equation}

Thus the natural invariant of higher-rank spectrum--geometry locking is the
fixed-spectrum fiber of $\pi_{\mathrm{spec}}^{(r)}$, not necessarily the
graph of a single-valued function.

%======================================================================

\subsection{Rank three: exact complete locking}

\label{subsec:HR-r3} 
%======================================================================

Rank three is the first case in which the relative geometry contains more
coordinates than the spectrum: 
\begin{equation}
\dim_{\mathbb{C}}\mathcal{M}_{\mathrm{spec}}^{(3)}=2, \qquad \dim_{\mathbb{C}%
}\mathcal{M}_{\mathrm{geom}}^{(3)}=4.
\end{equation}
Nevertheless, the generic fixed-spectrum fiber remains a single point.

Let 
\begin{equation}
u=\frac{\lambda_2}{\lambda_1}, \qquad v=\frac{\lambda_3}{\lambda_1}.
\label{eq:HR-uv}
\end{equation}
Using the canonical simple three-dimensional $B_3$ representative of Ref.~%
\cite{TubaWenzl2001}, diagonalizing both generators with the same ordered
eigenvalues, and evaluating Eq.~\eqref{eq:HR-xij} gives 
\begin{equation}
\begin{aligned} \mathfrak x_{22} &= \frac{(1+uv)(u^2+v)} {u(1+v)(u+v)}, \\
\mathfrak x_{23} = \mathfrak x_{32} &= \frac{1+uv}{u+v}, \\ \mathfrak x_{33}
&= \frac{(1+uv)(u+v^2)} {v(1+u)(u+v)}. \end{aligned}
\label{eq:HR-r3-SGL-map}
\end{equation}
Hence the four geometric coordinates are rational functions of the two
projective spectral coordinates.

\begin{theorem}[Generic rank-three complete locking]
\label{thm:HR-r3-SGL} Let $B_1,B_2\in GL(3,\mathbb{C})$ define a simple
regular-semisimple representation of $B_3$ with the same ordered distinct
nonzero eigenvalues. On the dense double-torus chart where Eq.~%
\eqref{eq:HR-xij} is defined, the relative ordered eigensplitting is
uniquely determined by the projective spectrum and is given by Eq.~%
\eqref{eq:HR-r3-SGL-map}. Consequently, 
\begin{equation}
\dim_{\mathbb{C}} \left( \pi_{\mathrm{spec}}^{(3)} \right)^{-1}(u,v) = 0,
\end{equation}
and, on the generic simple ordered branch, 
\begin{equation}
\# \left( \pi_{\mathrm{spec}}^{(3)} \right)^{-1}(u,v) = 1.
\end{equation}
\end{theorem}

\begin{proof}
The low-dimensional classification of simple $B_3$ modules implies that a
simple three-dimensional representation is uniquely determined, up to
equivalence, by its ordered generator eigenvalues
\cite{TubaWenzl2001}.  Suppose two braid-consistent pairs are written in the
same spectral gauge,
\begin{equation}
    (D,GDG^{-1}),
    \qquad
    (D,G'DG'^{-1}).
\end{equation}
Any equivalence $S$ between them must satisfy
\begin{equation}
    SDS^{-1}=D.
\end{equation}
Because $D$ has distinct eigenvalues, its centralizer is the diagonal torus.
Hence
\begin{equation}
    S\in T_3,
\end{equation}
and comparison of the second generators gives
\begin{equation}
    G'=SGR,
    \qquad
    R\in T_3.
\end{equation}
Therefore
\begin{equation}
    [G']=[G]
    \qquad
    \text{in }
    T_3\backslash PGL(3,\mathbb C)/T_3.
\end{equation}
For completeness, with $\lambda_1=1$, the matrices used to evaluate the
four rational coordinates are
\begin{equation}
 A=\begin{pmatrix}1&v/u+u&u\\0&u&u\\0&0&v\end{pmatrix},\qquad
 B=\begin{pmatrix}v&0&0\\-u&u&0\\u&-v/u-u&1\end{pmatrix}.
\end{equation}
They satisfy $ABA=BAB$. Taking eigenvector matrices $E_A,E_B$ in the
order $(1,u,v)$ and setting $G=E_A^{-1}E_B$ gives
Eq.~\eqref{eq:HR-r3-SGL-map} by direct substitution.
The fixed-spectrum relative geometry is consequently unique.
\end{proof}

%======================================================================

\subsection{Exceptional divisors and chart dependence}

\label{subsec:HR-r3-divisors} 
%======================================================================

For rank three, three kinds of exceptional loci must be distinguished.

Regular semisimplicity requires 
\begin{equation}
u\neq1, \qquad v\neq1, \qquad u\neq v.  \label{eq:HR-regular-divisors}
\end{equation}
The simple representation locus further excludes the Tuba--Wenzl divisors 
\begin{equation}
1+uv\neq0, \qquad u^2+v\neq0, \qquad v^2+u\neq0.  \label{eq:HR-simple-uv}
\end{equation}
Finally, the particular big-cell coordinates in Eq.~\eqref{eq:HR-r3-SGL-map}
require 
\begin{equation}
1+u\neq0, \qquad 1+v\neq0, \qquad u+v\neq0.  \label{eq:HR-chart-uv}
\end{equation}
The last conditions are chart boundaries rather than representation
singularities; a different anchor row and column provides another rational
chart.

Accordingly, 
\begin{equation}
\begin{gathered}
\text{regular-semisimple locus}\neq\text{simple locus},\\
\text{simple locus}\neq\text{coordinate-chart locus}.
\end{gathered}
\label{eq:HR-three-divisors}
\end{equation}

%======================================================================

\subsection{Ranks four and five: finite algebraic branching}

\label{subsec:HR-r45} 
%======================================================================

The first qualitative departure from complete single-valued locking occurs
at rank four.

For a simple four-dimensional $B_3$ representation with ordered spectrum $%
(\lambda_1,\ldots,\lambda_4)$, the classification introduces a projectively
nontrivial square-root branch parameter 
\begin{equation}
D_\pm^2 = \frac{\lambda_2\lambda_3} {\lambda_1\lambda_4} = \frac{q_1q_2}{q_3}%
.  \label{eq:HR-r4-D}
\end{equation}
Equivalently, 
\begin{equation}
D_\pm = \pm \sqrt{\frac{q_1q_2}{q_3}}.
\end{equation}
This branch label survives projectivization and cannot be removed by a
common scalar multiplication of the generator eigenvalues.

On every branch satisfying the simple-representation conditions of Ref.~\cite%
{TubaWenzl2001}, the representation is fixed up to equivalence. Thus 
\begin{equation}
\dim_{\mathbb{C}} \left( \pi_{\mathrm{spec}}^{(4)} \right)^{-1}(\mathbf{q})
= 0, \qquad \# \left( \pi_{\mathrm{spec}}^{(4)} \right)^{-1}(\mathbf{q})
\leq2.  \label{eq:HR-r4-fiber}
\end{equation}
When both square-root branches lie on the simple locus, the fiber has
cardinality two.

Rank five behaves analogously, but with fifth-root branching. Introduce 
\begin{equation}
\Gamma_5^5 = \prod_{i=1}^5\lambda_i, \qquad \eta=\frac{\Gamma_5}{\lambda_1}.
\end{equation}
In projective spectral coordinates, 
\begin{equation}
\eta^5=q_1q_2q_3q_4,
\end{equation}
so the possible branch parameters are 
\begin{equation}
\begin{aligned}
\eta_m&=\omega_5^m(q_1q_2q_3q_4)^{1/5},\qquad m=0,\ldots,4,\\
\omega_5&=e^{2\pi i/5}.
\end{aligned}
\label{eq:HR-r5-branches}
\end{equation}
Again the branch variable is projectively invariant. On each branch
satisfying the Tuba--Wenzl simplicity conditions, 
\begin{equation}
\dim_{\mathbb{C}} \left( \pi_{\mathrm{spec}}^{(5)} \right)^{-1}(\mathbf{q})
= 0, \qquad \# \left( \pi_{\mathrm{spec}}^{(5)} \right)^{-1}(\mathbf{q})
\leq5.  \label{eq:HR-r5-fiber}
\end{equation}
When all five simple branches are present, the fiber has cardinality five.

Ranks four and five therefore exhibit finite-branched, rather than
single-valued, spectrum--geometry locking.

%======================================================================

\subsection{Higher-rank fiber hierarchy and categorical scope}

\label{subsec:HR-hierarchy} 
%======================================================================

The low-rank results are summarized by 
\begin{equation}
\begin{array}{c|c|c|c}
r & \dim_{\mathbb C}\mathcal M_{\rm spec}^{(r)}
  & \dim_{\mathbb C}\mathcal M_{\rm geom}^{(r)}
  & \text{locking type} \\ \hline
2 & 1 & 1  & \text{single-valued} \\ 
3 & 2 & 4  & \text{single-valued} \\ 
4 & 3 & 9  & \text{finite-branched }(\leq2) \\ 
5 & 4 & 16 & \text{finite-branched }(\leq5)
\end{array}
\label{eq:HR-table}
\end{equation}

The important point is that the dimension of the kinematic geometric space
does not determine the dimension of the braid-consistent fixed-spectrum
fiber. At rank three, for example, four independent relative geometric
coordinates collapse to a unique geometry after the Artin relation and
simplicity are imposed.

For dimensions $r\geq6$, the cited low-dimensional classification does not
settle the fixed-spectrum fiber. Positive-dimensional fibers would
constitute partial spectrum--geometry locking, but their existence and
location require a separate representation-variety analysis.

The resulting hierarchy is therefore 
\begin{equation}
\begin{aligned} r=2,3 &:\quad \text{single-valued complete locking}, \\ r=4
&:\quad \text{finite-branched locking, degree }\leq2, \\ r=5 &:\quad
\text{finite-branched locking, degree }\leq5, \\ r\geq6 &:\quad \text{fiber
structure requires separate analysis}. \end{aligned}
\label{eq:HR-final-hierarchy}
\end{equation}

This shows that the rank-two law is not an isolated low-dimensional
coincidence. Its natural higher-rank generalization is the fibered
correspondence 
\begin{equation}
\mathfrak{C}_{\mathrm{SG}}^{(r)} \xrightarrow{\; \pi_{\rm spec}^{(r)}\;} 
\mathcal{M}_{\mathrm{spec}}^{(r)}.  \label{eq:HR-final-fibration}
\end{equation}

Finally, $\mathfrak{C}_{\mathrm{SG}}^{(r)}$ encodes only local braid
consistency. For a specified higher-rank fusion skeleton one must still
impose fusion sewing, Pentagon, and Hexagon constraints. Schematically, 
\begin{equation}
\mathfrak{C}_{\mathrm{coh}}^{(r)} = \mathfrak{C}_{\mathrm{SG}}^{(r)} \cap 
\mathfrak{C}_F^{(r)} \cap \mathfrak{C}_{\mathrm{Pent}}^{(r)} \cap \mathfrak{C%
}_{\mathrm{Hex}}^{(r)}.  \label{eq:HR-coh-locus}
\end{equation}
No universal form of the latter constraints exists without specifying the
fusion rules and multiplicity spaces. In particular, the finite spectral
branches at ranks four and five are braid-representation branches, not by
themselves distinct topological orders.

% ============================================================

\section{Kato Realization, Similarity Covariance, and Projective Flatness}
\setcounter{equation}{0}

\label{sup:kato} 
% ============================================================

This section collects the geometric ingredients that connect the abstract
statistical data to a non-Hermitian Hamiltonian family. We fix the
projector-defined Kato connection used throughout the microscopic analysis,
state its behavior under similarity transformations, compare it with common
non-Hermitian Berry conventions, and formulate the projective-flatness
criterion relevant for path-independent exchange transport.

% ============================================================

\subsection{Riesz projectors and Kato transport}

\label{sup:kato-basic} 
% ============================================================

Let $H(\lambda)$ be a smooth finite-dimensional non-Hermitian Hamiltonian
family~\cite{Kato1950,KatoBook,NenciuRasche1992}, and assume that a cluster of $r$ eigenvalues remains isolated from
the rest of the spectrum on the control region of interest. The associated
Riesz projector is 
\begin{equation}
P(\lambda) = \frac{1}{2\pi i} \oint_\Gamma \bigl(z-H(\lambda)\bigr)^{-1}\,dz
,  \label{eq:sup-Riesz}
\end{equation}
where $\Gamma$ remains in the resolvent set and encloses the same spectral
cluster throughout the deformation.

The images 
\begin{equation}
\mathcal{E}_\lambda=\func{Im}P(\lambda)
\end{equation}
define a rank-$r$ complex spectral bundle 
\begin{equation}
\mathcal{E}\longrightarrow \mathcal{M}_{\mathrm{ctrl}}.
\end{equation}
No orthogonality condition is assumed: 
\begin{equation}
P^2=P, \qquad P^\dagger \neq P
\end{equation}
in general.

The Kato connection is 
\begin{equation}
\nabla_K=P\,d .  \label{eq:sup-Kato}
\end{equation}
Along a smooth control path $\lambda(t)$, its parallel transport satisfies 
\begin{equation}
\frac{d}{dt}U_K(t) = [\dot P(t),P(t)]U_K(t), \qquad U_K(0)=I,
\label{eq:sup-Kato-generator}
\end{equation}
together with the intertwining identity 
\begin{equation}
P(t)U_K(t) = U_K(t)P(0).  \label{eq:sup-Kato-intertwining}
\end{equation}
Thus Kato transport follows the spectral subspace itself rather than a
chosen instantaneous eigenvector frame.

For a closed control loop $\gamma$, after any required endpoint
identification, the induced transport on the reference fiber is 
\begin{equation}
W_\gamma = \func{Hol}_{\nabla_K}(\gamma) \in GL(r,\mathbb{C}).
\label{eq:sup-Kato-holonomy}
\end{equation}

% ============================================================

\subsection{Similarity covariance}

\label{sup:similarity} 
% ============================================================

Consider two Hamiltonian families related by
\begin{equation}
H'(\lambda)=S(\lambda)H(\lambda)S^{-1}(\lambda).
\label{eq:sup-similarity-H}
\end{equation}
Their Riesz projectors satisfy
\begin{equation}
P'(\lambda)=S(\lambda)P(\lambda)S^{-1}(\lambda).
\label{eq:sup-similarity-P}
\end{equation}

If $S$ is constant on the control region, then 
\begin{equation}
\nabla^{\prime }_K = S\nabla_KS^{-1}, \qquad \mathcal{F}^{\prime }_K = S%
\mathcal{F}_KS^{-1},  \label{eq:sup-constant-similarity}
\end{equation}
and closed holonomies transform by conjugation, 
\begin{equation}
W^{\prime }_\gamma = SW_\gamma S^{-1}.  \label{eq:sup-similarity-holonomy}
\end{equation}
Hence a parameter-independent similarity changes only the representation of
the projective statistical data.

For a parameter-dependent similarity, define 
\begin{equation}
\Omega=S^{-1}dS.
\end{equation}
Pulling the primed connection back to the original spectral bundle gives, on 
$\func{Im}P$, 
\begin{equation}
S^{-1}\nabla^{\prime }_KS = \nabla_K+P\Omega P .
\label{eq:sup-variable-similarity}
\end{equation}
The additional bundle connection $P\Omega P$ can carry nontrivial projective
holonomy. Therefore 
\begin{equation}
\text{isospectrality} \not \Rightarrow \text{identical Kato transport}.
\label{eq:sup-isospectral-not-holonomy}
\end{equation}

% ============================================================

\subsection{Relation to other non-Hermitian connection conventions}

\label{sup:connection-choice} 
% ============================================================

The distinction above is relevant because several inequivalent geometric
connections are used in non-Hermitian systems~\cite{GarrisonWright1988,CuiZheng2012,Brody2014,Singhal2023}.

For a right eigenstate 
\begin{equation}
|R_\beta \rangle = S_\beta|R_0\rangle,
\end{equation}
the normalized-right Berry connection in the laboratory inner product is 
\begin{equation}
\mathcal{A}_R^{(\beta)} = i \frac{ \langle R_\beta|dR_\beta \rangle }{
\langle R_\beta|R_\beta \rangle }.  \label{eq:sup-right-connection}
\end{equation}
If $S_\beta$ is independent of the braid coordinate, then 
\begin{equation}
\mathcal{A}_R^{(\beta)} = i \frac{ \langle R_0| G_\beta \,dR_0\rangle }{
\langle R_0| G_\beta |R_0\rangle }, \qquad G_\beta=S_\beta^\dagger S_\beta.
\label{eq:sup-right-similarity}
\end{equation}
This quantity can vary with the metric deformation $G_\beta$ even when the
similarity map is constant along the transport coordinate.

By contrast, choosing the biorthogonal left state as 
\begin{equation}
\langle L_\beta| = \langle L_0|S_\beta^{-1},
\end{equation}
gives, for path-independent $S_\beta$, 
\begin{equation}
\mathcal{A}_{LR}^{(\beta)} = i\langle L_\beta|dR_\beta \rangle = i\langle
L_0|dR_0\rangle.  \label{eq:sup-biorthogonal-connection}
\end{equation}
Kato transport is likewise related by conjugation in this case.

Thus a continuously tunable normalized-right Berry phase need not imply a
continuously tunable Kato holonomy. Conversely, when the similarity
transformation varies along the path, the additional term in Eq.~%
\eqref{eq:sup-variable-similarity} must be retained and no general
identification of the three connection conventions is available.

For this reason, all microscopic braid and fusion data in this work are
defined using the single projector-based convention 
\begin{equation}
\nabla_K=P\,d.
\end{equation}

% ============================================================

\subsection{Projective curvature and path independence}

\label{sup:physical-flatness} 
% ============================================================

The curvature of the Kato connection is 
\begin{equation}
\mathcal{F}_K = P(dP\wedge dP)P.  \label{eq:sup-FK}
\end{equation}
Since an overall nonzero scalar is irrelevant to projective exchange data,
the relevant curvature is the traceless component 
\begin{equation}
\mathcal{F}_{\mathrm{proj}} = \mathcal{F}_K - \frac{1}{r} \func{Tr}(\mathcal{%
F}_K)I_r.  \label{eq:sup-Fproj}
\end{equation}

Local projective flatness is the condition 
\begin{equation}
\mathcal{F}_{\mathrm{proj}}=0.  \label{eq:sup-projective-flatness}
\end{equation}
On a regular simply connected control patch this is equivalent to local
flatness of the induced $PGL(r,\mathbb{C})$ connection. Consequently, the
projective transport between fixed endpoints depends only on the homotopy
class allowed within that patch.

The infinitesimal form of this statement is particularly useful. For a small
contractible loop spanning tangent directions $X,Y$ and oriented area $%
\delta S$ (orientation $X,Y,-X,-Y$), 
\begin{equation}
W_{\delta C} = I - \mathcal{F}_K(X,Y)\, \delta S + O(\delta S^{3/2}),
\label{eq:sup-small-loop}
\end{equation}
and therefore, after projectivization, 
\begin{equation}
[W_{\delta C}] = [I-\mathcal{F}_{\mathrm{proj}}(X,Y)\, \delta S +O(\delta
S^{3/2})].  \label{eq:sup-small-loop-projective}
\end{equation}
A nonzero $\mathcal{F}_{\mathrm{proj}}$ therefore gives an infinitesimal
obstruction to path-independent projective transport.

This condition is logically independent of categorical coherence. In
particular, 
\begin{equation}
W_{\mathrm{assoc}}=I
\end{equation}
tests discrete path independence in fusion-tree space, whereas 
\begin{equation}
\mathcal{F}_{\mathrm{proj}}=0
\end{equation}
tests continuous path independence in Hamiltonian control space.

For finite deformations, let $\gamma_1$ and $\gamma_2$ be two admissible
control paths representing the same braid word. A projective path residual
may be defined by 
\begin{equation}
\epsilon_{\mathrm{path}} = d_{\mathrm{proj}} \left( [W_{\gamma_1}],
[W_{\gamma_2}] \right).  \label{eq:sup-path-residual}
\end{equation}
A numerical path-independence certificate requires both individual mesh
convergence of the Kato holonomies and 
\begin{equation}
\epsilon_{\mathrm{path}}\longrightarrow0
\end{equation}
for the sampled braid-isotopic deformations.

The analytic complex-quadric calculation in Sec.~\ref{sup:quadric-flatness}
provides a stronger criterion for the five-arm model: there $\mathcal{F}_{%
\mathrm{proj}}$ is obtained explicitly and is generically nonzero, so
projective flatness fails independently of any finite set of sampled paths.

% ============================================================

\section{Five-Arm Microscopic Realization and Numerical Certificates}
\setcounter{equation}{0}

\label{sup:numerics} 
% ============================================================

The five-arm microscopic realization is specified by its finite-chain
Hamiltonian, control-space paths, common-fiber Kato holonomies, and
independent projective observables. Continuum convergence,
matched-spectrum controls, fusion sewing, finite-time dynamics and
control-path curvature furnish the corresponding numerical certificates.
The monitored implementation has its own conditional normalization and
event-probability observable.

Throughout, the numerical and experimental quantities are interpreted
projectively. Common nonzero complex amplitudes are removed whenever they do
not affect the induced element of $PGL(r,\mathbb{C})$.

% ============================================================

\subsection{Microscopic protocol and common-fiber conventions}

\label{sup:fivearm-protocol} 
% ============================================================

The microscopic Hamiltonian is built from finite Kitaev arms with
junction control of the wire-network type~\cite{Kitaev2001,Alicea2011}:
\begin{equation}
H_{5Y}(s) = \sum_{a=1}^{5}H_K^{(a)} + i\gamma_0 \sum_{a=1}^{5}
\lambda_a(s)\alpha_{a,1},  \label{eq:sup-H5Y}
\end{equation}
with complex junction controls $\lambda_a(s)$. Every exchange protocol is
implemented as a path in this control space while the same low-energy
spectral cluster is tracked continuously.

A reproducible realization requires the following conventions. The isolated
cluster is defined by a contour that remains in the resolvent set throughout
each protocol; transport is constructed using the same Riesz--Kato
connection; endpoint frames are closed onto a common reference fiber;
eigenline labels are continued from a fixed reference point rather than
independently reordered at every step; and projective residuals are used
whenever a central scalar normalization is not fixed.

The resulting neighboring exchange matrices are 
\begin{equation}
B_1^{\mathrm{micro}}, \qquad B_2^{\mathrm{micro}},
\end{equation}
represented on one common two-dimensional logical fiber.

The radial protocol has a direct, model-specific microscopic derivation:
the exact arm zero modes determine the dark-bundle projector and its exchange
holonomy. This is the point at which the abstract radial coordinate of the
main text becomes a function of an explicit Hamiltonian control path.

\paragraph{Exact finite-arm projection.}

For one arm, write 
\begin{equation}
H_K=\frac{i}{2}\! \left[-\mu \sum_{j=1}^{L}\alpha_j\beta_j
+2t\sum_{j=1}^{L-1}\beta_j\alpha_{j+1}\right] +iM\eta \beta_L .
\label{eq:sup-arm-HK}
\end{equation}
\paragraph{Matrix indices and Nambu convention.}
Order the Majoranas as $\gamma_0$, followed on each arm by
$(\alpha_{a,1},\beta_{a,1},\ldots,\alpha_{a,L},\beta_{a,L},\eta_a)$.
In one-based matrix indexing, define
\begin{equation}
\begin{aligned}
i_0&=1,\quad b_a=2+(a-1)(2L+1),\\
i_{\alpha_{a,j}}&=b_a+2(j-1),\quad
 i_{\beta_{a,j}}=b_a+2j-1,\\
i_{\eta_a}&=b_a+2L.
\end{aligned}
\end{equation}
For $H=(i/4)\boldsymbol\gamma^TA\boldsymbol\gamma$, all independent
nonzero entries are
\begin{equation}
\begin{aligned}
 A_{\alpha_{a,j},\beta_{a,j}}&=-\mu,&
 A_{\beta_{a,j},\alpha_{a,j+1}}&=2t,\\
 A_{\eta_a,\beta_{a,L}}&=2M,&
 A_{\gamma_0,\alpha_{a,1}}&=2\lambda_a(s),
\end{aligned}
\label{eq:sup-full-matrix-entries}
\end{equation}
with $A_{ji}=-A_{ij}$. The reference parameters
$(L,t,\mu,M,\Lambda)=(4,1,1,1.5,0.22)$ give a $46\times46$ matrix.
A fixed pairing and permutation of the Majoranas gives
$\boldsymbol\gamma=(\mathbf x,\mathbf y)^T$, where
$x_j=f_j+f_j^\dagger$ and $y_j=-i(f_j-f_j^\dagger)$. In the Nambu basis
$(\mathbf f,\mathbf f^\dagger)^T$,
\begin{equation}
U_N=\frac1{\sqrt2}\begin{pmatrix}I&I\\-iI&iI\end{pmatrix},
\quad h_{\rm BdG}=U_N^\dagger(iA)U_N.
\end{equation}
The algebraic particle--hole constraint is
$\tau_xh_{\rm BdG}^T\tau_x=-h_{\rm BdG}$, including complex controls.
The transpose in this constraint preserves the non-Hermitian pairing
structure. The same convention applies to the even-dimensional honeycomb
Majorana matrix in Sec.~\ref{sup:2d-kitaev}.

With $r=-\mu/(2t)$, an exact zero Majorana is 
\begin{equation}
\zeta=\mathcal{N}_L \left[ \sum_{j=1}^{L}r^{j-1}\alpha_j -\frac{t}{M}r^L\eta %
\right],  \label{eq:sup-arm-zero-mode}
\end{equation}
where 
\begin{equation}
\mathcal{N}_L^{-2} =\frac{1-r^{2L}}{1-r^2} +\frac{t^2}{M^2}r^{2L}.
\label{eq:sup-arm-zero-normalization}
\end{equation}
For five identical arms the endpoint projection $\alpha_{a,1}\mapsto 
\mathcal{N}_L\zeta_a$ gives 
\begin{equation}
H_{\mathrm{low}} =i\mathcal{N}_L\gamma_0 \sum_{a=1}^{5}\lambda_a\zeta_a
=i\Delta \gamma_0\,n^T\boldsymbol{\zeta},  \label{eq:sup-Hlow}
\end{equation}
with 
\begin{equation}
n=\frac{\boldsymbol{\lambda}} {\sqrt{\boldsymbol{\lambda}^T\boldsymbol{%
\lambda}}}, \qquad n^Tn=1, \qquad \Delta=\mathcal{N}_L\sqrt{\boldsymbol{%
\lambda}^T\boldsymbol{\lambda}}.  \label{eq:sup-nlambda}
\end{equation}
The scalar $\mathcal{N}_L$ is common to all arms and therefore does not
change the normalized control path. One outer-Majorana direction parallel to 
$n$ is bright, while $n^Tv=0$ defines the four-dimensional dark space, 
\begin{equation}
P_0(n)=I-nn^T.  \label{eq:sup-dark-projector-origin}
\end{equation}
The regular control manifold is consequently the complex quadric $n^Tn=1$.
For real $n$ it reduces to the compact Hermitian sphere; complex $n$
supplies the noncompact geometric directions.

% BEGIN finite_arm_theorem.tex

\begin{proposition}[Exact finite-arm dark-bundle embedding]
\label{prop:finite-arm-projector} Consider five identical arms with the
real, fixed parameters in Eq.~\eqref{eq:sup-arm-HK}, with $tM\ne0$ and $%
|r|<1 $, coupled to the central Majorana by $i\gamma_0\sum_a\lambda_a%
\alpha_{a,1}$. Let $Z$ be the constant matrix whose five columns are the
normalized arm modes in Eq.~\eqref{eq:sup-arm-zero-mode}, including zero
central component. Assume $\boldsymbol{\lambda}^T\boldsymbol{\lambda}\ne0$
and that zero is an isolated, semisimple eigenvalue of the full
antisymmetric matrix $A$ with multiplicity four. With the fusion resolvers
switched off, 
\begin{equation}
P_{\mathrm{full}}(n)=Z(I-nn^T)Z^T,\qquad Z^TZ=I_5.
\label{eq:exact-finite-projector}
\end{equation}
Consequently its Kato connection and curvature are the constant embeddings
of those of $P_0=I-nn^T$: 
\begin{equation}
P_{\mathrm{full}}(dP_{\mathrm{full}}\wedge dP_{\mathrm{full}})P_{\mathrm{full%
}} =ZP_0(dP_0\wedge dP_0)P_0Z^T.  \label{eq:exact-finite-curvature}
\end{equation}
\end{proposition}

\begin{proof}
The uncoupled arm matrix annihilates each column of $Z$. The central
couplings give
$AZv=2\mathcal N_L(\boldsymbol \lambda^Tv)e_0$.
Thus $Z(n^\perp)$ lies in $\ker A$ and has dimension four. The assumed
multiplicity identifies it with the full kernel. The bilinear form on
$n^\perp$ is nondegenerate because $n^Tn=1$, and its projector is
$I-nn^T$. Antisymmetry implies $r^TA=0$ for every $r\in \ker A$;
semisimplicity then identifies the transpose-orthogonal kernel projector
with the Riesz spectral projector. Differentiation of the constant embedding
$Z$ proves Eq.~\eqref{eq:exact-finite-curvature}.
\end{proof}

This identity is exact at finite arm length. It concerns the unsplit dark
bundle and its geometric transport; the energy-dependent bright sector and
the physical fusion resolvers retain their full Hamiltonian definitions. For
the calculation in this paper, $\mu=t=1$, $M=1.5$, and $\lambda_a=0.22n_a$.
Direct full-matrix Schur projectors at $L=2,4,8$ and $%
\kappa=0,0.05,0.10,0.20 $, with 25 nodes on each radial triangle, give 
\begin{equation}
\max \frac{\|P_{\mathrm{Riesz}}-P_{\mathrm{full}}\|_F}{\|P_{\mathrm{full}%
}\|_F} =1.55\times10^{-14}.
\end{equation}
The corresponding minimum external gaps are $0.38734$, $0.37850$, and $%
0.37781$ for $L=2,4,8$, respectively. The same finite-arm identity therefore
supports the radial solid-angle calculation and the control-curvature
calculation with a common microscopic projector.

% END finite_arm_theorem.tex

\paragraph{Complex solid angle of the radial loop.}

The first exchange can be confined to the active subspace $\func{span}%
\{e_1,e_2,e_5\}$. Put $\beta=2\pi \kappa$ and choose the geodesic triangle 
\begin{equation}
a=e_5, \qquad b=e_1, \qquad c=\cosh \beta \,e_2+i\sinh \beta \,e_5.
\label{eq:sup-radial-triangle}
\end{equation}
The identity $\cosh^2\beta-\sinh^2\beta=1$ gives $c^Tc=1$, so the entire
triangle lies on the analytically continued quadric. The complex
spherical-excess formula is 
\begin{equation}
\tan \frac{\Omega_C}{2} = \frac{a\cdot(b\times c)} {1+a\cdot b+b\cdot
c+c\cdot a}.  \label{eq:sup-solid-angle-formula}
\end{equation}
For Eq.~\eqref{eq:sup-radial-triangle}, 
\begin{equation}
\begin{aligned}
\tan\frac{\Omega_C}{2}
&=\frac{\cosh\beta}{1+i\sinh\beta}
=\func{sech}\beta-i\tanh\beta,\\
&=\tan\left(\frac{\pi}{4}-\frac{i\beta}{2}\right).
\end{aligned}
\label{eq:sup-solid-angle-eval}
\end{equation}
Continuity from the Hermitian loop at $\beta=0$ fixes 
\begin{equation}
\boxed{\Omega_C=\frac{\pi}{2}-i\beta =\frac{\pi}{2}-2i\pi \kappa.}
\label{eq:sup-OmegaC-radial}
\end{equation}
Thus the imaginary part of the exchange angle is not an independent gate
parameter: it is the imaginary part of the geometric solid angle generated
by the microscopic complex coupling path.

\paragraph{Spin lift and projective modulus.}

The dark-Majorana transformation is a complex orthogonal rotation. Its Spin
lift acts on the two exchange channels with multipliers 
\begin{equation}
\lambda_\pm=e^{\pm i\Omega_C/2} =e^{\pm \beta/2}e^{\pm i\pi/4}.
\label{eq:sup-spin-multipliers}
\end{equation}
Equivalently, 
\begin{equation}
U_{12}^{\mathrm{NH}} =\exp \! \left[\frac{\Omega_C}{2}\gamma_1\gamma_2\right]
=\exp \! \left[ \left(\frac{\pi}{4}-i\pi \kappa \right)\gamma_1\gamma_2 %
\right].  \label{eq:sup-UNH-derived}
\end{equation}
The projective ratio is therefore 
\begin{equation}
\boxed{ q=e^{i\Omega_C}=i e^{2\pi \kappa}, \qquad |q|=e^{2\pi \kappa},
\qquad |\lambda_\pm|=e^{\pm \pi \kappa}.}
\label{eq:sup-single-exchange-opening}
\end{equation}
For a generic complex loop the same relation reads 
\begin{equation}
q=e^{i\Omega_C}, \qquad \ln|q|=-\func{Im}\Omega_C.
\label{eq:sup-q-Omega-general}
\end{equation}
Hence on the radial Ising slice 
\begin{equation}
\boxed{ \kappa[\boldsymbol \lambda(\cdot)]
=-\frac{\operatorname{Im}\Omega_C[n(\cdot)]}{2\pi}.}
\label{eq:sup-kappa-functional}
\end{equation}
For the triangular family, the apex couplings are $\lambda_2=\Lambda \cosh
\beta$ and $\lambda_5=i\Lambda \sinh \beta$. They give the closed
microscopic parameterization 
\begin{equation}
\boxed{ \kappa= \frac{1}{2\pi} \operatorname{artanh}\!
\left(-i\frac{\lambda_5}{\lambda_2}\right) = \frac{1}{4\pi} \log \!
\left(\frac{\lambda_2-i\lambda_5} {\lambda_2+i\lambda_5}\right),}
\label{eq:sup-kappa-lambda}
\end{equation}
with the branch chosen continuously from $\kappa=0$. If the arm endpoint
weights are unequal, the same formula applies to the projected couplings $%
\widetilde \lambda_a=w_a\lambda_a$.

\paragraph{Both neighboring exchange paths.}
For bilinear-unit vectors $v,w$, define the geodesic segment
\begin{equation}
g(u;v,w)=\frac{\sin[(1-u)\vartheta]}{\sin\vartheta}v
+\frac{\sin(u\vartheta)}{\sin\vartheta}w,
\quad \cos\vartheta=v^Tw.
\label{eq:sup-geodesic-segment}
\end{equation}
All branches are continued from the Hermitian reference. The first path is
\begin{equation}
\mathcal C_1:e_5\longrightarrow e_1\longrightarrow
\cosh\beta\,e_2+i\sinh\beta\,e_5\longrightarrow e_5.
\end{equation}
Its segment boundaries are $s=(0,1/3,2/3,1)$ and
$\boldsymbol\lambda(s)=\Lambda g((s-s_j)/(s_{j+1}-s_j);v_j,v_{j+1})$.
For the second path, take the continuous square-root branches of
\begin{equation}
c_q^2=-\frac{q}{(q-1)^2},\qquad
s_q^2=\frac{q^2-q+1}{(q-1)^2},\qquad q=ie^\beta,
\end{equation}
with $c_q=s_q=1/\sqrt2$ at $\beta=0$. Set
\begin{equation}
\begin{aligned}
a_q&=c_q^2-s_q^2,& b_q&=2c_qs_q,\\
h_q&=-a_qe_1+b_qe_3,& m_q&=b_qe_1+a_qe_3.
\end{aligned}
\end{equation}
Then $h_q^Th_q=m_q^Tm_q=1$, $h_q^Tm_q=0$, and the neighboring path is
\begin{equation}
\begin{aligned}
\mathcal C_2:e_5&\longrightarrow m_q\longrightarrow e_2\\
&\longrightarrow\cosh\beta\,h_q+i\sinh\beta\,m_q
\longrightarrow m_q\longrightarrow e_5.
\end{aligned}
\label{eq:sup-second-control-path}
\end{equation}
The segment boundaries are $(0,1/8,3/8,5/8,7/8,1)$, using the same
interpolation and amplitude $\Lambda$. At $\beta=0$, $h_q=e_3$ and
$m_q=e_1$. For finite-time propagation each local segment coordinate is
replaced by $\rho(u)=3u^2-2u^3$; this fixes the timing while preserving the
oriented control path and its Kato holonomy.

\paragraph{Common endpoint frame and logical lift.}
Let $Q_0^TQ_0=I_4$ be one initial dark frame used for both paths. Endpoint
closure and Kato transport give
\begin{equation}
O_r=Q_0^T\mathcal G_rU_{K,r}(1)Q_0\in SO(4,\mathbb C).
\label{eq:sup-common-closure}
\end{equation}
For the closed five-arm controls, $\mathcal G_r=I$; for honeycomb exchanges
$\mathcal G_r$ implements the endpoint site-sign gauge closure. Select the
Spin lift continuously from the identity with
\begin{equation}
\mathcal S_r^{-1}\widehat\gamma_a\mathcal S_r
=\sum_b(O_r)_{ab}\widehat\gamma_b,
\qquad B_r^{\rm micro}=\left.\mathcal S_r\right|_{\rm even}.
\end{equation}
A single simultaneous logical basis change relates the pair to the
algebraic $B_1,B_2$ representative. This prescription retains the relative
eigenline geometry. Independent diagonalizations are used to extract the
ordered eigenvectors, with both sets expressed in that same logical frame.

\paragraph{Full-Hamiltonian numerical realization.}

The numerical calculation extracts the logical operation from the full
Hamiltonian. At each point of the prescribed $\lambda_a(s)$
path it builds the complete finite-chain complex-antisymmetric Majorana
matrix, reconstructs the isolated rank-four zero-mode projector, and
transports the retained frame with the transpose metric $Q^TQ=I$. The closed
dark-Majorana transformation is projected to its complex-orthogonal factor,
lifted to $\func{Spin}(4,\mathbb{C})$, and restricted to a
fixed-total-parity logical doublet. Diagonalizing the resulting $B^{\mathrm{%
micro}}$ gives $q_{\mathrm{micro}}$ independently of the analytic
solid-angle formula. The mesh, size, path-shape, and reparameterization
scans therefore test the full sequence 
\begin{equation}
\begin{gathered}
\lambda_a(s)\longrightarrow H_{5Y}(s)\longrightarrow P(s)
\longrightarrow O_K,\\
O_K\longrightarrow B^{\mathrm{micro}}
\longrightarrow q_{\mathrm{micro}}.
\end{gathered}
\label{eq:sup-micro-modulus-chain}
\end{equation}
The Hermitian Ising point $\kappa=0$ has $|q|=1$, while the radial
non-Hermitian coordinate is the geometric relative modulus $\ln|q|=2\pi
\kappa $. The full-Hamiltonian Kato scan independently recovers this law;
the numerical certificate is reported in Sec.~\ref{sup:micro-coherence} and
Table~\ref{tab:sup-hex}.

For comparisons between an SGL protocol and a control protocol, both the
projective exchange spectrum and the minimum isolation gap are matched: 
\begin{equation}
q_A\simeq q_B, \qquad \Delta_{\mathrm{iso}}^{A} \simeq \Delta_{\mathrm{iso}%
}^{B}.  \label{eq:sup-matched-spectrum-gap}
\end{equation}
The remaining difference can then be attributed to the relative geometry of
the neighboring exchange eigenspaces.

% ============================================================

\subsection{Independent projective invariants}

\label{sup:micro-invariants} 
% ============================================================

For a $2\times2$ exchange matrix $B$, define the branch-independent spectral
invariant 
\begin{equation}
\mathfrak{x}_{\mathrm{spec}} = \frac{(\func{Tr}B)^2}{\det B}-3.
\label{eq:sup-xspec}
\end{equation}
If the ordered projective eigenvalue ratio is 
\begin{equation}
q=\frac{\lambda_+}{\lambda_-},
\end{equation}
then 
\begin{equation}
\mathfrak{x}_{\mathrm{spec}} = q+q^{-1}-1.
\end{equation}

Let $v_i^\pm$ span the ordered right eigenlines of $B_i$ and define 
\begin{equation}
[u,v]=\det(u,v).
\end{equation}
The relative geometric invariant is 
\begin{equation}
\mathfrak{x}_{\mathrm{geom}} = \frac{ [v_1^+,v_2^+]\,[v_1^-,v_2^-] }{
[v_1^+,v_2^-]\,[v_1^-,v_2^+] }.  \label{eq:sup-xgeom}
\end{equation}
It is invariant under independent eigenvector rescalings and under a common $%
GL(2,\mathbb{C})$ basis transformation.

The microscopic SGL mismatch is therefore 
\begin{equation}
\delta_{\mathrm{SGL}} = \left| \mathfrak{x}_{\mathrm{geom}} - \mathfrak{x}_{%
\mathrm{spec}} \right|.  \label{eq:sup-deltaSGL}
\end{equation}

For the Artin relation, define 
\begin{equation}
A=B_1B_2B_1, \qquad C=B_2B_1B_2.
\end{equation}
The projective braid residual is 
\begin{equation}
\epsilon_{\mathrm{braid}}^{\mathrm{proj}} = \inf_{\zeta \in \mathbb{C}%
^\times} \frac{ \|A-\zeta C\|_F }{ \|A\|_F }.  \label{eq:sup-proj-braid}
\end{equation}
Thus $\mathfrak{x}_{\mathrm{spec}}$ and $\mathfrak{x}_{\mathrm{geom}}$ are
extracted independently, while $\epsilon_{\mathrm{braid}}^{\mathrm{proj}}$
tests the full multi-generator relation.

% ============================================================

\paragraph{Scalar minimization and frame convention.}
For nonzero matrices $X,Y$ in the common logical frame, the Frobenius
projective distance has the closed form
\begin{equation}
\begin{aligned}
\zeta_*&=\frac{\Tr(Y^\dagger X)}{\Tr(Y^\dagger Y)},\\
d_{\rm proj}(X,Y)^2
&=1-\frac{|\Tr(Y^\dagger X)|^2}{\|X\|_F^2\|Y\|_F^2}.
\end{aligned}
\label{eq:sup-projective-distance-explicit}
\end{equation}
If $\zeta_*=0$, the expression is the infimum over nonzero scalars.
The zero set expresses projective equality. Finite Frobenius distances are
reported in the specified common frame; their numerical magnitudes depend
on a general nonunitary similarity. In contrast, $q$, $\mathfrak x$, and
vanishing of the projective relations are invariant under that similarity.

\paragraph{Hamiltonian-to-invariant numerical extraction protocol.}
For each stored exchange path $\mathcal C_r$, assemble the full complex
antisymmetric matrix $A(s)$ and form the isolated Riesz projector
\begin{equation}
 P(s)=\frac{1}{2\pi i}\oint_{\Gamma}
 [zI-iA(s)]^{-1}\,\dd z ,\qquad
 \Gamma\cap\operatorname{Spec}(iA(s))=\varnothing .
\end{equation}
The Kato frame is propagated in the physical spectral bundle using
$\dot U_K=[\dot P,P]U_K$ and endpoint-identified by the same base-frame
$Q_0$. Its four-Majorana holonomy is
$O_r=Q_0^T G_r U_{K,r}(1)Q_0$; the continuous even-parity Spin lift
provides the two-channel $B_r^{\rm micro}$. The endpoint sewing matrix
$G_r$ accounts for the gauge closure fixed by the protocol; for the
five-arm loops it is the identity.

The numerical invariants are extracted \emph{after} both matrices have
been transported to the common logical frame:
\begin{align}
 q_r&=\lambda_{r,+}/\lambda_{r,-},
 &\mathfrak x_{\rm spec}^{(r)}
 &=\frac{[\operatorname{Tr}B_r]^2}{\det B_r}-3,\\
 \mathfrak x_{\rm geom}
 &=\frac{\det(v_1^+,v_2^+)\det(v_1^-,v_2^-)}
 {\det(v_1^+,v_2^-)\det(v_1^-,v_2^+)} .
\end{align}
At every sampled parameter, eigenvalue labels are continued from the
reference frame and the same ordering is used in both generators.
The scalar normalization of either matrix cancels from $q_r$ and the
trace/determinant ratio; eigenvector rescaling and common logical
similarity cancel from the determinant quotient. The extraction pipeline
therefore tests two algebraically independent functions of the stored
microscopic matrices.

The recorded residuals are
\begin{align}
 \delta_r^{\rm SGL}&=|\mathfrak x_{\rm geom}
 -\mathfrak x_{\rm spec}^{(r)}|,\\
 \epsilon_{\rm Artin}^{\rm proj}
 &=d_{\rm proj}(B_1B_2B_1,B_2B_1B_2).
\end{align}
For the five-arm common-ratio branch, the reference
$\mathfrak x_{\rm spec}^{(1)}=\mathfrak x_{\rm spec}^{(2)}$ is evaluated
independently from each generator. The $\kappa$-scan, $N_s$-mesh scan, and
fixed-spectrum geometry-detuning control probe distinct conditions:
spectral continuity, continuum accuracy, and sensitivity to relative
eigenframe placement. In particular, fitting
$\delta_{\rm SGL}(N_s)=\delta_\infty+aN_s^{-p}$ keeps the continuum
intercept separate from the finite-mesh values.

\subsection{Continuum convergence and matched-spectrum controls}

\label{sup:mesh-controls} 
% ============================================================

Along the radial family 
\begin{equation}
q(\kappa) = i e^{2\pi \kappa},
\end{equation}
the analytic SGL trajectory is 
\begin{equation}
\mathfrak{x}_{\mathrm{SGL}}(\kappa) = -1+2i\sinh(2\pi \kappa).
\end{equation}

The sequence in Table~\ref{tab:sup-convergence} uses the ordered
projector product with normalization at the closed endpoint. With
$V_0=Q_0$ it is
\begin{equation}
\begin{aligned}
 V_{j+1}&=P_{j+1}V_j,\qquad M_N=Q_0^TV_N,\\
 O_N&=M_N(M_N^TM_N)^{-1/2}.
\end{aligned}
\label{eq:sup-projector-product-audit}
\end{equation}
The square root is continued from the regular complex-orthogonal branch,
and the even Spin block of $O_N$ supplies the tabulated invariants.
The per-step transpose-polar update in
Eq.~\eqref{eq:sup-kato-step-corrected} is a second discretization with the
same Kato limit. On the geodesic unsplit dark bundle it supplies the exact
parallel rotation between successive normals up to numerical roundoff;
the final-normalization projector product instead gives the tabulated
finite-mesh corrections. The accompanying
\texttt{verify\_fullchain\_sgl.py} evaluates both schemes from the same
full-chain spectral projectors and saves their logical matrices separately.

At $\kappa=0.10$, the Kato mesh sequence gives

\begin{center}
\refstepcounter{table}\label{tab:sup-convergence} {\small \textbf{TABLE %
\thetable.} Radial Kato mesh data at $\kappa=0.10$.}

\medskip {\small 
\begin{tabular}{rcc}
\toprule $N_s$ & $\delta_{\mathrm{SGL}}$ & $\epsilon_{\mathrm{braid}}^{%
\mathrm{proj}}$ \\ 
\midrule 12 & $8.4548\times10^{-3}$ & $4.0979\times10^{-3}$ \\ 
20 & $3.0266\times10^{-3}$ & $1.4654\times10^{-3}$ \\ 
32 & $1.1800\times10^{-3}$ & $5.7113\times10^{-4}$ \\ 
48 & $5.2409\times10^{-4}$ & $2.5363\times10^{-4}$ \\ 
70 & $2.4636\times10^{-4}$ & $1.1922\times10^{-4}$ \\ 
100 & $1.2070\times10^{-4}$ & $5.8407\times10^{-5}$ \\ 
150 & $5.3640\times10^{-5}$ & $2.5957\times10^{-5}$ \\ 
220 & $2.4935\times10^{-5}$ & $1.2066\times10^{-5}$ \\ 
\bottomrule &  & 
\end{tabular}%
}
\end{center}

with 
\begin{equation}
\delta_{\mathrm{SGL}} \propto N_s^{-2.00}, \qquad \epsilon_{\mathrm{braid}}^{%
\mathrm{proj}} \propto N_s^{-2.00}.  \label{eq:sup-convfits}
\end{equation}
The nearly identical exponents show that the geometric mismatch and the
projective braid defect vanish on the same discretization scale.

The stronger test is obtained by changing the relative generator geometry
while keeping the local spectral data fixed. In the fixed-axis control, the
projective spectrum follows the radial family while 
\begin{equation}
\mathfrak{x}_{\mathrm{ctrl}}=-1
\end{equation}
is held at its Hermitian Ising value. The archived data are

\begin{center}
\refstepcounter{table}\label{tab:sup-fixed} {\small \textbf{TABLE \thetable.}
Reported fixed-axis controls along the radial family.}

\medskip {\small 
\begin{tabular}{ccc}
\toprule $\kappa$ & $\delta_{\mathrm{geom}}$ & $\epsilon_{\mathrm{braid}}^{%
\mathrm{proj}}$ \\ 
\midrule 0.02 & 0.251989 & 0.124762 \\ 
0.04 & 0.507964 & 0.244295 \\ 
0.06 & 0.771970 & 0.354208 \\ 
0.08 & 1.048184 & 0.451528 \\ 
0.10 & 1.340973 & 0.534873 \\ 
0.12 & 1.654968 & 0.604270 \\ 
0.16 & 2.366859 & 0.705933 \\ 
0.20 & 3.229093 & 0.769536 \\ 
\bottomrule &  & 
\end{tabular}%
}
\end{center}

so that, at $\kappa=0.10$, the control retains a finite braid residual $%
\simeq0.535$ while the SGL branch converges to zero.

A continuous geometry-detuning test gives the same conclusion. At fixed $q$, 
\begin{equation}
\epsilon_{\mathrm{braid}}^{\mathrm{proj}} = O(\delta_{\mathrm{geom}})
\end{equation}
close to the SGL locus, in agreement with the exact Artin-defect
factorization of Sec.~\ref{sup:artindefect}.

A full-Hamiltonian small-detuning scan resolves this local linear regime.
Write $a_{12}=\cos \vartheta$, $a_{23}=\sin \vartheta$ for the locked
relative frame and vary $\vartheta \mapsto \vartheta+\varepsilon$ at fixed $%
\kappa=0.10$. The transfer axis and exchange-plane axis change together as $%
m=\sin \vartheta \,e_1+\cos \vartheta \,e_3$ and $h=-\cos \vartheta
\,e_1+\sin \vartheta \,e_3$; each control path retains $n^Tn=1$ and the same
complex triangle angle. With 240 steps per triangle edge and transpose-polar
Kato transport, the scan $\varepsilon=10^{-4},2\times10^{-4},5%
\times10^{-4},10^{-3},\ldots,0.1$ gives a maximum generator-ratio mismatch
below $10^{-13}$. For $10^{-4}\leq \varepsilon \leq10^{-3}$, the geometric
mismatch ranges from $3.11\times10^{-4}$ to $3.12\times10^{-3}$ and the
Artin residual from $1.51\times10^{-4}$ to $1.51\times10^{-3}$. Their
log--log slope is $1.00$, directly resolving the linear Artin-defect
factorization. The unsplit identical-arm spectrum is unchanged by these
complex-orthogonal frame variations. Thus the scan holds the spectrum and
external gap fixed while controlling the physical eigenline geometry.

Consequently, 
\begin{equation}
\begin{gathered}
\text{same projective spectrum + same protecting gap}\\
\not\Rightarrow\ \text{same braid action}
\end{gathered}
\label{eq:sup-negative-control}
\end{equation}

% ============================================================

\subsection{%
\texorpdfstring{Complex-$q$
scan
and
channel
labeling}{Complex-q
scan
and channel labeling}%
}

\label{sup:complex-scan} 
% ============================================================

The radial trajectory resolves a one-dimensional subset of the complex
statistical space. We therefore also sample 
\begin{equation}
q=e^{2\chi+2i\theta}.
\end{equation}
The analytic SGL relation gives 
\begin{align}
\func{Re}\mathfrak{x }&= 2\cosh(2\chi)\cos(2\theta)-1, \\
\func{Im}\mathfrak{x }&= 2\sinh(2\chi)\sin(2\theta).
\end{align}

At each point, the ordered ratio of $B_1$ is fixed first and the channel
ordering of $B_2$ is obtained by continuous projective matching. This is
necessary because a common central scalar can invalidate naive matching by
absolute eigenvalue distance.

After projective labeling, the full 56-point patch follows the same SGL
surface within the numerical transport resolution. Finite-mesh transport and SGL discrepancies are of order
$10^{-7}$--$10^{-4}$; explicit refinement separates these
finite-mesh corrections from projector idempotency and eigensolver
accuracy and determines the continuum trend.

% ============================================================

\subsection{Resolver settings across the microscopic certificates}
The three uses of $J_{\rm split}$ specify different operations on the same
finite-chain platform. The geometric exchange and curvature certificates
use $J_{\rm split}=0$, preserving the exact four-dimensional dark bundle.
Fusion sewing prepares resolver-defined endpoint charts and transports the
full isolated four-mode cluster along reassociation paths; the braid probes
switch the resolver off during their exchange segment. The persistent-splitting
dynamics keeps $J_{\rm split}=5\times10^{-4}$ throughout the full protocol.
This last setting resolves the competition between bulk adiabaticity and
accumulated in-manifold dynamical evolution. Consequently the duration
window in Sec.~\ref{sup:full-finite-time} and the quadratic resolver
extrapolation in Sec.~\ref{sup:micro-coherence} test complementary aspects
of the microscopic realization.
\section{Full Finite-Time Dynamics and Monitored Implementation}
\setcounter{equation}{0}
\label{sup:dynamics}
The full finite-arm Gaussian calculation connects the Kato operation to
finite-duration evolution. The monitored embedding at the end of this
section specifies a complementary conditional realization and its event
yield, with its own physical normalization.

\subsection{Full finite-arm finite-time Gaussian dynamics}

\label{sup:full-finite-time} 
% ============================================================

This section records the full finite-time calculation underlying Fig.~3 of the main text. It uses the same finite-arm Hamiltonian and
exchange paths as the Kato calculation, but the microscopic degrees of
freedom are propagated directly in physical time. The calculation therefore
provides an independent dynamical realization of the geometric exchange map
before the monitored open-system embedding of Sec.~\ref%
{sup:open-system-realization}.

\subsubsection{Full $46$-Majorana time evolution}

\label{sup:full-time-propagation}

For $L=4$ the five-arm model contains $N=46$ Majoranas and may be written as 
\begin{equation}
\widehat H(t)=\frac{i}{4}\Gamma^T A(t)\Gamma, \qquad A^T=-A.
\label{eq:sup-full-Ht}
\end{equation}

\paragraph{Persistent-splitting Hamiltonian and matrix convention.}
The finite-splitting run adds the static physical pair-$12$ term
\begin{equation}
\begin{aligned}
 \widehat H_J(t)&=H_{5Y}[n(t)]+\widehat H_{\rm split},\\
 \widehat H_{\rm split}&=\frac{iJ_{\rm split}}{2}
 \alpha_{1,1}\alpha_{2,1},\qquad J_{\rm split}=5\times10^{-4}.
\end{aligned}
\label{eq:sup-persistent-split-H}
\end{equation}
Its matrix is
\begin{equation}
 A_J(t)=A_{5Y}(t)+J_{\rm split}
 \bigl(E_{\alpha_{1,1},\alpha_{2,1}}
       -E_{\alpha_{2,1},\alpha_{1,1}}\bigr).
\label{eq:sup-persistent-split-A}
\end{equation}
Here $E_{IJ}$ has one nonzero entry at $(I,J)$.
The coefficient of the many-body bilinear is $J_{\rm split}/2$ in the
$\widehat H=(i/4)\Gamma^TA\Gamma$ convention. In the arm-contiguous
indexing of Sec.~\ref{sup:fivearm-protocol}, the added entries are
$(2,11)$ and $(11,2)$. The supplied implementation places the five
terminal $\eta_a$ modes after all chain sites, so its corresponding
indices are $(2,10)$ and $(10,2)$; a fixed simultaneous permutation
connects these two matrix orderings.

The same static pair-$12$ perturbation is retained throughout both
exchange protocols and across every join of the uninterrupted words.
The junction paths and their smooth segment schedules retain their
previous definitions. At the reference point $n=e_5$, the endpoint
projection gives
\begin{equation}
 \widehat H_{\rm split}^{\rm low}
 =\frac{iJ_{\rm split}\mathcal N_L^2}{2}\zeta_1\zeta_2,
 \qquad \mathcal N_L^2=0.751958224543.
\end{equation}
The characteristic logical energy separation is therefore
$\delta E_{\rm split}\simeq |J_{\rm split}|\mathcal N_L^2$.
Full-matrix diagonalization gives the reference cluster radius
$3.759791067\times10^{-4}$ and cluster-to-bulk separation
$0.3781260341$, consistent with the two time scales below.
The separation is the minimum distance between a cluster eigenvalue
and a complementary eigenvalue; the smallest bulk distance from zero
at this reference point is $0.3785020132$.

The source configuration fixes
$\kappa\in\{0,0.01,0.05,0.10\}$,
$T\in\{40,80,160,320,640,1280,2560\}$,
$\Delta t\in\{0.05,0.025\}$, and words $1,2,121,212$.
Its Kato/adiabatic reference uses 200 and 400 steps per segment for
the paired mesh comparison. This static perturbation specifies the
finite-time run; the fusion-chart resolver is varied along its separate
reassociation paths in Sec.~\ref{sup:fusion-channel-audit}.

The associated Gaussian columns obey 
\begin{equation}
\dot Y=A(t)Y.  \label{eq:sup-full-Ydot}
\end{equation}
For a step $h$, the midpoint Cayley update is 
\begin{equation}
Y_{n+1}= \left(I-\frac{h}{2}A_{n+1/2}\right)^{-1} \left(I+\frac{h}{2}%
A_{n+1/2}\right)Y_n.  \label{eq:sup-cayley}
\end{equation}
Because $A^T=-A$, this update preserves the complex-orthogonal bilinear
structure up to solver roundoff. No instantaneous Riesz projection is
applied to $Y$ during the propagation. The propagated columns may therefore
leave the instantaneous low-energy subspace, explore the full finite-arm
Majorana space, and return coherently at the endpoint. The finite-time scan
uses $\kappa=0,0.01,0.05,0.10$ and $T=40,80,160,320,640,1280,2560$
with $\Delta t=0.05,0.025$ for the exact-dark and
$J_{\rm split}=5\times10^{-4}$ data sets, for each of the four
words $1,2,121,212$. The separate $\Delta t=0.1$ runs provide the
three-step-size convergence comparison.

\subsubsection{Biorthogonal Gaussian logical encoding}

\label{sup:gaussian-encoding}

The initial isolated four-mode Majorana frame is denoted $%
Q=(Q_1,Q_2,Q_3,Q_4) $ with $Q^TQ=I_4$. The two logical complex-fermion
directions are chosen as 
\begin{equation}
v_1=\frac{Q_1+iQ_2}{\sqrt2},\qquad v_2=\frac{Q_3+iQ_4}{\sqrt2},
\label{eq:sup-logical-v}
\end{equation}
with transpose-dual partners 
\begin{equation}
w_1=\frac{Q_1-iQ_2}{\sqrt2},\qquad w_2=\frac{Q_3-iQ_4}{\sqrt2}.
\label{eq:sup-logical-w}
\end{equation}
The remaining $21$ complex fermions are obtained from the bulk eigenmodes
and biorthogonalized with the transpose metric. Writing the complete
annihilation and creation frames as $V_n,V_p\in \mathbb{C}^{46\times23}$,
the canonical conditions are 
\begin{equation}
V_n^TV_p=I, \qquad V_n^TV_n=V_p^TV_p=0.  \label{eq:sup-biorth-canonical}
\end{equation}
The finite-time calculation propagates all $23$ annihilation directions and
the two logical creation directions, 
\begin{equation}
Y(0)=\bigl[V_n,\,V_p(:,1{:}2)\bigr]\in \mathbb{C}^{46\times25}.
\label{eq:sup-Yinitial}
\end{equation}
Thus the many-body logical information is retained without constructing the
full $2^{23}$-dimensional Fock-space evolution operator.

\subsubsection{Exact logical-block reconstruction}

\label{sup:gaussian-block}

Let $Y_n$ denote the propagated first $23$ columns and $Y_p$ the propagated
two logical creation columns. In the initial biorthogonal chart define 
\begin{equation}
D=V_p^TY_n, \qquad B=V_{n,L}^TY_n, \qquad C=V_p^TY_p,  \label{eq:sup-DBC}
\end{equation}
where $V_{n,L}$ contains the first two logical annihilation directions. With 
$E_L$ the $23\times2$ selector of these logical modes, form 
\begin{equation}
\begin{aligned}
Z&=E_L^TD^{-1}C,\qquad X=BD^{-1}E_L,\\
E&=(E_L^TD^{-1}E_L)^T.
\end{aligned}
\label{eq:sup-ZXE}
\end{equation}
The Gaussian Wick/Pfaffian reduction on the even-parity logical basis $%
\{|00\rangle,|11\rangle \}$ then gives the projective block 
\begin{equation}
B_{\mathrm{dyn}} \sim 
\begin{pmatrix}
1 & -Z_{12} \\ 
X_{12} & \det E-X_{12}Z_{12}%
\end{pmatrix}%
.  \label{eq:sup-gaussian-block}
\end{equation}

\paragraph{Vacuum overlap and four-point Wick reduction.}
We fix the ordered even-parity basis as
$|00\rangle=|\Omega_0\rangle$ and
$|11\rangle=a_1^\dagger a_2^\dagger|\Omega_0\rangle$,
with dual bra $\langle11|=\langle\Omega_0|a_2a_1$.
Here $a_j$ and $a_j^\dagger$ are canonical operators in the initial
transpose-biorthogonal chart. Let $\widehat{\mathcal U}$ denote the
many-body Gaussian lift of the propagated Majorana transformation, with its
Spin branch continued from the identity. On a nonsingular vacuum chart,
the vacuum-to-vacuum overlap is a single complex scalar
\begin{equation}
\Omega=\langle\Omega_0|\widehat{\mathcal U}|\Omega_0\rangle,
\qquad \Omega^2=\det D,
\label{eq:sup-vacuum-overlap}
\end{equation}
where the square-root sign is fixed by $\Omega(0)=1$ and continuous
transport. The determinant formula refers to the same canonical ordering as
Eq.~\eqref{eq:sup-DBC}; a chart change alters the individual overlap
representation but not the reconstructed logical operator.
Gaussian contractions normalized by this common overlap are
\begin{align}
\Omega^{-1}\langle00|\widehat{\mathcal U}|11\rangle
 &=-Z_{12}, &
\Omega^{-1}\langle11|\widehat{\mathcal U}|00\rangle
 &=X_{12},\\
\Omega^{-1}\langle00|a_i\widehat{\mathcal U}a_j^\dagger|00\rangle
&=E_{ij}.\label{eq:sup-gaussian-contractions}
\end{align}
For the doubly occupied matrix element, Wick's theorem amounts to the
Pfaffian of the antisymmetric four-contraction matrix (the two outgoing
annihilators are ordered $a_2,a_1$, followed by the two incoming creators):
\begin{equation}
\Omega^{-1}\langle11|\widehat{\mathcal U}|11\rangle
=\operatorname{Pf}\!\begin{pmatrix}
0&X_{12}&E_{21}&E_{22}\\
-X_{12}&0&E_{11}&E_{12}\\
-E_{21}&-E_{11}&0&-Z_{12}\\
-E_{22}&-E_{12}&Z_{12}&0
\end{pmatrix}.
\label{eq:sup-four-point-pfaffian}
\end{equation}
Using $\operatorname{Pf}M=M_{12}M_{34}-M_{13}M_{24}
+M_{14}M_{23}$ gives
\begin{equation}
\frac{\langle11|\widehat{\mathcal U}|11\rangle}{\Omega}
=\det E-X_{12}Z_{12}.
\label{eq:sup-four-point-expansion}
\end{equation}
The minus sign in the vacuum-to-pair entry and the sign of the
$X_{12}Z_{12}$ term follow from the stated Fock ordering; reversing the
order of either fermion pair changes the corresponding Pfaffian signs.
The four contractions therefore reconstruct the full $2\times2$
return block in Eq.~\eqref{eq:sup-gaussian-block} with the common
prefactor $\Omega$. Removing this prefactor is a projective operation,
while the final common
Frobenius normalization does not change any relative complex channel
weight. Physical event probabilities require the unremoved amplitude and
the corresponding return/no-jump projection. At a zero of $\Omega$, the
four particle--hole charts described below provide a regular overlapping
coordinate system for the same Gaussian map.

Equation~\eqref{eq:sup-gaussian-block} provides the even-parity
projective logical operation. Absolute return and monitored-event probabilities
retain the common vacuum amplitude, as specified below.

The logical block is the return amplitude to the chosen bulk-vacuum
sector, retaining both logical occupations. Even for a Hermitian full
Hamiltonian, a finite-time return block can have unequal singular values
or unequal eigenvalue moduli because amplitude occupies other bulk sectors.
The geometric limit is identified by comparison with the independently
transported low-energy frame. Absolute event yields require the common
vacuum amplitude and, for monitored dynamics, the specified physical
no-jump normalization.

A single vacuum chart can become ill-conditioned when its reference vacuum
amplitude crosses zero even though the Gaussian transformation remains
regular. The implementation therefore evaluates four equivalent
particle--hole charts: unswapped, incoming-logical swap, outgoing-logical
swap, and a double swap. Each valid chart is transformed back to the same $%
\{|00\rangle,|11\rangle \}$ basis before comparison. Direct small-system
Fock-space self-tests of the same reconstruction verify the logical block to
machine precision, including pure-pairing transformations for which the
original vacuum chart is singular.

\subsubsection{Kato and full-adiabatic references}

\label{sup:finite-time-references}

The finite-time block is compared with two independently constructed
references. $B_K$ is the logical Spin lift of the Kato transport. The full
adiabatic reference additionally includes the projected Hamiltonian inside
the transported four-mode frame, 
\begin{equation}
h_\parallel(t)=Q(t)^T h(t)Q(t), \qquad D_\parallel(T)=\mathcal{T }%
e^{-i\int_0^T h_\parallel(t)dt},  \label{eq:sup-hparallel}
\end{equation}
where $Q(t)=U_K(t)Q_0$ is the Kato-parallel frame, so
$Q^T\dot Q=0$. In an arbitrary moving frame the projected generator also
contains $-iQ^T\dot Q$. The Majorana adiabatic map in the parallel frame is
$O_KD_\parallel(T)$ and defines $B_{\mathrm{ad}}(T)$ after the same endpoint identification and
logical lift. The three projective distances are 
\begin{align}
\epsilon_{\mathrm{NA}}&=d_{\mathrm{proj}}(B_{\mathrm{dyn}},B_{\mathrm{ad}}),
\\
\epsilon_K&=d_{\mathrm{proj}}(B_{\mathrm{dyn}},B_K), \\
\epsilon_{\mathrm{split}}&=d_{\mathrm{proj}}(B_{\mathrm{ad}},B_K).
\label{eq:sup-three-errors}
\end{align}
For composite words, the Kato and adiabatic logical references are formed by
the chronological products of the generator-wise lifted blocks. This keeps
the continuous Spin branch fixed across multi-generator compositions.

\subsubsection{Time-step convergence and canonical diagnostics}

\label{sup:finite-time-diagnostics}

The Cayley propagation is second order in $\Delta t$. Across the finite-time
scan, halving the step from $0.1$ to $0.05$ and from $0.05$ to $0.025$
reduces both the raw-frame and projective logical differences by the
expected factor of four in the matched-case comparison [Fig.~\ref{fig:sup-final-audit}(a)]. At the finest step
the largest transpose-Gram defect of the complete propagated Gaussian frame
is $5.75\times10^{-12}$ for $J_{\mathrm{split}}=0$ and $2.90\times10^{-11}$
in the persistent-splitting scan. Every reported Gaussian logical block
passes the chart-validity test. The smallest chart reciprocal condition
number is $3.24\times10^{-2}$ in the exact-dark scan and $2.56\times10^{-3}$
in the extended split scan. These values are monitored as conditioning
diagnostics rather than physical observables.

\subsubsection{Finite splitting and the dynamical braid window}

\label{sup:finite-time-splitting}

The duration $T$ refers to one elementary exchange. Each uninterrupted
$121$ or $212$ word has total duration $3T$ and is propagated as one
full-system history. Gaussian reconstruction is applied at its final
endpoint. Products of projected single-exchange blocks provide a separate
logical-composition diagnostic, since the uninterrupted evolution retains
bulk excursions between successive exchanges.

At $J_{\mathrm{split}}=0$ the four low-energy modes are dark to numerical
precision, so the full adiabatic operation is projectively the Kato braid.
The finite-time error is then controlled by nonadiabatic departure from that
moving manifold. At $T=640$ and $\Delta t=0.025$, the extracted projective
ratios and cross-ratio coordinates reproduce the non-Hermitian radial family
throughout the sampled $\kappa$ interval, while the direct
uninterrupted-word defect lies between $1.29\times10^{-4}$ and $%
1.48\times10^{-4}$.

\begin{figure*}[t]
\centering
\includegraphics[width=0.93\textwidth]{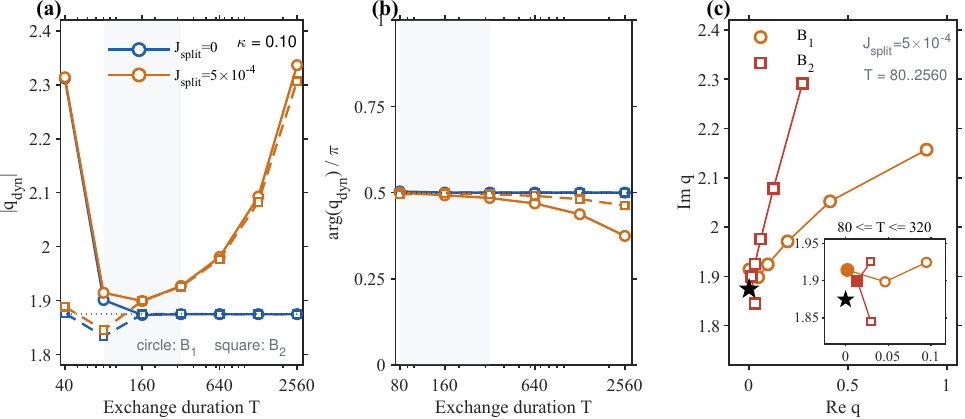}
\caption{\textbf{Finite-time complex exchange spectrum.}
(a) Absolute projective ratio $|q_{\rm dyn}|$ from full $46$-Majorana
Gaussian propagation, with the Kato reference $|q_K|=e^{2\pi\kappa}$.
(b) Principal spectral phase at $\kappa=0.10$ (displayed for $T\ge80$);
the $T=40$ result remains in the raw archive and in panel (a).
(c) Complex spectral ratios for the persistent-splitting exchanges,
with a local inset around the sampled intermediate-duration points and
the target $q_K=ie^{2\pi\kappa}$. Color distinguishes splitting status;
marker shape distinguishes generators. Connecting segments guide the eye
between discrete propagated durations, and are not interpolated
Hamiltonian solutions. The shaded range highlights the sampled finite-time
operating region, not an asymptotic splitting-induced convergence.}
\label{fig:sup-final-spectrum}
\end{figure*}

For the persistent-splitting test, $J_{\mathrm{split}}=5\times10^{-4}$ is
kept on during the full exchange. The sampled isolation gap is $0.378126$
and the low-energy cluster radius is $3.75979\times10^{-4}$, giving a clean
separation between the bulk and in-manifold time scales. Figure~\ref%
{fig:sup-final-competition} shows the resulting decomposition. At short duration $%
\epsilon_{\mathrm{NA}}$ dominates. In the intermediate regime the
microscopic dynamics closely follows $B_{\mathrm{ad}}$. At large $T$, $%
\epsilon_{\mathrm{NA}}$ continues to decrease while $\epsilon_{\mathrm{split}%
}$ grows because the in-manifold energy splitting accumulates dynamical
phase. The same structure appears directly in the exchange-spectrum modulus
and phase, as well as in the complex-$q$ trajectories [Fig.~\ref{fig:sup-final-spectrum}]. For the uninterrupted three-exchange words,
the minimum over the discrete scan occurs at $T=160$ or $320$ for all four
values of $\kappa$, with projective defects between $4.75\times10^{-2}$ and $%
5.11\times10^{-2}$.

\subsubsection{Channel-resolved phase and amplification}

\label{sup:finite-time-channels}

\begin{figure*}[t]
\centering
\includegraphics[width=0.93\textwidth]{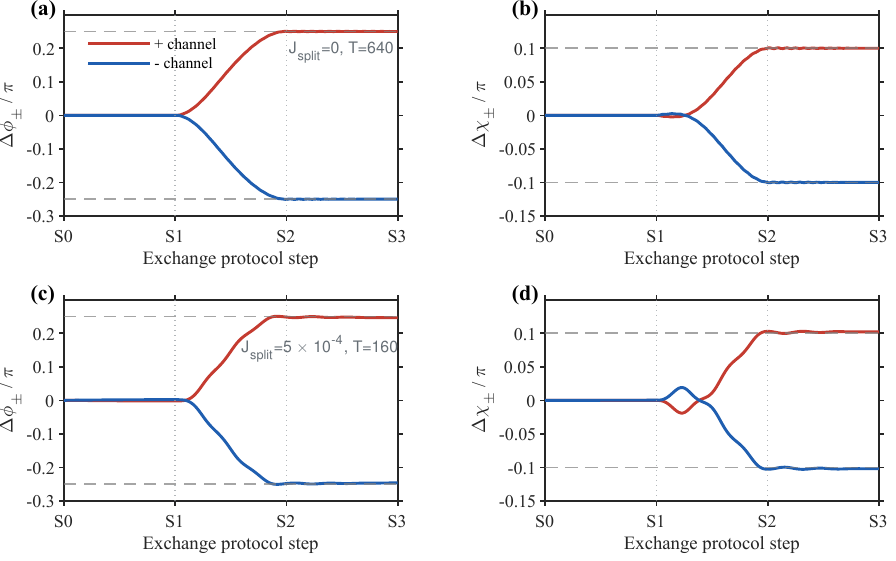}
\caption{\textbf{Channel-resolved mechanism of exchange softening.}
The full Gaussian history determines the complex ratio $\rho(s)$; its
symmetric-projective-gauge components $\Delta\phi_\pm/\pi$
and $\Delta\chi_\pm/\pi$ are displayed for (a,b) the exact-dark
$T=640$ exchange and (c,d) the persistent-splitting $T=160$
exchange at $\kappa=0.10$. Vertical guides mark the prescribed
exchange segments. Dashed horizontal guides are the separately evaluated
Kato channel targets, whereas continuous colored histories come from
finite-time propagation. The reflection between the two colors is the
chosen symmetric gauge, not an independently imposed channel equality.}
\label{fig:sup-final-channels}
\end{figure*}

For a fixed final Kato channel frame, let $a_\pm(s)$ denote the two diagonal
channel amplitudes of the propagated projective logical block. Their
relative ratio 
\begin{equation}
\rho(s)=\frac{a_+(s)}{a_-(s)}  \label{eq:sup-channel-rho}
\end{equation}
contains the gauge-invariant projective information. We display it in the
symmetric projective gauge, 
\begin{equation}
\Delta \phi_\pm(s)=\pm \frac12\arg \rho(s), \qquad \Delta \chi_\pm(s)=\pm
\frac12\ln|\rho(s)|.  \label{eq:sup-channel-symmetric-gauge}
\end{equation}
For $\kappa=0.10$, $J_{\mathrm{split}}=0$, and $T=640$, the endpoint values
are $\Delta \phi_\pm/\pi=\pm0.249998706$ and $\Delta
\chi_\pm/\pi=\pm0.100000774$. The fixed-channel off-diagonal mixing remains
below $1.72\times10^{-15}$ along the recorded protocol. With the persistent
splitting at $T=160$, the corresponding endpoint values are $\pm0.246127$
and $\pm0.102060$, and the intermediate amplitude trace resolves the small
dynamical transient visible in Fig.~\ref{fig:sup-final-channels}. The strict
red/blue mirror symmetry in that figure is the chosen symmetric projective
gauge; the directly computed invariant is the relative quantity $\rho(s)$.

\paragraph{Finite-time endpoint records and selection rule.}
The archived exact-dark and persistent-splitting calculations retain
$|q_1|$, $\arg(q_1)/\pi$, spectral--geometric comparisons,
generator-level Artin residuals, and uninterrupted exchange-word
distances across the specified $(\kappa,T,\Delta t)$ scans.
Endpoint summaries and the underlying propagation matrices are
preserved separately.
At $\kappa=0$ and $\Delta t=0.025$, the archived word distances are
$d_{121,212}=4.113\times10^{-4}$ for $J_{\rm split}=0$, $T=320$, and
$d_{121,212}=5.241\times10^{-2}$ for $J_{\rm split}=5\times10^{-4}$,
$T=160$.
The time-step comparison in Fig.~\ref{fig:sup-final-audit}, the channel histories
in Fig.~\ref{fig:sup-final-channels}, and the Kato-versus-dynamical distances
provide the complementary convergence and physical diagnostics. The total
duration of an uninterrupted word is $3T$.

 % ============================================================

\begin{figure*}[t]
\centering
\includegraphics[width=0.93\textwidth]{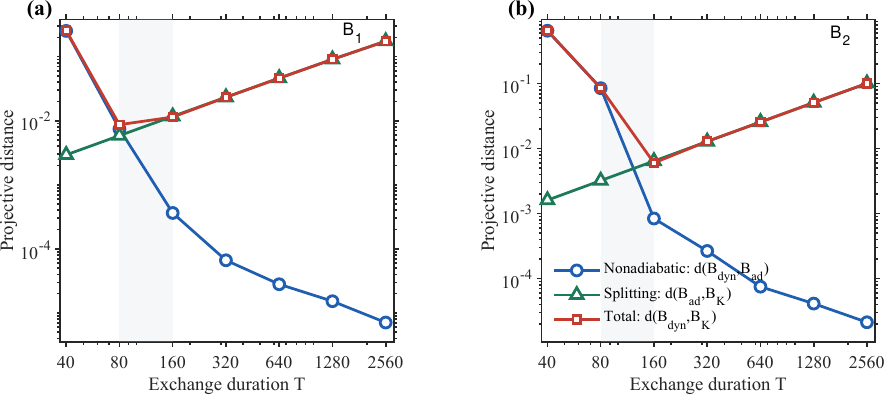}
\caption{\textbf{Competing finite-time mechanisms.}
For persistent $J_{\rm split}=5\times10^{-4}$ and $\kappa=0.10$,
(a) $B_1$ and (b) $B_2$ show the independently computed projective
distances between the dynamical, full-adiabatic and Kato logical blocks.
The decreasing nonadiabatic distance and accumulating in-manifold
splitting response define a finite operation window.
Symbols are the actual sampled durations; connecting lines serve as guides.}
\label{fig:sup-final-competition}
\end{figure*}

\subsection{Supplementary monitored open-system embedding}

 \label{sup:open-system-realization} 
% ============================================================

 The full finite-arm calculation above establishes the
Hamiltonian-level time-domain realization used in the main text. As an
additional implementation route, not required for any of the main-text
certificates, we now connect the same complex junction controls to a standard
monitored open-system dynamics. This extension adds the accepted-event
probability as a complementary operational observable. 

 Consider a Markovian master equation~\cite%
{GKSL1976,Lindblad1976} 
\begin{equation}
\dot \rho = -i[H_{\mathrm{coh}},\rho] + \sum_a \left( L_a\rho L_a^\dagger -
\frac12 \{L_a^\dagger L_a,\rho \} \right).  \label{eq:sup-lindblad}
\end{equation}
Conditioning on trajectories with no detected quantum jump gives~\cite%
{Dalibard1992} 
\begin{equation}
H_{\mathrm{eff}} = H_{\mathrm{coh}} - \frac{i}{2} \sum_aL_a^\dagger L_a.
\label{eq:sup-Heff}
\end{equation}

 For arm $a$, choose the monitored junction channel 
\begin{equation}
L_a = \sqrt{\frac{\Gamma_a}{2}} \left( \gamma_0 + e^{i\phi_a}\alpha_{a,1}
\right),  \label{eq:sup-jump}
\end{equation}
with $\Gamma_a\geq0$. Using 
\begin{equation}
\gamma_0^2=\alpha_{a,1}^2=1, \qquad \{ \gamma_0,\alpha_{a,1}\}=0,
\end{equation}
one finds 
\begin{equation}
L_a^\dagger L_a = \Gamma_a \left[ 1 + i\sin \phi_a\, \gamma_0\alpha_{a,1} %
\right].  \label{eq:sup-LdagL}
\end{equation}

 Adding the coherent junction term 
\begin{equation}
H_{\mathrm{coh}}^{(a)} = iJ_a\gamma_0\alpha_{a,1}
\end{equation}
gives 
\begin{equation}
H_{\mathrm{eff}}^{(a)} = i\lambda_a \gamma_0\alpha_{a,1} - \frac{i\Gamma_a}{2%
},
\end{equation}
where 
\begin{equation}
\lambda_a = J_a - i\frac{\Gamma_a}{2}\sin \phi_a.
\label{eq:sup-complex-lambda}
\end{equation}

 Thus 
\begin{equation}
\func{Re}\lambda_a = J_a, \qquad \func{Im}\lambda_a = -\frac{\Gamma_a}{2}%
\sin \phi_a,
\end{equation}
and coordinated variation of $J_a$, $\Gamma_a$, and $\phi_a$ generates the
complex control trajectories required in Eq.~\eqref{eq:sup-H5Y}. 

 Summing over the active arms yields 
\begin{equation}
H_{\mathrm{eff}} = \sum_{a=1}^{5}H_K^{(a)} + i\gamma_0 \sum_{a=1}^{5}
\lambda_a(s)\alpha_{a,1} - \frac{i}{2} \sum_a\Gamma_a(s).
\label{eq:sup-fivearm-open-Heff}
\end{equation}
The jump operators are odd in fermion parity, whereas the no-jump
Hamiltonian is even and preserves the conditional fixed-parity sector.
Recorded jumps connect parity sectors; the conditional no-jump branch
provides the fixed-parity logical operation considered here. 

 The final term is proportional to the identity and therefore
contributes only a common attenuation factor. It does not change the
projective quantities $q$, $\mathfrak{x}$, or $\epsilon_{\mathrm{braid}}^{%
\mathrm{proj}}$. 

 The attenuation determines the probability of obtaining the
conditional operation. Survival and projective action are therefore reported
as complementary operational observables. 

 % ============================================================

\subsubsection{Conditional evolution and projective
observables}

 \label{sup:open-projective} 
% ============================================================

 For a prescribed control trajectory $\Gamma$, the conditional
evolution is 
\begin{equation}
K_\Gamma(T) = \mathcal{T }\exp \left[ -i \int_0^T H_{\mathrm{eff}}(t)\,dt %
\right].  \label{eq:sup-Kraus}
\end{equation}
The physical conditional map $K_\Gamma$ is not identical to the Kato
parallel transporter: the latter isolates the geometric transport of the
instantaneous Riesz subspace. An experimental comparison with Kato holonomy
therefore requires removal of dynamical contributions and a common endpoint
convention. 

 After that calibration, the relevant logical object is its
projective class 
\begin{equation}
[K_\Gamma] \in PGL(r,\mathbb{C}).
\end{equation}

 For a two-channel logical sector with eigenvalues $%
\Lambda_+,\Lambda_-$, define 
\begin{equation}
q_{\mathrm{exp}} = \frac{\Lambda_+}{\Lambda_-}.  \label{eq:sup-qexp}
\end{equation}
For two neighboring reconstructed operations 
\begin{equation}
K_i v_i^\pm = \Lambda_i^\pm v_i^\pm,
\end{equation}
the relative geometry is 
\begin{equation}
\mathfrak{x}_{\mathrm{exp}} = \frac{ [v_1^+,v_2^+]\,[v_1^-,v_2^-] }{
[v_1^+,v_2^-]\,[v_1^-,v_2^+] }.  \label{eq:sup-xexp}
\end{equation}
The direct experimental SGL residual is 
\begin{equation}
\delta_{\mathrm{SGL}}^{\mathrm{exp}} = \left| \mathfrak{x}_{\mathrm{exp}} -
\left( q_{\mathrm{exp}} + q_{\mathrm{exp}}^{-1} -1 \right) \right|.
\label{eq:sup-exp-SGL}
\end{equation}

 A process-level braid test is obtained from 
\begin{equation}
K_{121}=K_1K_2K_1, \qquad K_{212}=K_2K_1K_2,
\end{equation}
with projective residual 
\begin{equation}
\epsilon_B^{\mathrm{exp}} = \inf_{c\in \mathbb{C}^\times} \frac{
\|K_{121}-cK_{212}\|_F }{ \|K_{121}\|_F }.  \label{eq:sup-exp-braid-residual}
\end{equation}

 The experimentally relevant set of observables is therefore 
\begin{equation}
\left( q_{\mathrm{exp}}, \mathfrak{x}_{\mathrm{exp}}, \epsilon_B^{\mathrm{exp%
}}, \Delta_{\mathrm{iso}} \right).  \label{eq:sup-exp-observables}
\end{equation}

 A particularly stringent test reproduces the microscopic
negative control: engineer two protocol families satisfying 
\begin{equation}
q_A\simeq q_B, \qquad \Delta_{\mathrm{iso}}^{A} \simeq \Delta_{\mathrm{iso}%
}^{B},
\end{equation}
while imposing 
\begin{equation}
\mathfrak{x}_A\neq \mathfrak{x}_B.
\end{equation}
The SGL branch should approach the experimental noise floor in both $\delta_{%
\mathrm{SGL}}^{\mathrm{exp}}$ and $\epsilon_B^{\mathrm{exp}}$, whereas the
geometry-detuned control should retain a finite braid residual. 

 A complementary test inserts a small contractible detour into
an otherwise fixed exchange trajectory. A resulting projective change of the
logical map would directly probe the curvature-induced path dependence
derived in Sec.~\ref{sup:quadric-flatness}. 

 % ============================================================

\subsubsection{Adiabaticity and monitored operation
window}

 \label{sup:open-window} 
% ============================================================

 The monitored protocol must satisfy two competing constraints.
First, the control must remain slow compared with the isolated spectral
dynamics. A useful scale estimate is 
\begin{equation}
\epsilon_{\mathrm{ad}} \sim \max_s \frac{ \| \partial_s H_{\mathrm{eff}%
}(s)\| }{ T\, \Delta_{\mathrm{iso}}(s)^2 } \ll1,
\label{eq:sup-adiabatic-condition}
\end{equation}
supplemented by resolvent bounds, projector conditioning, and dynamical
amplification control. This estimate sets the operational scale used for the
monitored adiabatic protocol. 

 Second, excessively slow protocols suppress the probability of
remaining in the no-jump sector. For an initial state $|\Psi_0\rangle$, 
\begin{equation}
P_0(T) = \left \| K_\Gamma(T)|\Psi_0\rangle \right \|^2,
\label{eq:sup-nojump-prob}
\end{equation}
or equivalently 
\begin{equation}
P_0(T) = \exp \left[ - \int_0^Tdt\, \left \langle \sum_aL_a^\dagger L_a
\right \rangle_c \right].  \label{eq:sup-nojump-exp}
\end{equation}
A useful operational window is therefore 
\begin{equation}
T_{\mathrm{ad}} \ll T_{\mathrm{loop}} \ll T_{\mathrm{loss}}.
\label{eq:sup-time-window}
\end{equation}

 The open-system construction embeds the effective five-arm
control manifold directly into conditional Lindblad dynamics. It therefore
supplies the microscopic control principle for a monitored realization, with
platform-level reservoir parameters, detection architecture, heating
budgets, and postselection rates providing the natural experimental
optimization layer. 

 Its scope is precisely 
\begin{equation}
\begin{gathered} \text{coherent control + engineered monitoring} \\
\longrightarrow \lambda_a(s)\in \mathbb C \longrightarrow H_{\rm eff}(s)
\longrightarrow P(s) \longrightarrow [W_\gamma] \\ \longrightarrow
(q,\mathfrak x). \end{gathered}  \label{eq:sup-experimental-chain}
\end{equation}
Thus the non-Hermitian control manifold used in the microscopic calculation
has a concrete monitored open-system interpretation and a direct route to
platform-specific experimental engineering. 

 % ============================================================
% BEGIN monitored_dynamics.tex

\subsection{Finite-time conditional exchange and event
yield}

 \label{sup:monitored-dynamics} 

 The monitored construction admits an explicit finite-time
realization on the six-Majorana junction of Eq.~\eqref{eq:sup-Hlow}. In the
even-parity Fock sector, of dimension four, use 
\begin{equation}
\begin{gathered}
H(s)=i\Delta \gamma_0\sum_{a=1}^{5}n_a(s)\zeta_a,\\
\Delta=0.1907741546,\qquad \kappa=0.01.
\end{gathered}
\end{equation}
The rank-two eigenspace of energy $-\Delta$ is the logical subspace. Its
spectral projector is $P(s)=[I-H(s)/\Delta]/2$, since $H(s)^2=\Delta^2I$.
The control path is the radial triangle of Eq.~\eqref{eq:sup-radial-triangle}%
, with equal segment durations. Along each segment, the geodesic parameter
is $v(u)=3u^2-2u^3$; the control velocity therefore vanishes at the
vertices. Time is measured with $\hbar=t=1$. This finite-time benchmark
specifies the effective junction Hamiltonian directly, while Proposition~\ref%
{prop:finite-arm-projector} applies to the geometric dark sector of the
finite-arm Hamiltonian. 

 For the effective-junction coefficients $\lambda_a^{\mathrm{j}%
}=\Delta n_a=h_a+ig_a$, choose 
\begin{equation}
L_a^{\mathrm{j}}=\sqrt{\Gamma_a/2}\, [\gamma_0-i\func{sgn}%
(g_a)\zeta_a],\qquad \Gamma_a=2|g_a|.
\end{equation}
Together with $i\sum_a h_a\gamma_0\zeta_a$, these jumps give the conditional
Hamiltonian $H(s)-i\sum_a\Gamma_a(s)/2$. Their common scalar attenuation
gives 
\begin{equation}
\mathcal{I}=\int_0^1\!ds\sum_a\Gamma_a(s)=0.0096765987.
\end{equation}
Let $Q$ be an orthonormal frame of $P(0)=P(1)$ and let $X_T$ solve $%
\partial_sX_T=-iT H(s)X_T$ with $X_T(0)=Q$. The full no-jump output is $e^{-T%
\mathcal{I}/2}X_T(1)$. The independent Kato reference solves $%
\partial_sW=[\partial_sP,P]W$, $W(0)=Q$. With $B_T=Q^\dagger X_T(1)$ and $%
B_K=Q^\dagger W(1)$, the reported observables are 
\begin{align}
\epsilon_T&=\min_z\frac{\|B_T-zB_K\|_F}{\|B_T\|_F}, \\
\ell_T&=\frac{\|(I-P(0))X_T(1)\|_F}{\|X_T(1)\|_F}, \\
[p_{\min},p_{\max}]&=\func{range}\func{Spec} \! \left(e^{-T\mathcal{I}%
}X_T(1)^\dagger X_T(1)\right).  \label{eq:finite-time-observables}
\end{align}
Here $\ell_T$ is an aggregate leakage amplitude and the survival interval
covers all normalized logical input states. Thus projective accuracy,
logical-subspace retention, and event yield are separately resolved. 

 At $T=120$, the projective error is $5.06\times10^{-6}$, the
leakage amplitude is $1.89\times10^{-3}$, and survival ranges from $0.296$
to $0.336$. At $T=320$, the corresponding values are $9.02\times10^{-10}$, $%
2.10\times10^{-5}$, and $0.0426$--$0.0483$. These operating points quantify
the duration-dependent balance between conditional geometric accuracy and
accepted-event yield. The separately archived full-duration monitored scan also resolves coherent nonadiabatic interference, so the control duration is
an operational parameter rather than a monotone accuracy proxy. The
Schr\"odinger and Kato equations are integrated independently with DOP853
and relative tolerances $2\times10^{-10}$ and $2\times10^{-11}$,
respectively. Tightening both tolerances by a factor of 100 changes the
quoted $\epsilon_T$ values at $T=120,320$ by less than $1.6\times10^{-13}$
in absolute terms.

 % END monitored_dynamics.tex

\section{Independent Frozen Two-Dimensional Kitaev
Vortex Certificate}
\setcounter{equation}{0}

 \label{sup:2d-kitaev} 
% ============================================================

 This section records the microscopic definition,
reproducibility conventions, projective diagnostics, and numerical controls
underlying Fig.~4 of the main text. The two-dimensional calculation is used
as an independent finite-size response certificate in a microscopic
Hamiltonian geometrically distinct from the five-arm junction. The exact
rank-two SGL relation remains the structural benchmark, but is not assumed to
be realized exactly at finite vortex separation; the frozen
scan resolves the single-exchange spectral responses and the
neighboring-eigenframe response independently and quantifies their
finite-geometry closure. 

 % ============================================================

\subsection{Fixed-flux honeycomb model and frozen
realization}

 \label{sup:2d-model} 
% ============================================================

 In a fixed $\mathbb{Z}_2$ flux sector of the chiral Kitaev
honeycomb model~\cite{Kitaev2006}, the matter Majoranas are described by 
\begin{equation}
H_{\mathrm{2D}}(\gamma)=\frac{i}{4}\mathbf{c}^T A[u;\gamma]\mathbf{c},
\qquad A^T=-A.  \label{eq:sup-2d-H}
\end{equation}
The non-Hermitian deformation is implemented on a frozen set of real-space
bonds as 
\begin{equation}
J_e(\gamma)=J\bigl(1+i\gamma \xi_e\bigr),  \label{eq:sup-2d-texture}
\end{equation}
with one common texture $\xi_e$ used for both neighboring generators. A
texture that merely rescales the entire Hamiltonian would leave the
instantaneous projectors unchanged and therefore cannot produce the effect
studied here. 

 The publication realization is fixed before the non-Hermitian
continuation: 
\begin{equation}
\begin{aligned} L_x\times L_y&=43\times52,\\ R_{\rm vortex}&=9,\\ d_s&=18,\\
N_s&=7,\\ J&=1,\\ \kappa_{\rm ch}&=0.20. \end{aligned}
\label{eq:sup-2d-frozen-parameters}
\end{equation}
The four vortex plaquettes, in zero-based brick-wall coordinates, are $%
p_1=(12,30)$, $p_2=(21,30)$, $p_3=(30,30)$, and $p_4=(21,12)$. Vortices $%
1,2,3$ define the two adjacent exchanges, and vortex $4$ is the remote
spectator. Gauge strings are closed after each exchange, and all generators
are represented on one common logical reference fiber. 

 % ============================================================
% BEGIN kitaev_protocol_details.tex

\paragraph{Bond texture and exchange manifest.}

 The brick-wall lattice has two sites per cell and periodic
boundary conditions. Its oriented nearest-neighbor bonds are $A(x,y)\to
B(x,y),B(x-1,y),B(x,y-1)$. For a bond separating plaquettes $(x_1,y_1)$ and $%
(x_2,y_2)$, let $(x_e,y_e)$ be their midpoint in the non-wrapping coordinate
patch. The common texture is 
\begin{equation}
\xi_e=\cos^2\! \frac{\pi(x_e-x_c)}{2h_x} \cos^2\! \frac{\pi(y_e-y_c)}{2h_y}
\tanh \! \frac{y_e-y_c}{w},  \label{eq:explicit-frozen-texture}
\end{equation}
inside $|x_e-x_c|<h_x$, $|y_e-y_c|<h_y$, and zero outside this window and on
wrap-seam bonds. Here $(x_c,y_c)=(21,34.5)$, $(h_x,h_y)=(10.5,6.5)$, and $%
w=1.5$. The oriented nearest-neighbor matrix entry is $A_{ij}=2J
u_{ij}(1+i\gamma \xi_e)$. For each length-two path $i\to c\to k$, the chiral
entry is $2\kappa_{\mathrm{ch}}\func{sgn}[(\mathbf{r}_i-\mathbf{r}_c) \times(%
\mathbf{r}_k-\mathbf{r}_c)]_z u_{ic}u_{ck}$, with the orientation convention
implemented in the accompanying bond list. These chiral terms use the same
interpolating real gauge variables as the nearest-neighbor terms. Initial
gauge strings join $p_1$ to $p_2$ and $p_3$ to $p_4$ by sorted shortest
dual-lattice paths; the second string avoids the first pair's vortex
plaquettes. 

 For adjacent plaquettes $A=(x_A,30)$ and $B=(x_B,30)$, the
exchange moves $A$ up to $(x_A,39)$ and across to $(x_B,39)$, moves $B$ to $%
A $, and moves the top vortex down to $B$. This sequence contains $4R=36$
elementary dual-edge moves. A moved bond is interpolated as $%
u_e(f)=u_e(0)\cos(\pi f)$ for $f=1/N_s,\ldots,1$. The endpoint site-sign
gauge transformation is applied before comparison with the common initial
fiber. 

 The interpolating $u_e(f)$ specifies a continuous externally controlled
bond amplitude between the endpoint $\mathbb Z_2$ flux configurations.
It enters both nearest-neighbor and chiral terms. At the endpoint a
diagonal site-sign matrix $\mathcal G_r$ satisfies
\begin{equation}
\mathcal G_r A_r(1)\mathcal G_r^T=A(0).
\end{equation}
The common-fiber matrix is then obtained from
Eq.~\eqref{eq:sup-common-closure}. Both exchanges start from the same
four-vortex configuration and use the same frozen texture.

At each step, eight eigenpairs of $iA$ near the shift $%
10^{-6}(1+i)$ locate the isolated rank-four cluster. With a right frame $R$,
the transpose-dual projector is $P=R(R^TR)^{-1}R^T$. Frame transport first projects
$\widetilde Q_{j+1}=P_{j+1}Q_j$ and then uses
\begin{equation}
Q_{j+1}=\widetilde Q_{j+1}
(\widetilde Q_{j+1}^T\widetilde Q_{j+1})^{-1/2},
\label{eq:sup-kato-step-corrected}
\end{equation}
with the square-root branch continuous from the identity. The endpoint orthogonal map is lifted
to the four-Majorana Fock space and restricted to even parity. Matrices,
cluster radii, external gaps, conditioning diagnostics, and eigenpair
residuals are saved for each parameter point. 

 % END kitaev_protocol_details.tex

\subsection{Blind Kato stage and independent projective
invariants}

 \label{sup:2d-blind} 
% ============================================================

 For each $\gamma$, the microscopic stage first constructs the
isolated rank-four vortex projector, transports it with the same Riesz--Kato
connection used throughout the paper, closes the endpoint gauge, and lifts
the resulting complex-orthogonal transformation to the fixed-parity logical
block. This stage outputs the raw matrices $B_1$ and $B_2$; the projective
spectrum and eigenline geometry are then reconstructed from these saved
microscopic holonomies. 

 Let $q_i$ be the ordered projective eigenvalue ratio of $B_i$.
The two branch-independent spectral coordinates are 
\begin{equation}
\mathfrak{x}_{\mathrm{spec}}^{(i)} =\frac{(\func{Tr} B_i)^2}{\det B_i}-3
=q_i+q_i^{-1}-1.  \label{eq:sup-2d-xspectra}
\end{equation}
The independent geometric coordinate $\mathfrak{x}_{\mathrm{micro}}$ is the
cross ratio of the four ordered right eigenlines of $B_1$ and $B_2$. The
common-spectrum mismatch is 
\begin{equation}
\delta_{\mathrm{spec}} =\frac{|\mathfrak{x}_{\mathrm{spec}}^{(1)}-\mathfrak{x%
}_{\mathrm{spec}}^{(2)}|} {1+\max_i|\mathfrak{x}_{\mathrm{spec}}^{(i)}|},
\label{eq:sup-2d-dspec}
\end{equation}
and the finite-size SGL residual is 
\begin{equation}
\delta_{\mathrm{SGL}} =\max_{i=1,2} \frac{|\mathfrak{x}_{\mathrm{micro}}-%
\mathfrak{x}_{\mathrm{spec}}^{(i)}|} {1+|\mathfrak{x}_{\mathrm{spec}}^{(i)}|}%
.  \label{eq:sup-2d-dsgl}
\end{equation}
For the projective Artin residual, with 
\begin{equation}
X=B_1B_2B_1,\qquad Y=B_2B_1B_2,
\end{equation}
we minimize over a common scalar, 
\begin{equation}
\epsilon_{\mathrm{braid}} =\frac{\|X-\zeta_*Y\|_F}{\max \{ \|X\|_F,\|Y\|_F\}}%
, \qquad \zeta_*=\frac{\func{Tr}(Y^\dagger X)}{\func{Tr}(Y^\dagger Y)}.
\label{eq:sup-2d-ebraid}
\end{equation}
The two-dimensional diagnostics in this section use the displayed
dimensionless denominators. The five-arm $\delta_{\mathrm{SGL}}$ in Eq.~%
\eqref{eq:sup-deltaSGL} is an absolute cross-ratio difference. Cross-model
comparisons use the separately reported response scales and their explicit
normalizations. 

 These three diagnostics resolve complementary spectral,
geometric, and Artin closure information; $\delta_{\mathrm{SGL}}$ and $%
\epsilon_{\mathrm{braid}}$ are computed independently. 

 % ============================================================

\subsection{Non-Hermitian continuation and numerical
conditioning}

 \label{sup:2d-continuation} 
% ============================================================

 The frozen continuation contains the weak-deformation points 
\begin{equation}
\gamma=0,0.01,0.02,0.04,0.06,0.08,0.10,
\end{equation}
and the extended-amplitude points 
\begin{equation}
\gamma=0.12,0.16,0.20,0.24,0.28,0.32,0.36,0.40,
\end{equation}
with the same geometry, texture, paths, gauge prescription, and mesh. The
diagnostic amplitude level $|\ln|q_i||=10^{-3}$ provides a common scale for
comparing the two smooth noncompact responses. Adjacent-point interpolation
places the two crossings at $\gamma \simeq0.311$ and $0.346$. 

 Across the combined $15$-point scan, 
\begin{equation}
\begin{aligned} \max \delta_{\rm spec}&=2.13\times10^{-4},\\ \max
\delta_{\rm SGL}&=2.27\times10^{-3},\\ \max \epsilon_{\rm
braid}&=2.16\times10^{-3},\\ \max_i|\ln|q_i||&=1.27\times10^{-3},\\ \min
\Delta_{\rm iso}&=0.618,\\ \min s_{\rm Kato}&=0.918. \end{aligned}
\label{eq:sup-2d-audit-extrema}
\end{equation}
The correlation of the two independently defined defects over the same
points is 
\begin{equation}
\func{corr}(\delta_{\mathrm{SGL}},\epsilon_{\mathrm{braid}})=0.9999008.
\label{eq:sup-2d-defect-correlation}
\end{equation}
Both residuals are functions of the same generator pair. Their correlated
finite-geometry trend is consistent with the local algebraic relation
between cross-ratio mismatch and Artin closure; the differential response
and its mesh resolution are quantified separately below. 

 For the differential trajectory in Fig.~4(c) of the main text,
we define 
\begin{equation}
\Delta \mathfrak{x}(\gamma)=\mathfrak{x}(\gamma)-\mathfrak{x}(0).
\label{eq:sup-2d-delta-x}
\end{equation}
This subtraction isolates the NH-induced displacement, while the absolute
residual in Eq.~\eqref{eq:sup-2d-dsgl} is retained independently. The two
quantities therefore display response and closure accuracy side by side.

 % ============================================================

\subsection{Finite-geometry and numerical controls}

 \label{sup:2d-controls} 
% ============================================================

 The frozen $R=9$ realization is supported by complementary
Hermitian finite-geometry and numerical controls summarized in
Fig.~\ref{fig:sup-final-audit}(b,c). These panels use the
$\gamma=0$ baseline, while the nonzero-$\gamma$ transport-mesh comparison
below is a separate frozen-manifest computation. Across the large-$R$ candidates the SGL and Artin
diagnostics enter the $10^{-3}$--level closure regime while the physical
rank-four vortex bundle remains localized and well conditioned. The
transport-mesh sequence shows convergence of $B_1$, $B_2$, and the eigenline
cross ratio toward the finest odd reference. Contractible gauge-string
deformations preserve the projective observables to numerical precision, and
the spectator-angle scan maps the finite-geometry anisotropy at fixed
separation.

 Together these controls resolve the independent non-Hermitian spectral
and relative-geometric responses in a geometrically distinct microscopic
Hamiltonian with an isolated, numerically stable vortex bundle. They also
quantify how the finite-size realization approaches the SGL/Artin benchmark,
without identifying the finite-separation response with an exact second
realization of the five-arm trajectory. 

 % BEGIN kitaev_response_audit.tex

\subsection{Differential non-Hermitian response, mesh
resolution, and separation}

 \label{sup:2d-response-audit} 

 The frozen $15$-point scan supports an explicit comparison of
response and closure scales. With the same-size, same-mesh Hermitian
baseline, 
\begin{equation}
\Delta \mathfrak{x}_\nu(\gamma)=\mathfrak{x}_\nu(\gamma)-\mathfrak{x}%
_\nu(0), \quad \nu \in \{ \mathrm{micro},\mathrm{spec1},\mathrm{spec2}\},
\end{equation}
the differential discrepancy is 
\begin{equation}
E_{\mathrm{NH}}=\max_{i=1,2}|\Delta \mathfrak{x}_{\mathrm{micro}} -\Delta%
\mathfrak{x}_{\mathrm{spec}}^{(i)}|.
\end{equation}
At $\gamma=0.4$ and $N_s=7$, 
\begin{equation}
|\Delta \mathfrak{x}_{\mathrm{micro}}|=2.37\times10^{-3},\qquad E_{\mathrm{NH%
}}=3.79\times10^{-4}.
\end{equation}
The differential discrepancy is $16.0\%$ of the measured geometric
displacement. This reports how the common finite-geometry eigenframe
responds relative to each generator's spectral prediction, while retaining
both distinct spectral trajectories in the comparison. 

\paragraph{Refinement at nonzero deformation.}

 The identical $R=9$ manifest is evaluated at $N_s=7,11,15$ for $%
\gamma=0,0.2,0.4$. Define the differential mesh resolution by 
\begin{equation}
m_N(\gamma)=\max_\nu \left| \Delta \mathfrak{x}_\nu^{(N)}(\gamma) -\Delta%
\mathfrak{x}_\nu^{(15)}(\gamma)\right|.  \label{eq:NH-mesh-resolution}
\end{equation}
The $N_s=15$ values are the finite-mesh reference; their zero self-distance
is kept in the table and omitted from the logarithmic trace. The paired
Hermitian runs remove the mesh-dependent baseline at each resolution. An
independent Python implementation reproduces the archived $N_s=7$ projective
observables to $10^{-10}$ absolute agreement or better at all three
comparison points. At $\gamma=0.4$, $m_7=1.54\times10^{-6}$ and $%
m_{11}=3.74\times10^{-7}$, compared with a geometric response of $%
2.37\times10^{-3}$. Thus the frozen response exceeds the measured $N_s=7$--$%
15$ mesh difference by a factor of $1.54\times10^3$. Across the complete
mesh and separation set, the minimum external gap is $0.6181$, the minimum
Kato-metric singular value is $0.9183$, and the maximum relative eigenpair
residual is $1.86\times10^{-11}$.

\begin{figure*}[!t]
\centering
\includegraphics[width=0.98\textwidth]{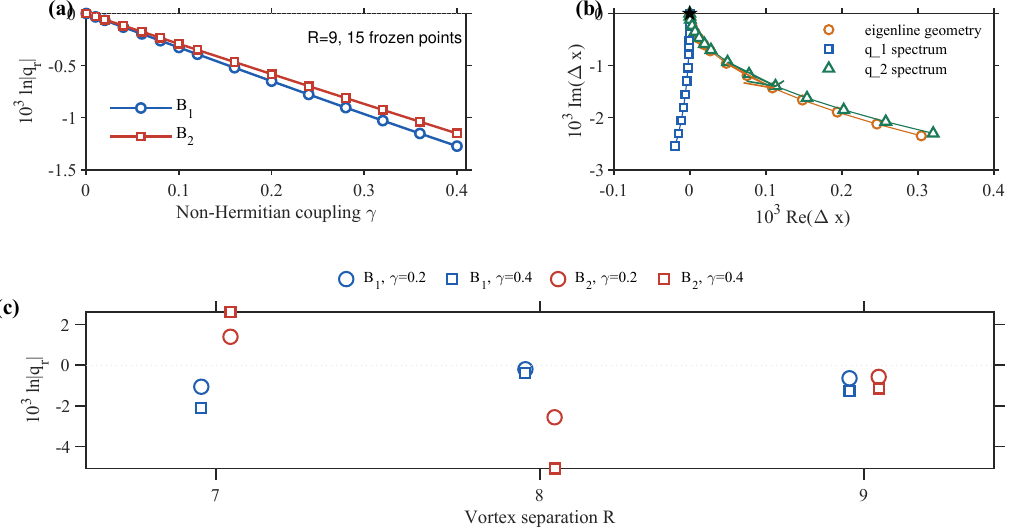}
\caption{\textbf{Independent two-dimensional spectral and relative-geometric response.}
(a) Signed $\ln|q_1|$ and $\ln|q_2|$ from the $15$-point frozen
$R=9$ honeycomb scan. (b) Independently extracted four-eigenline
$\Delta\mathfrak x_{\rm geom}$ and the two separate spectral SGL
predictions, all referenced to $\gamma=0$; their finite-geometry
differences are quantified in the text. (c) Discrete $R=7,8,9$
finite-geometry comparison at sampled $\gamma=0.2,0.4$.
This is a finite-geometry comparison, not a thermodynamic-size fit.
Markers are microscopic outputs; connections in (a,b) are visual guides.}
\label{fig:sup-final-two-d}
\end{figure*}

\paragraph{Separation and texture geometry.}

 A prescribed scaling family increases the vortex separation and
torus simultaneously: 
\begin{equation}
\begin{aligned}
L_x&=4R+7,\quad L_y=5R+7,\quad y_{\rm top}=4R+3,\\
p_i&=(iR+3,3R+3),\quad i=1,2,3,\\
p_4&=(2R+3,R+3),\\
(x_c,y_c)&=(2R+3,3.5R+3),\\
(h_x,h_y,w)&=(R+1.5,0.5R+2,1.5).
\end{aligned}
\end{equation}
Thus $R=9$ reproduces the frozen manifest, while $R=10,11$ give $47\times57$
and $51\times62$ tori. The common algorithm, $N_s=11$, $J=1$, and $\kappa_{%
\mathrm{ch}}=0.2$ are retained. Each size has its own $\gamma=0$ baseline
and $\gamma=0.4$ response. This family resolves the combined dependence on
separation, finite torus, and the specified smooth texture profile.

 The $R=9,10,11$ family in this paragraph is the prescribed
simultaneous torus-and-separation scan at $N_s=11$; it is distinct from
the archived $R=7,8,9$ finite-geometry comparison displayed in
Fig.~\ref{fig:sup-final-two-d}(c). The two groups use their own
frozen baseline and are not combined into a single finite-size fit.
At $\gamma=0.4$ the $R=9,10,11$ signed logarithmic pairs
$(10^3\ln|q_1|,10^3\ln|q_2|)$ are respectively
$(-1.2728,-1.1516)$, $(-0.6304,1.1471)$ and
$(0.4673,0.2843)$, as recorded in the size-scan output.

The larger geometries retain an isolated vortex bundle and
finite noncompact spectral responses. Their signs and relative amplitudes
resolve how the moving vortices sample the common texture. For example, the
two $R=10$ generators acquire oppositely directed logarithmic moduli,
whereas both $R=11$ ratios open in the positive direction. In the same
comparison, the absolute normalized SGL residual is $2.11\times10^{-3}$ at $%
R=9$, $1.14\times10^{-3}$ at $R=10$, and $1.27\times10^{-3}$ at $R=11$. The
separate common-spectrum diagnostic identifies the generator asymmetry of
each geometry. These observables distinguish noncompact channel action,
finite-geometry braid closure, and the geometry required for their alignment
on the exact SGL locus. 

 % END kitaev_response_audit.tex

\section{Microscopic Fusion Sewing, Categorical Selection, and Curvature}
\setcounter{equation}{0}
\label{sup:fusion-curvature}
Fusion-chart transport, categorical selection, and control-space curvature
resolve three distinct structures of the same microscopic statistical
bundle. We give their normalized observables and independent certificates
in that order.

\subsection{Microscopic fusion and fixed-Ising coherence}

\label{sup:micro-coherence} 
% ============================================================

The microscopic fusion analysis uses independently prepared fusion-resolved
Hamiltonians inside the same isolated four-mode manifold as the five-arm
braid calculation. A weak resolver $J_{\mathrm{split}}$ distinguishes the
relevant fusion channels while remaining parametrically below the bulk
isolation scale. At $J_{\mathrm{split}}=5\times10^{-4}$, the resolved
cluster radius is $3.760\times10^{-4}$ compared with $\Delta_{\mathrm{iso}%
}=0.378502013189$.

\subsubsection{Independent microscopic sewing map}

\begin{figure*}[t]
\centering
\includegraphics[width=0.98\textwidth]{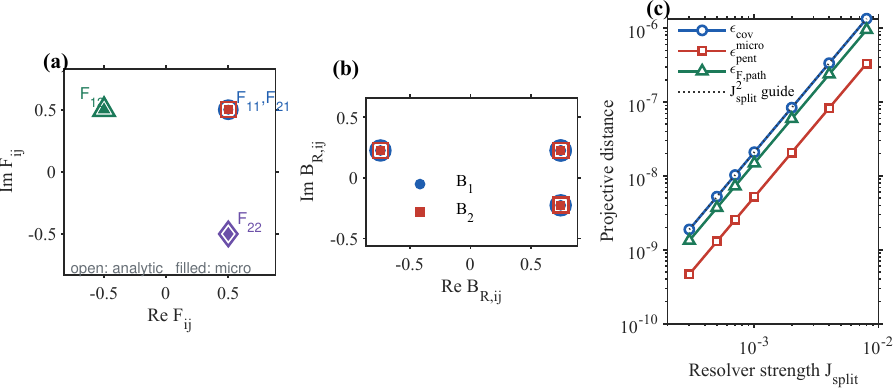}
\caption{\textbf{Microscopic fusion sewing and exchange channels.}
(a) Complex entries of the analytic inverse-chart reference
$G_{L\leftarrow R}$ (open symbols) and the microscopic
inverse-chart matrix (filled symbols), in the fixed plotting gauge.
(b) The two right-chart exchange generators obtained from microscopic
transport are compared with $G_{L\leftarrow R}^{-1}B_LG_{L\leftarrow R}$.
Equation~\eqref{eq:sup-fusion-gauge-match} gives the equivalent
forward-map relation used in the main text. Both comparisons use the
same ordered logical basis and projective normalization.
(c) Resolver-dependent covariance, microscopic Pentagon and
fixed-endpoint path responses, with the $J_{\rm split}^2$ guide.
These are distinct microscopic tests of sewing, reassociation and
control-path geometry, not substitutions for the exact categorical
Pentagon and Hexagon identities.}
\label{fig:sup-final-fusion}
\end{figure*}

Let $\mathcal{E}_L$ and $\mathcal{E}_R$ denote the physical $12$- and $23$%
-resolved fusion charts. Each endpoint Hamiltonian is diagonalized
independently, and an open full-Hamiltonian Kato reassociation transports
the isolated logical bundle from $L$ to $R$. In the working gauge the
resulting map is 
\begin{equation}
F_{\mathrm{micro}}\simeq 
\begin{pmatrix}
0.5+0.5i & 0.5+0.5i \\ 
-0.5+0.5i & 0.5-0.5i%
\end{pmatrix}%
.  \label{eq:sup-Fmicro-matrix}
\end{equation}
The figure-data convention also retains the inverse-chart map
$G_{L\leftarrow R}$, whose action carries right-chart coordinates into
the left chart. In the associated raw chart gauges,
\begin{equation}
B_{i,\rm raw}^{(R)}\simeq
G_{L\leftarrow R}^{-1}B_i^{(L)}G_{L\leftarrow R}.
\label{eq:sup-fusion-raw-orientation}
\end{equation}
The forward map in Eq.~\eqref{eq:sup-Fmicro-matrix} is related to this
inverse-chart representative by the fixed rephasing
\begin{equation}
F_{\rm micro}=iD_R G_{L\leftarrow R}^{-1},\qquad
D_R=\operatorname{diag}(1,-1),
\label{eq:sup-fusion-gauge-match}
\end{equation}
up to the recorded numerical precision. Hence
$B_{i,\rm pub}^{(R)}=D_R B_{i,\rm raw}^{(R)}D_R^{-1}$ and
$B_{i,\rm pub}^{(R)}\simeq F_{\rm micro}B_i^{(L)}F_{\rm micro}^{-1}$.
The left-chart common phase is immaterial to its braid matrices;
the right-chart rephasing must be applied to both the sewing map
and its exchange probes. This identifies the inverse-chart convention
used for Fig.~\ref{fig:sup-final-fusion} with the forward convention
of the main text.

The forward/reverse reassociation inverse residual is $1.356\times10^{-15}$.
The gauge- and row/column-rescaling invariant cross ratio 
\begin{equation}
\Xi_F=\frac{F_{12}F_{21}}{F_{11}F_{22}}
\end{equation}
takes the value 
\begin{equation}
\Xi_F=-0.999999985189, \qquad |\Xi_F+1|=1.481\times10^{-8}.
\label{eq:sup-XiF-final}
\end{equation}
The standard Ising associator enters separately as an exact categorical
benchmark; Eq.~\eqref{eq:sup-Fmicro-matrix} is reconstructed from the
microscopic reassociation path.

The braid probes in the two fusion charts are recorded in their fixed gauges.
Using the published, rephased chart convention and the
fixed-normalization covariance residual defined below, 
\begin{equation}
B_i^{(R)}\simeq F_{\mathrm{micro}}B_i^{(L)}F_{\mathrm{micro}}^{-1},
\end{equation}
gives the two recorded generator comparisons with the common maximum value 
\begin{equation}
\epsilon_{\rm cov}^{\max}=5.236\times10^{-9}.
\label{eq:sup-cov-final}
\end{equation}

In the published forward-map gauge, the fixed-normalization
covariance residual is
\begin{equation}
\epsilon_{\rm cov}=
\max_i\frac{\|B_i^{(R)}-F_{\rm micro}B_i^{(L)}F_{\rm micro}^{-1}\|_F}
{\max(1,\|B_i^{(R)}\|_F)}.
\label{eq:sup-cov-normalization}
\end{equation}
Equivalently, the archived raw-chart comparison uses
$G_{L\leftarrow R}^{-1}B_i^{(L)}G_{L\leftarrow R}$;
its normalized Frobenius residual is identical after the
unitary diagonal rephasing $D_R$. The common scalar and logical
frame are fixed in both descriptions.
For two reassociation paths, the archived loop convention closes in the
target frame: $O_{\rm rel}=O_{\rm short}O_{\rm long}^{-1}$.
Its continuous even-parity lift gives
\begin{equation}
\epsilon_{\rm loop}=\|\mathcal S_+(O_{\rm rel})-I_2\|_F/\sqrt2.
\label{eq:sup-loop-normalization}
\end{equation}
The values $\epsilon_{\rm pent}^{\rm micro}$ and
$\epsilon_{F,\rm path}$ below use this normalization. The source-frame
loop $P_{\rm long}^{-1}P_{\rm short}$ is conjugate to the target-frame
loop and expresses the same closure condition; the stated norm refers to
the target frame. Scalar-minimized tests are defined separately in
Sec.~\ref{sup:coherence}. Thus the numerical sewing certificate fixes a
representative of the projective fusion map.

Reassociation consistency is probed with three geometrically distinct
microscopic paths between the same fusion charts: direct interpolation, $%
L\to H_0\to R$, and a constant-norm arc in the $(J_{12},J_{23})$ plane.
Their continuously lifted relative-loop residuals are 
\begin{equation}
\epsilon_{F,\mathrm{path}}^{H_0}=3.703\times10^{-9}, \qquad \epsilon_{F,%
\mathrm{path}}^{\mathrm{arc}}=2.113\times10^{-9},  \label{eq:sup-Fpath-final}
\end{equation}
and the corresponding $\Xi_F$ values are approximately $-0.9999999852$, $%
-1.0000000000$, and $-0.9999999767$.

\subsubsection{Microscopic associahedral Pentagon}

Five fusion-resolved microscopic Hamiltonians $T_1,\ldots,T_5$ define the
vertices of the associahedral Pentagon. Each edge map is obtained by Kato
reassociation of the full Hamiltonian. Let $P_{\mathrm{short}}$ and $P_{%
\mathrm{long}}$ denote the two ordered products connecting the same initial
and final fusion charts and define 
\begin{equation}
W_{\mathrm{assoc}}=P_{\mathrm{long}}^{-1}P_{\mathrm{short}}.
\end{equation}
The continuously lifted target-frame loop in the even-fusion sector yields 
\begin{equation}
\epsilon_{\rm pent}^{\rm micro}=1.296\times10^{-9}.
\label{eq:sup-micro-pent-final}
\end{equation}
The short and long products are order-one reassociation transformations, so
Eq.~\eqref{eq:sup-micro-pent-final} measures discrete fusion-space holonomy
rather than proximity of each path to the identity. Floating-point evaluation of the exact
categorical Ising benchmark gives 
\begin{equation}
\epsilon_{\mathrm{pent}}^{\mathrm{cat}}=1.110\times10^{-16}, \qquad \|W_{%
\mathrm{cat}}-I\|=2.220\times10^{-16}.  \label{eq:sup-cat-pent-final}
\end{equation}

The fusion-coherence data are summarized in Table~\ref%
{tab:sup-coherence-final}.

\begin{center}
\refstepcounter{table}\label{tab:sup-coherence-final} {\small \textbf{TABLE %
\thetable.} Microscopic fusion-coherence certificate at $J_{\mathrm{split}%
}=5\times10^{-4}$, together with the floating-point categorical Pentagon benchmark.}

\medskip {\small 
\begin{tabular}{lc}
\toprule Quantity & Value \\ 
\midrule $\Delta_{\mathrm{iso}}$ & $0.378502013189$ \\ 
resolved cluster radius & $3.760\times10^{-4}$ \\ 
$\Xi_F$ & $-0.999999985189$ \\ 
$F$ inverse residual & $1.356\times10^{-15}$ \\ 
$\epsilon_{\mathrm{cov}}^{\max}$ & $5.236\times10^{-9}$ \\ 
$\epsilon_{\mathrm{pent}}^{\mathrm{micro}}$ & $1.296\times10^{-9}$ \\ 
$\epsilon_{F,\mathrm{path}}^{H_0}$ & $3.703\times10^{-9}$ \\ 
$\epsilon_{F,\mathrm{path}}^{\mathrm{arc}}$ & $2.113\times10^{-9}$ \\ 
$\epsilon_{\mathrm{pent}}^{\mathrm{cat}}$ & $1.110\times10^{-16}$ \\ 
\bottomrule & 
\end{tabular}%
}
\end{center}

\subsubsection{Two-scale convergence of fusion coherence}

Resolver removal and Kato-mesh refinement separate cleanly.
In the main-text Fig.~5(a), the horizontal axis is $J_{\rm split}$ and the
quadratic guide has log--log slope $+2$. Plotting the same residual against
$J_{\rm split}^2$ would give slope $+1$. The publication
resolver scan uses 
\begin{equation}
J_{\mathrm{split}}\in \{8,4,2,1,0.7,0.5,0.3\} \times10^{-3}
\end{equation}
and gives 
\begin{align}
\epsilon_{\mathrm{cov}}&\propto J_{\mathrm{split}}^{2}, \\
\epsilon_{\mathrm{pent}}&\propto J_{\mathrm{split}}^{2}, \\
\epsilon_{F,\mathrm{path}}&\propto J_{\mathrm{split}}^{2}.
\label{eq:sup-resolver-powers}
\end{align}
At fixed $J_{\mathrm{split}}=5\times10^{-4}$, the mesh series $%
N_s=16,\ldots,160$ approaches 
\begin{equation}
\epsilon_\infty=5.236438124116\times10^{-9}
\end{equation}
with plateau-subtracted slope 
\begin{equation}
p_N=-1.97\simeq-2.
\end{equation}
The independent split scan gives $p_J\simeq2.00$ and quadratic coefficient $%
A=0.0209461913$. At the publication resolver, 
\begin{equation}
A J_{\mathrm{split}}^2=5.236547831014\times10^{-9},
\end{equation}
which agrees with the mesh-extrapolated plateau to relative difference $%
2.10\times10^{-5}$. The combined asymptotic structure is therefore 
\begin{equation}
 \epsilon_{\rm coh}(N_s,J_{\rm split}) =A J_{\rm split}^2+B
N_s^{-2}+\cdots . \label{eq:sup-two-scale-coherence}
\end{equation}

\subsubsection{Exact planar-Ising Hexagon selection and microscopic $R$ ratio%
}

The Hexagon step uses the conventional Ising fusion rules and the
Pentagon-closed associator gauge, while treating all nontrivial $R$ symbols
as unknowns. Among the nonredundant clockwise/anticlockwise equations one
obtains 
\begin{equation}
r_{\psi \sigma}^{2}+1=0, \qquad r_1-r_\psi r_{\psi \sigma}=0, \qquad q
r_1-r_\psi=0.
\end{equation}
For nonzero admissible $R$ symbols these equations eliminate to 
\begin{equation}
q^2+1=0,\qquad q=\pm i.  \label{eq:sup-hex-exact-selection-final}
\end{equation}
The elimination is explicit: $r_1\ne0$ gives
$q r_{\psi\sigma}=1$, and
\begin{equation}
q^2+1=q^2(r_{\psi\sigma}^2+1)
 -(q r_{\psi\sigma}-1)(q r_{\psi\sigma}+1).
\end{equation}
Thus the required polynomial belongs to the denominator-cleared constraint
ideal. The quantity $\eta_{\rm hex}$ evaluates this exact spectral selector
on the microscopic $R$ ratio; matrix-valued Hexagon residuals additionally
retain the specified central $R$-symbol normalization.

The floating-point equation residuals are $0$ for the clockwise set and $%
2.220\times10^{-16}$ for the denominator-cleared anticlockwise set.

The microscopic projective $R$ ratio is extracted independently from forward
and reverse physical Kato exchanges, 
\begin{equation}
q_{\mathrm{micro}} =\frac{R_{\mathrm{micro}}^{\sigma \sigma \to \psi}} {R_{%
\mathrm{micro}}^{\sigma \sigma \to1}}.  \label{eq:sup-qmicro-def}
\end{equation}
The symmetric publication scan uses 
\begin{equation}
\begin{aligned} \kappa \in
\{&0,\pm0.0025,\pm0.005,\pm0.01,\pm0.015,\pm0.02,\\
&\pm0.03,\pm0.05,\pm0.075,\pm0.10\}. \end{aligned}
\end{equation}
Across all 19 points, 
\begin{align}
\max|q_+q_--1|&=4.599\times10^{-14}, \\
\max \frac{\bigl||q_+|-e^{2\pi \kappa}\bigr|}{e^{2\pi \kappa}}
&=2.959\times10^{-14}.
\end{align}
Thus the microscopic branch follows 
\begin{equation}
q_{\mathrm{micro}}(\kappa)=i e^{2\pi \kappa}
\end{equation}
at the quoted numerical precision. Evaluating the exact selection polynomial
on the microscopic ratio gives the normalized projective obstruction 
\begin{equation}
\eta_{\mathrm{hex}} =\frac{|q_{\mathrm{micro}}^2+1|}{1+|q_{\mathrm{micro}}|^2%
} =|\tanh(2\pi \kappa)|,  \label{eq:sup-micro-hex-prediction}
\end{equation}
with 
\begin{equation}
\max \left| \eta_{\mathrm{micro}}-|\tanh(2\pi \kappa)|
\right|=2.300\times10^{-14}.  \label{eq:sup-hex-diagnostics}
\end{equation}
The projective ratio $q$ is exactly the spectral coordinate entering SGL and
the Hexagon selection; the common Abelian $R$ factor belongs to the central
determinant-line sector and drops out of these projective relations.

Representative points are

\begin{center}
\refstepcounter{table}\label{tab:sup-hex} {\small \textbf{TABLE \thetable.}
Microscopic projective $R$ ratio and planar-Ising Hexagon obstruction.}

\medskip {\small 
\begin{tabular}{rccc}
\toprule $\kappa$ & $|q_+|$ & $\eta_{\mathrm{CW}}$ & $\eta_{\mathrm{CCW}}$
\\ 
\midrule $-0.1000$ & 0.533488091 & 0.5568933069 & 0.5568933069 \\ 
$-0.0500$ & 0.730402691 & 0.3042161929 & 0.3042161929 \\ 
$-0.0200$ & 0.881911378 & 0.1250063906 & 0.1250063906 \\ 
$-0.0100$ & 0.939101367 & 0.0627493000 & 0.0627493000 \\ 
$-0.0025$ & 0.984414763 & 0.0157066715 & 0.0157066715 \\ 
$0$ & 1.000000000 & $2.96\times10^{-14}$ & $1.64\times10^{-14}$ \\ 
$+0.0025$ & 1.015831982 & 0.0157066715 & 0.0157066715 \\ 
$+0.0100$ & 1.064847773 & 0.0627493000 & 0.0627493000 \\ 
$+0.0200$ & 1.133900780 & 0.1250063906 & 0.1250063906 \\ 
$+0.0500$ & 1.369107771 & 0.3042161929 & 0.3042161929 \\ 
$+0.1000$ & 1.874456088 & 0.5568933069 & 0.5568933069 \\ 
\bottomrule &  &  & 
\end{tabular}%
}
\end{center}

Near the Hermitian point, 
\begin{equation}
\eta_{\mathrm{hex}} =2\pi|\kappa|-\frac{8\pi^3}{3}|\kappa|^3+O(|\kappa|^5),
\end{equation}
so that 
\begin{equation}
\boxed{ \lim_{\kappa \to0}\frac{\eta_{\rm hex}}{|\kappa|}=2\pi.}
\end{equation}
A fit of $\eta_{\mathrm{hex}}/|\kappa|$ versus $\kappa^2$ over $%
0<|\kappa|\le0.02$ gives intercept $6.283163951674$, compared with $%
2\pi=6.283185307180$, and slope $-82.162246$, compared with $%
-8\pi^3/3=-82.683404$. The radial noncompact family is therefore
braid-consistent and equipped with projective fusion sewing across its
continuous $\kappa$ range, while the conventional planar Ising Hexagon
selects the discrete categorical locus $q=\pm i$.

% ============================================================
% BEGIN fusion_channel_audit.tex

\subsection{Physical fusion channels and resolver-plane geometry}

\label{sup:fusion-channel-audit}

The microscopic atlas is specified by physical resolver Hamiltonians. Write $%
Q_{ab}=(i/2)\alpha_{a,1}\alpha_{b,1}$, so that a coefficient $J$ in the
antisymmetric matrix corresponds to $JQ_{ab}$ in the many-body Hamiltonian.
At the reference junction $n=e_5$, the five vertices are 
\begin{equation}
\begin{array}{ccl}
T_1 & (((12)3)4) & H_0+JQ_{12}, \\ 
T_2 & ((12)(34)) & H_0+J(Q_{12}+\eta Q_{34}), \\ 
T_3 & (1(2(34))) & H_0+JQ_{34}, \\ 
T_4 & (1((23)4)) & H_0+J(Q_{23}+\eta Q_{14}), \\ 
T_5 & ((1(23))4) & H_0+JQ_{23},%
\end{array}
\qquad \eta=0.35.  \label{eq:five-physical-trees}
\end{equation}
Their endpoint frame seeds use arm orders $(1,2,3,4)$ for $T_1,T_2$, $%
(3,4,1,2)$ for $T_3$, and $(2,3,1,4)$ for $T_4,T_5$. Projecting these seeds
and applying the transpose-polar normalization fixes the oriented frames.
Each directed edge $E_{ji}$ maps the $T_i$ frame to the $T_j$ frame through
linear interpolation of the physical resolver coefficients, with 80 Kato
steps. The two products are $E_{32}E_{21}$ and $E_{34}E_{45}E_{51}$. The
Spin lift is continued from the identity and restricted to the even sector.
With this scalar and sign convention, the near-identity loop norm $\|U_{%
\mathrm{loop}}-I\|_F/\sqrt2$ gives the displayed Pentagon residual; its
leading traceless contribution agrees with the projective comparison.

\paragraph{Channel alignment.}

A fusion chart supplies both a transported frame and a physical channel
observable. Diagonalizing its projected resolver in the even sector gives $%
V_{ab}$. Switching that resolver off, performing the radial exchange in $%
\func{span}\{e_a,e_b,e_5\}$, and restoring it defines $B_{ab}$. At $%
\kappa=0.10$ the direct channel diagnostic is 
\begin{equation}
d_{ab}=\frac{\|V_{ab}^{-1}B_{ab}V_{ab} -\func{diag}(V_{ab}^{-1}B_{ab}V_{ab})%
\|_F} {\|V_{ab}^{-1}B_{ab}V_{ab}\|_F}.
\end{equation}
For $ab=12,23$ and $J=0.0005,0.001,0.002,0.004,0.008$, $\max
d_{ab}=2.27\times10^{-15}$. This identifies the local fusion outputs with
the microscopic exchange eigenchannels. The separately reported
chart-covariance calculation uses the $12$ and $34$ probes.

\paragraph{Fixed-endpoint deformation.}

On the edge $T_1\to T_5$, a transverse resolver term $J\sin(\pi u)Q_{13}$, $%
0\leq u\leq1$, preserves both endpoints and changes the enclosed
resolver-plane area. The two loop families give 
\begin{equation}
\epsilon_{\mathrm{pent}}\simeq0.0051838\,J^2,\qquad \epsilon_{\mathrm{detour}%
}\simeq0.0271648\,J^2.  \label{eq:resolver-detour-scaling}
\end{equation}
These coefficients summarize the displayed finite-mesh sequence. At $%
J=5\times10^{-4}$ the respective values are $1.296\times10^{-9}$ and $%
6.791\times10^{-9}$. The factor $5.24$ between coefficients resolves the
dependence on the physical resolver path at fixed endpoints. Both families
converge to the same projective reassociation structure as the resolver is
removed, while the finite-$J$ coefficient measures its continuous geometric
embedding. The conventional Ising associator supplies the exact discrete
categorical benchmark used in the Hexagon calculation.

% END fusion_channel_audit.tex

% Pre-position curvature float for two-column layout.
\begin{figure*}[t]
\centering
\includegraphics[width=0.93\textwidth]{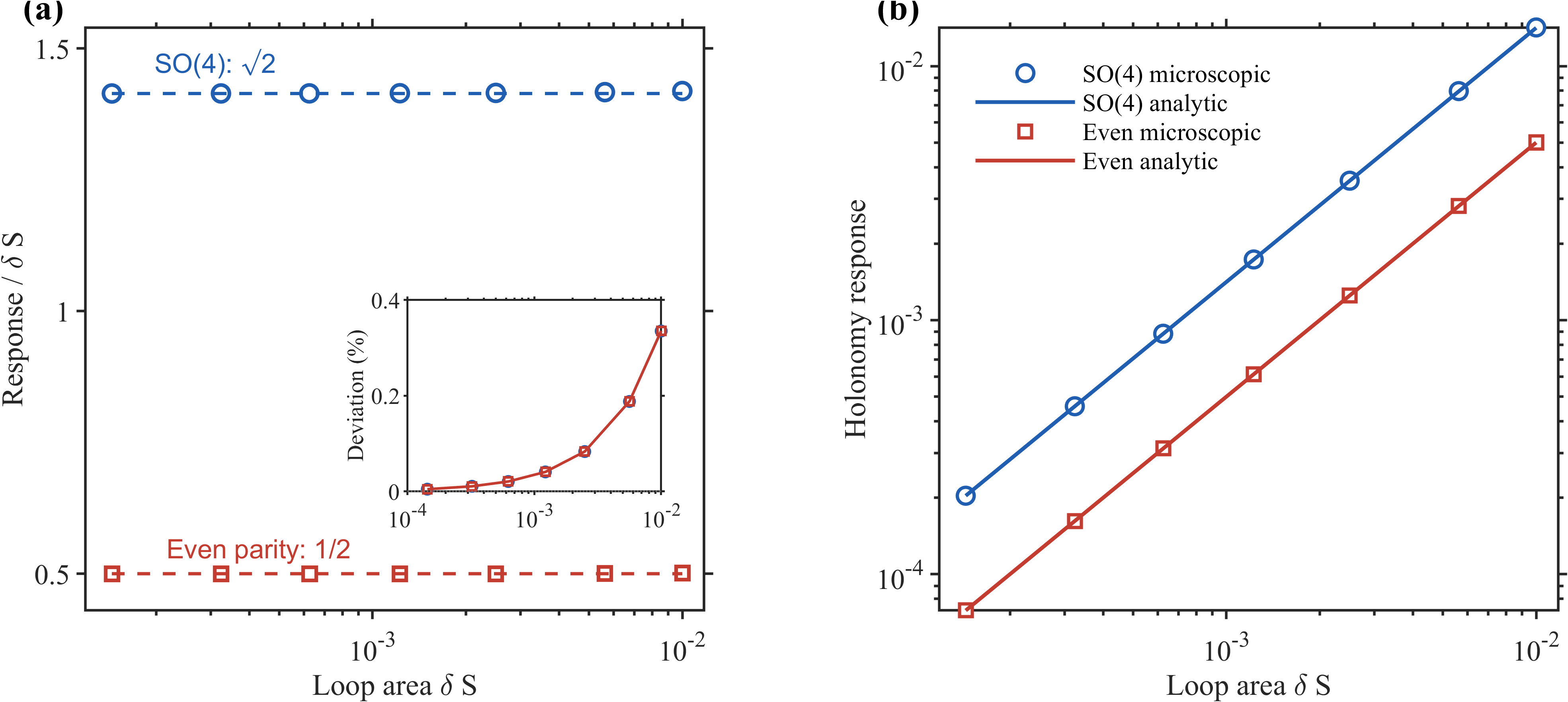}
\caption{\textbf{Analytic curvature and full-chain microscopic holonomy.}
For seven square loops on the real control patch, the exact
$46\times46$ Majorana Hamiltonian supplies the four-mode Kato frame
and its even-parity Spin lift. (a) The normalized coefficients approach
$\epsilon_{SO}/\delta S=\sqrt2$ and
$\epsilon_+/\delta S=1/2$; the inset displays finite-area deviations.
(b) The microscopic loop responses agree with the independently
evaluated linear-area predictions. Here
$\epsilon_{SO}=\|O_{\delta C}-I_4\|_F$ and
$\epsilon_+=\|W^{(+)}_{\delta C}-I_2\|_F/\sqrt2$.
Symbols denote full-chain computations and lines the local analytic laws.}
\label{fig:sup-final-curvature}
\end{figure*}

\subsection{Exact projective-curvature obstruction}

\label{sup:quadric-flatness} 
% ============================================================

For identical finite arms with the resolver switched off,
Proposition~\ref{prop:finite-arm-projector} gives the exact embedding
$P_{\rm full}=ZP_0Z^T$ with constant $Z$. The control-curvature calculation
therefore applies directly to the full finite-chain dark bundle under the
spectral assumptions of that proposition. The normalized projector is 
\begin{equation}
P_0(n) = I-nn^T, \qquad n^Tn=1.  \label{eq:sup-quadric-P}
\end{equation}
For 
\begin{equation}
X,Y\in T_nQ_{m-1}, \qquad n^TX=n^TY=0,
\end{equation}
one has 
\begin{equation}
\partial_XP_0 = -Xn^T-nX^T.
\end{equation}
A direct calculation gives 
\begin{equation}
P_0 [\partial_XP_0,\partial_YP_0] P_0 = XY^T-YX^T,
\end{equation}
with 
\begin{equation}
\func{Tr}(XY^T-YX^T)=0.
\end{equation}
Hence 
\begin{equation}
\mathcal{F}_{\mathrm{proj}}(X,Y) = \mathcal{F}_K(X,Y) = XY^T-YX^T.
\label{eq:sup-quadric-F}
\end{equation}
For linearly independent tangent vectors this endomorphism is nonzero.
Therefore no full-dimensional regular open patch of the present
complex-quadric control manifold is projectively flat.

The obstruction survives the physical parity reduction. At 
\begin{equation}
n=e_5, \qquad X=e_1, \qquad Y=e_2,
\end{equation}
one obtains 
\begin{equation}
\mathcal{F}_{\mathrm{proj}}(e_1,e_2) = E_{12}-E_{21}.
\end{equation}
Its Spin lift is represented by 
\begin{equation}
\frac12\gamma_1\gamma_2,
\end{equation}
which acts on the fixed-parity logical doublet as 
\begin{equation}
\pm \frac{i}{2}\sigma_z
\end{equation}
up to basis and sign conventions, and is therefore not projectively scalar.

For a contractible square of side $\epsilon$ in the $(e_1,e_2)$ tangent
plane, 
\begin{equation}
W_\square-I = -\epsilon^2(E_{12}-E_{21}) + O(\epsilon^3),
\end{equation}
so that 
\begin{equation}
\frac{ \|W_\square-I\|_F }{ \epsilon^2 } \longrightarrow \sqrt2.
\label{eq:sup-square-limit}
\end{equation}
The ordered-projector calculation gives

\begin{center}
\refstepcounter{table}\label{tab:sup-curvature} {\small \textbf{TABLE %
\thetable.} Reported small-loop curvature diagnostic in the dark-vector
representation.}

\medskip {\small 
\begin{tabular}{cc}
\toprule $\epsilon$ & $\|W_\square-I\|_F/\epsilon^2$ \\ 
\midrule 0.10 & 1.41894 \\ 
0.05 & 1.41539 \\ 
0.02 & 1.41440 \\ 
0.01 & 1.41426 \\ 
\bottomrule & 
\end{tabular}%
}
\end{center}

in agreement with $\sqrt2=1.41421356\ldots$.

The finite path-shape dependence observed in the microscopic controls is the
finite-loop manifestation of the exact nonzero projective curvature, and the
small-loop sequence converges to the analytic $\sqrt2$ coefficient.

% ============================================================

% Pre-position final audit float to keep it ahead of references.
\begin{figure*}[!t]
\centering
\includegraphics[width=0.93\textwidth]{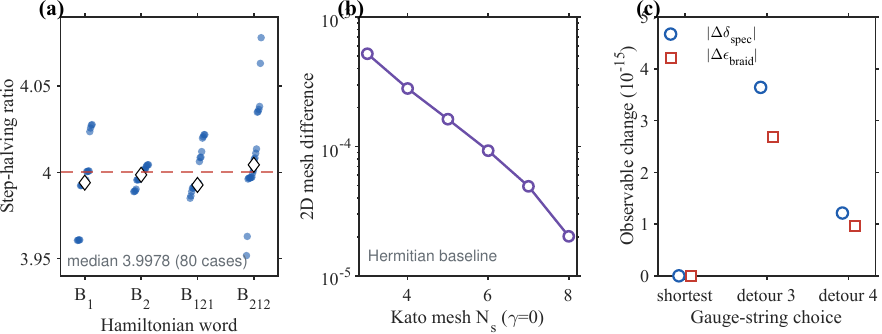}
\caption{\textbf{Numerical convergence and gauge controls.}
(a) Time-step convergence of the finite-time exchange
calculations, with the second-order reference ratio $4$.
(b) Kato-mesh convergence for the Hermitian
($\gamma=0$) honeycomb baseline.
(c) Invariance of the computed exchange observables
under contractible gauge-string transformations.}
\label{fig:sup-final-audit}
\end{figure*}

\subsection{Full finite-arm small-loop curvature certificate}
\label{sup:fullchain-curvature}
The full finite-chain calculation uses the $46$-Majorana matrix of
Eq.~\eqref{eq:sup-full-matrix-entries}, with the same reference parameters
and the resolver switched off. At the base point $n_0=e_5$, the local
control patch is
\begin{equation}
n(x,y)=(x,y,0,0,\sqrt{1-x^2-y^2})^T.
\end{equation}
A positively oriented square has vertices
$(0,0)\to(\ell,0)\to(\ell,\ell)\to(0,\ell)\to(0,0)$ and coordinate area
$\delta S=\ell^2$. Its full-chain sequence is
\begin{equation}
\delta C\longrightarrow A[n(x,y)]\longrightarrow P(x,y)
\longrightarrow O_{\delta C}\longrightarrow W^{(+)}_{\delta C}.
\end{equation}
At every node, diagonalization of $iA$ selects the four eigenvalues nearest
zero and constructs their spectral projector. Since this patch is real,
$iA$ is Hermitian and the orthonormal spectral projector is also the Riesz
projector. Equation~\eqref{eq:sup-kato-step-corrected} transports the frame;
closure uses the common base frame $Q_0$ obtained by projecting the first
site of arms $1$--$4$. The calculation builds the complete matrix at each
node; the analytic dark projector supplies an independent comparison.

For the small loops, the logarithm $K=\log O_{\delta C}$ is continued from
zero and projected to its antisymmetric part. With
\begin{equation}
\begin{aligned}
\gamma_1&=\sigma_x\otimes I,&\gamma_2&=\sigma_y\otimes I,\\
\gamma_3&=\sigma_z\otimes\sigma_x,&
\gamma_4&=\sigma_z\otimes\sigma_y,
\end{aligned}
\end{equation}
the even-parity logical lift is the $\{|00\rangle,|11\rangle\}$ block of
$\exp[\tfrac12\sum_{a<b}K_{ab}\gamma_a\gamma_b]$.
The response norms are
\begin{equation}
\epsilon_{SO}=\|O_{\delta C}-I_4\|_F,\qquad
\epsilon_+=\|W^{(+)}_{\delta C}-I_2\|_F/\sqrt2.
\end{equation}
The analytic curvature predicts
\begin{equation}
\epsilon_{SO}/\delta S\to\sqrt2,\qquad
\epsilon_+/\delta S\to1/2.
\label{eq:sup-two-curvature-coefficients}
\end{equation}
The small-loop response shown in main-text Fig.~5(c) is the full-chain
holonomy test; the $SO(4,\mathbb C)$ and even-parity logical norms are specified
separately above and in Fig.~\ref{fig:sup-final-curvature}.
Their leading area coefficients are respectively $\sqrt2$ and $1/2$.
The numerical record retains both frame-level and even-parity
responses, with their distinct normalizations fixed by the equations
above and the corresponding figure axes.
The leading coefficient is fixed locally, while the finite-area correction
records the specified embedding of the square in the control quadric.

A log--log fit over $\ell\leq0.05$ gives slopes
$p_{SO}=1.00026433$ and
$p_+=1.00026439$ with respect to area.
The minimum external gap is $0.378502013189$,
the maximum cluster radius is $1.03\times10^{-14}$,
and the maximum projector idempotency residual is
$3.37\times10^{-13}$.
The maximum relative 40/80 mesh difference is
$7.87\times10^{-7}$ (archived independently of the plotted response).
These separately recorded diagnostics resolve the linear area response
throughout the scan. The mesh and roundoff diagnostics refer to the 40/80-node calculation.
The saved matrices fix the common-frame and Spin conventions for this
finite-chain certificate.

\section{Data Provenance and Reproducibility}
\setcounter{equation}{0}
\label{sup:provenance}

The microscopic calculations use the five-arm Majorana
network and the two-dimensional Kitaev model, with
finite-time propagation and fusion-sewing protocols.
We extract exchange spectra, relative eigenline geometry,
and logical operations from the microscopic transport
matrices and compare them with analytic predictions.
The calculations specify the isolated spectral cluster,
endpoint gauge, common logical basis, continuous Spin
branch, eigenline ordering, and residual normalization.
Finite-time calculations distinguish uninterrupted
exchange words from products of logical return blocks.

Analytic identities, microscopic geometric transport,
and finite-time propagation provide distinct checks
of the statistical structure. The five-arm realization
uses analytically designed controls; the honeycomb
bond texture is fixed before invariant extraction.
The code and numerical data supporting the results
are available from the corresponding author upon
reasonable request.